\documentclass[onecolumn]{emulateapj}
\usepackage{longtable}
\usepackage{graphicx}
\usepackage{graphics}
\usepackage{color}
\usepackage{epsfig}
\usepackage{rotating}
\usepackage{subfigure}
\usepackage{float}
\shortauthors{}
\shorttitle{}
\begin{document}

\title{\emph{Chandra} Survey of Nearby Highly-Inclined Disk Galaxies I:\\X-ray Measurements of Galactic Coronae}

\author{Jiang-Tao Li\altaffilmark{1} and Q. Daniel Wang\altaffilmark{1}} \altaffiltext{1}{Department of Astronomy, University of
Massachusetts, 710 North Pleasant Street, Amherst, MA 01003, U.S.A.}


\nonumber

\begin{abstract}
We present a systematical analysis of the \emph{Chandra} observations of 53 nearby highly-inclined ($i\gtrsim60^\circ$) disk galaxies to study the coronae around them. This sample covers a broad range of galaxy properties: e.g., about three orders of magnitude in the SFR and more than two orders of magnitude in the stellar mass. The \emph{Chandra} observations of the \emph{diffuse} soft X-ray emission from 20 of these galaxies are presented for the first time. The data are reduced in a uniform manner, including the excision/subtraction of both resolved and unresolved stellar contributions. Various coronal properties, such as the scale height and luminosity, are characterized for all the sample galaxies. For galaxies with high enough counting statistics, we also examine the thermal and chemical states of the coronal gas. We note on galaxies with distinct multi-wavelength characteristics which may affect the coronal properties. The uniformly processed images, spectra, and brightness profiles, as well as the inferred hot gas parameters, form a large X-ray database for studying the coronae around nearby disk galaxies. We also discuss various complications which may cause biases to this database and their possible corrections or effects, such as the uncertainty in the thermal and chemical states of hot gas, the different galactic disk inclination angles, the presence of AGN, and the contribution of the emission from charge exchange at interfaces between hot and cool gases. Results from a detailed correlation analysis are presented in a companion paper, to gain a more comprehensive statistical understanding of the origin of galactic coronae.
\end{abstract}

\section{Introduction}\label{PaperIsec:Introduction}

Large-scale diffuse soft X-ray emission (galactic coronae) has been detected around various types of nearby galaxies (e.g., \citealt{Wang10} and references therein). They include elliptical galaxies containing active galactic nuclei (AGN) (e.g., \citealt{Kraft00,Forman05}) or not (e.g., \citealt{OSullivan03,David06,Boroson11}), normal disk galaxies with little star formation (SF) (e.g., \citealt{LiZ06,Rasmussen09,Li11}), galaxies with active SF spread widely over their disk (e.g., \citealt{Bregman94,Tullmann06a,Li08}), nuclear starburst galaxies (e.g., \citealt{Strickland04a,Grimes05}), dwarf galaxies (e.g., \citealt{Martin02,Hartwell04}), and interacting or merging galaxies (e.g., \citealt{Fabbiano01,Zezas03,Machacek04}), etc. The presence of such galactic coronae is a manifestation of the energetic interplay between galaxies and their surroundings, although the exact origin of the X-ray emission remains very uncertain.

In X-ray observations of \emph{active SF} galaxies, the coronal luminosity is observed to be tightly (nearly linearly) correlated with the SF rate (SFR; e.g., \citealt{Strickland04b,Grimes05}). Furthermore, the coronal gas often shows metallicity patterns consistent with being enriched by massive stellar objects [e.g., massive stellar winds and core collapsed (CC) SNe] (e.g., \citealt{Martin02,Yamasaki09,Konami11}). These strongly suggest that the coronal gas in star forming galaxies is mainly generated by SF feedback. In addition, deep observations of some nuclear starburst galaxies have revealed a high-temperature ($T\gtrsim10^7\rm~K$) thermal component, which is usually not detected because of its low density and hence low emissivity, compared to the dominant soft X-ray-emitting gas of typical temperature of a few $\times10^6\rm~K$ \citep{Strickland09}. This high-temperature component is thought to be produced by the SN-driven superwinds themselves, which carry the bulk of the SN feedback energy and interact with the cool interstellar medium (ISM), producing much of the soft X-ray emission at the interfaces \citep{Strickland00a,Strickland00b}, although this view is still far from being widely accepted (e.g., see \citealt{Bauer07}).

In contrast, the diffuse soft X-ray luminosity ($L_X$) of SF-quiescent \emph{elliptical} galaxies is shown to be closely correlated with the stellar mass ($M_*$). Furthermore, the slope of the $L_X-M_*$ relation is shown to be far larger than unity, especially for massive galaxies (e.g., \citealt{Forman85,Canizares87,Helsdon01,Mathews03,OSullivan03,Boroson11}), strongly suggesting that gravity plays an important role in heating and/or confining the coronal gas. Metallicity measurements further indicate that the hot gas in elliptical galaxies (maybe except for those located in cluster environments) is primarily enriched by old stars (via Type~Ia SNe and stellar mass loss, e.g., \citealt{Kim04,Humphrey04,Humphrey06,Ji09}). The modeling of this stellar feedback shows that galactic scale outflow are most likely responsible for removing a large fraction of the SN energy, and chemically-enriched mass from central regions of low- and intermediate-mass elliptical galaxies (e.g., \citealt{Tang10}). But both the dynamics and ultimate fate of the outflows are sensitive to the feedback history, as well as to the gravitational confinement of the galaxies (e.g., \citealt{David06,Tang09a,Tang09b}).

The origin of the coronal gas remains very uncertain for relatively \emph{quiescent disk} galaxies (such as the Milky Way and M31). Indeed, the hierarchical galaxy formation scenario definitively predicts the presence of coronae around galaxies with individual halo masses greater than a few $\times10^{11}\rm~M_\odot$ (e.g., \citealt{White91,Benson00,Crain10a}). Such coronae are formed primarily from the (virial) shock-heating and subsequent gravitational compression of gas accreted from the intergalactic medium (IGM). But non-gravitational heating is also expected to be important. Observationally, non-nuclear-starburst galaxies, for example, seem to follow the same $L_X-{\rm SFR}$ relation as nuclear starburst ones \citep{Tullmann06b,Li08}. Even in passively evolving bulge-dominated disk galaxies (e.g., S0 galaxies) which are often quiescent, non-gravitational feedback can still be significant, primarily in forms of Ia SNe and mass-loss of evolved low-mass stars (e.g., \citealt{LiZ07a,Li09,Li11}). It is thus of great interest to check if the coronae around quiescent disk galaxies represent the interplay between gravitational and non-gravitational effects, and what roles different heating mechanisms play in producing galactic coronae in and around various types of galaxies.

We herein present a comprehensive \emph{Chandra} data analysis for a sample of 53 nearby highly-inclined \emph{disk} galaxies, together with multi-wavelength measurements of other galaxy properties, such as the SFR, stellar mass, and dynamical mass. Our sample includes all kinds of disk galaxies and covers a large range of galaxy properties. In comparison, similar existing studies of nearby highly-inclined disk galaxies, based on \emph{Chandra} and \emph{XMM-Newton} observations, all have sample sizes less than about 10 galaxies \citep{Strickland04a,Strickland04b,Tullmann06a,Tullmann06b} (Fig.~\ref{fig:sample}). Elliptical galaxies are relatively well studied (e.g., \citealt{Boroson11}) and the results will be compared with those from the present study in the companion paper (\citealt{Li12b}, hereafter Paper~II). The large sample size of the disk galaxies in the present study will enable us to examine the similarities and differences between various types of galaxies. By conducting a multi-variable statistical analysis, we will also examine how the coronal properties are affected by other processes, such as the environmental effects (e.g., \citealt{Mulchaey10}) and the cool-hot gas interaction (e.g., \citealt{Strickland02,Li11}).

This analysis of archival \emph{Chandra} data forms part of a long-term multi-wavelength investigation of galactic halos and their interplay with galactic disks. In particular, an extensive radio survey of 35 nearby highly-inclined disk galaxies, which represents a volume-limited sample, is being carried out with the \emph{EVLA} (the \emph{CHANG-ES} project; \citealt{Irwin12a,Irwin12b}). 16 of these galaxies are included in the present \emph{Chandra} analysis, which represents the first step to provide the X-ray data coverage of the \emph{CHANG-ES} galaxies. This archival study represents a necessary preparation for proposing new X-ray observations for deep exposures or for additional galaxies. With the superb spatial resolution, \emph{Chandra} observations are ideal for detecting and subtracting point-like sources, mapping out galactic coronae, particularly in regions close to the galactic disks. An analysis program, using existing and new \emph{XMM-Newton} observations, is also planned, which can provide complementary capabilities (e.g., greater counting statistics and field coverage) and are better suited for spectral analysis of large-scale X-ray emission.

The present paper utilizes mainly the \emph{Chandra} data for the detection and excision of X-ray point sources and the characterization of the galactic coronae  in and around the sample galaxies. We report the X-ray measurements of the coronae in this paper. In \S\ref{PaperIsec:SampleSelection} we introduce the sample selection criteria and the definitions of subsamples. The data reduction and analysis are detailed in \S\ref{PaperIsec:DateReduction}, providing uniform measurements of the diffuse soft X-ray properties for all the sample galaxies. Notes on individual galaxies of distinct properties are given in \S\ref{PaperIsec:Individual}. We discuss uncertainties related to the data analysis in \S\ref{PaperIsec:Discussion}. The main results are summarized in \S\ref{PaperIsec:summary}. 

Paper II presents (1) a detailed correlation analysis between the X-ray properties (e.g., luminosity, temperature, and abundance ratio) and multi-wavelength galaxy properties (e.g., SF properties, galaxy mass, and morphology) for different types of galaxies; (2) a comparison of our measurements (e.g., the coronal luminosity and/or temperature) with those from cosmological simulations or \emph{Chandra} observations of elliptical galaxies; (3) an estimate of the X-ray radiation efficiency of the galactic feedback; and (4) an investigation of the origin of the coronal emission.

\section{Sample Selection and Multi-wavelength Galaxy Properties}\label{PaperIsec:SampleSelection}

\subsection{Selection Criteria}\label{PaperIsubsec:criteria}

We are aiming at understanding the relative importance of various physical processes in generating galactic coronae in and around different types of disk galaxies. We thus need a large sample of galaxies with all types of SF properties, masses, morphologies, and formation environments. Limited by the availability of existing \emph{Chandra} observations, our sample selection for the present study is thus related to, but not limited by, the \emph{CHANG-ES} sample \citep{Irwin12a,Irwin12b}, with the following criteria:

(1) \emph{Morphological classification.} The present study includes both spiral and lenticular galaxies with optical morphological type codes of $-3\lesssim TC\lesssim9$ (Table~\ref{table:SampleSelectionPara}). This means that some interacting galaxies may be included. These galaxies are sometimes observed to have quite different X-ray properties from the isolated ones (e.g., \citealt{Machacek04,Read05}), and may provide us with a laboratory to study how galaxy interaction could regulate the coronal properties.

(2) \emph{Inclination angle.} With the inclination angle restricted to $i\gtrsim60^\circ$, the sample includes not only nearly perfect edge-on galaxies, but also some moderately inclined ones. For the latter ones, additional inclination corrections may be needed to uniformly compare them with other galaxies in the sample. But in general, the selection of only highly inclined disk galaxies minimizes the confusion of the soft X-ray emission from discrete sources in galactic disks (young stars, supernova remnants, etc.). The present study thus preferentially samples the diffuse soft X-ray emission from regions outside galactic disks with dense cool gas.

(3) \emph{Distance.} All the selected galaxies have distances $d\lesssim30\rm~Mpc$ so that the bulk of luminous point sources can be individually detected and that the disk/halo emission can be reasonably separated. The distances of the galaxies are obtained based on various redshift independent methods, as listed in Table~\ref{table:distance}.

(4) \emph{Extinction.} All the galaxies are selected to have the foreground \ion{H}{1} column density $N_H\lesssim8\times10^{20}\rm~cm^{-2}$, minimizing the Galactic extinction in soft X-ray.

(5) \emph{Size.} All the galaxies have optical diameters $D_{25}\lesssim16^\prime$ so that a single \emph{Chandra} observation could typically cover the bulk of the coronal emission. $D_{25}$ is also limited to $\gtrsim1^\prime$ so that the disk and halo can be well separated.

(6) \emph{Chandra data.} Each of the selected galaxies has a total non-grating \emph{Chandra}/ACIS exposure $t_{exp}\gtrsim10\rm~ks$. We only include observations with individual exposures $\gtrsim10\rm~ks$ (Table~\ref{table:ChandraData}). Galaxies with only multiple snapshot observations, even with a total exposure $\gtrsim10\rm~ks$, are not included, except for NGC~660 ($\sim7.1\rm~ks$) and NGC~4666 ($\sim5\rm~ks$), which are known to have high SFRs and are expected to have strong diffuse X-ray emission.

(7) \emph{AGN contamination.} We do \emph{not} select our sample against galaxies containing AGN. Nevertheless, we have excluded one galaxy (NGC~2992) from our original sample, because the scattered X-ray light from its AGN (wings of point-spread function, as well as a severe CCD readout steak) is too bright to allow for an effective study of the galactic corona (the thermal component is $<5\%$ of the total diffuse X-ray emission in the near-nuclear region of $3^{\prime\prime}<r<18^{\prime\prime}$; \citealt{Colbert05}). Other galaxies may contain AGN; but their presence does not seriously affect the data reduction.

%\clearpage
\begin{deluxetable}{lcccccccccccccc}
\centering
\tiny %\tiny\scriptsize\footnotesize\small\normalsize\large\Large\LARGE\huge\Huge
%\ptlandscape
  \tabletypesize{\tiny}
  \tablecaption{Parameters of the Sample Galaxies (I)}
  \tablewidth{0pt}
  \tablehead{
 \colhead{Name} & \colhead{Type} & \colhead{TC} & \colhead{i} & \colhead{d} & \colhead{$N_H$} & \colhead{$D_{25}$} & \colhead{$r_{25}$} & \colhead{B-V} & \colhead{$v_{rot}$} & \colhead{$\rho$} & \colhead{$M_{H_2}$} & \colhead{$M_{HI}$} \\
   &  &  & (deg) & (Mpc) & ($10^{20}\rm cm^{-2}$) & ($\prime$) &  & (mag) & ($\rm km~s^{-1}$) & ($\rm Mpc^{-3}$) & ($10^{8}\rm M_\odot$) & ($10^{8}\rm M_\odot$) \\
   & (1) & (2) & (3) & (4) & (5) & (6) & (7) & (8) & (9) & (10) & (11) & (12)
}
\startdata
IC2560      & SBb  & $3.4\pm0.6$  & 65.63 & 29.2 & 6.51  & 3.55 & 2.13 & -     & $196\pm3$  & -    & -     & -      \\
M82$^a$     & Sd   & $8.1\pm3.5$  & 79.4  & 3.53 & 3.98  & 11.0 & 2.15 & 0.674 & $100\pm10^*$& 0.16& -     & -      \\
NGC24       & Sc   & $5.1\pm0.4$  & 70.11 & 9.08 & 2.27  & 6.18 & 2.57 & 0.480 & $94\pm1$   & 0.12 & -     & -      \\
NGC520$^b$  & Sa   & $0.8\pm2.7$  & 77.45 & 27.8 & 3.27  & 4.09 & 2.55 & 0.699 & $72\pm2$   & 0.25 & -     & -      \\
NGC660      & SBa  & $1.2\pm1.1$  & 78.85 & 14.7 & 4.86  & 4.57 & 2.71 & 0.710 & $141\pm3$  & 0.12 & -     & -      \\
NGC891$^a$  & Sb   & $3.0\pm0.3$  & 88    & 9.95 & 7.64  & 13.0 & 4.31 & 0.697 & $212\pm5$  & 0.55 & 71.12 & 81.66  \\
NGC1023$^c$ & E-S0 & $-2.7\pm0.6$ & 76.7  & 11.6 & 7.16  & 7.40 & 2.42 & 0.911 & $113\pm5$  & 0.57 & -     & -      \\
NGC1380     & S0   & $-2.3\pm0.7$ & 90    & 21.2 & 1.31  & 4.58 & 2.07 & 0.882 & -          & 1.54 & -     & -      \\
NGC1386$^d$ & S0-a & $-0.8\pm0.9$ & 90    & 15.3 & 1.37  & 3.59 & 2.67 & 0.776 & -          & 1.36 & 2.93  & $<$0.40\\
NGC1482$^a$ & S0-a & $-0.9\pm0.5$ & 63.59 & 19.6 & 3.69  & 2.46 & 1.72 & 0.860 & $121\pm8$  & 0.31 & 31.82 & 5.16   \\
NGC1808$^e$ & SABa & $1.2\pm0.5$  & 83.91 & 12.3 & 2.70  & 5.42 & 2.96 & 0.690 & $122\pm5$  & 0.30 & -     & -      \\
NGC2787$^f$ & S0-a & $-1.1\pm0.8$ & 66.21 & 13.0 & 4.32  & 3.24 & 1.79 & 0.891 & $182\pm14$ & 0.06 & 0.24  & 7.70   \\
NGC2841     & Sb   & $3.0\pm0.4$  & 68    & 14.1 & 1.45  & 6.90 & 2.08 & 0.792 & $319\pm9$  & 0.13 & -     & -      \\
NGC3079$^a$ & SBcd & $6.7\pm0.9$  & 82.5  & 16.5 & 0.789 & 8.18 & 6.34 & 0.512 & $210\pm5$  & 0.29 & 93.56 & 89.35  \\
NGC3115$^f$ & E-S0 & $-2.9\pm0.5$ & 81.6  & 9.77 & 4.32  & 7.10 & 2.36 & 0.899 & $108\pm6$  & 0.08 & 0.05  & 5.74   \\
NGC3198     & Sc   & $5.2\pm0.6$  & 70    & 14.5 & 1.02  & 6.46 & 3.51 & 0.421 & $148\pm4$  & 0.15 & -     & -      \\
NGC3384$^c$ & E-S0 & $-2.7\pm0.8$ & 90    & 11.8 & 2.73  & 5.24 & 2.21 & 0.875 & 17         & 0.54 & -     & -      \\
NGC3412     & S0   & $-2.0\pm0.5$ & 71.92 & 11.5 & 2.59  & 3.96 & 1.83 & 0.850 & -          & 0.52 & 0.06  &$<$0.22 \\
NGC3521     & SABb & $4.0\pm0.2$  & 65.5  & 11.2 & 4.06  & 8.32 & 1.86 & 0.698 & $233\pm7$  & 0.19 & 77.23 & 90.74  \\
NGC3556$^g$ & SBc  & $6.0\pm0.3$  & 67.51 & 10.7 & 0.794 & 3.98 & 2.40 & 0.568 & $153\pm3$  & 0.15 & 14.58 & 49.39  \\
NGC3628$^a$ & Sb   & $3.1\pm0.4$  & 79.29 & 13.1 & 2.22  & 11.0 & 3.21 & 0.675 & $215\pm4$  & 0.39 & -     & -      \\
NGC3877$^h$ & Sc   & $5.1\pm0.5$  & 83.24 & 14.1 & 2.22  & 5.36 & 4.36 & 0.654 & $155\pm4$  & 1.53 & 10.28 & 12.64  \\
NGC3955     & S0-a & $0.2\pm0.7$  & 90    & 20.6 & 4.74  & 4.15 & 3.44 & 0.522 & 86         & 0.08 & -     & -      \\
NGC3957     & S0-a & $-1.1\pm0.6$ & 90    & 27.5 & 3.63  & 3.24 & 5.32 & -     & -          & 0.48 & -     &$<$43.32\\
NGC4013     & Sb   & $3.0\pm0.3$  & 90    & 18.9 & 1.39  & 4.89 & 3.96 & 0.829 & 182        & 1.34 & -     & -      \\
NGC4111     & S0-a & $-1.4\pm0.7$ & 84.2  & 15.0 & 1.40  & 1.78 & 2.79 & 0.815 & $72\pm4$   & 1.09 & 0.17  & 7.21   \\
NGC4217     & Sb   & $3.0\pm0.3$  & 81.05 & 19.5 & 1.23  & 5.43 & 3.37 & 0.750 & $188\pm4$  & 0.95 & -     & -      \\
NGC4244$^a$ & Sc   & $6.1\pm0.5$  & 88    & 4.37 & 1.67  & 16.2 & 2.24 & 0.412 & $89\pm2$   & 0.39 & 0.60  & 26.77  \\
NGC4251$^d$ & S0   & $-1.9\pm0.7$ & 74.56 & 19.6 & 1.84  & 2.34 & 1.91 & 0.800 & -          & 1.20 &$<$0.35&$<$0.57 \\
NGC4342$^r$ & E-S0 & $-3.4\pm0.9$ & 90    & 16.8 & 1.60  & 1.25 & 1.96 & 0.964 & -          & 2.64 & -     &$<$0.76 \\
NGC4388$^i$ & Sb   & $2.8\pm0.6$  & 82    & 17.1 & 2.60  & 5.38 & 4.18 & 0.575 & $173\pm5$  & 1.56 & 6.07  & 5.67   \\
NGC4438$^j$ & Sa   & $0.6\pm1.4$  & 73.29 & 14.4 & 2.66  & 9.16 & 2.28 & 0.756 & $167\pm10$ & 2.67 & -     & -      \\
NGC4501     & Sb   & $3.4\pm0.7$  & 61    & 15.7 & 2.48  & 8.65 & 1.98 & 0.623 & $276\pm8$  & 2.04 & 83.00 & 23.39  \\
NGC4526$^c$ & S0   & $-1.9\pm0.4$ & 82.8  & 17.2 & 1.65  & 6.95 & 2.81 & 0.890 & $152\pm9$  & 2.45 & 6.03  & 24.56  \\
NGC4565$^h$ & Sb   & $3.2\pm0.5$  & 90    & 11.1 & 1.30  & 16.7 & 5.79 & 0.677 & $245\pm6$  & 1.00 & 38.94 & 109.75 \\
NGC4569     & SABa & $2.4\pm0.6$  & 66    & 9.86 & 2.49  & 9.12 & 2.39 & 0.605 & $177\pm9$  & 1.15 & -     & -      \\
NGC4594$^k$ & Sa   & $1.1\pm0.3$  & 78.5  & 9.77 & 3.77  & 8.45 & 1.72 & 0.878 & $358\pm10$ & 0.32 &$<$15.65& 3.50  \\
NGC4631$^l$ & SBcd & $6.5\pm0.7$  & 85    & 7.62 & 1.27  & 14.5 & 6.58 & 0.386 & $139\pm4$  & 0.41 & -     & -      \\
NGC4666     & SABc & $5.1\pm0.8$  & 69.58 & 15.7 & 1.74  & 4.97 & 2.52 & 0.641 & $193\pm2$  & 0.54 & 79.70 & 43.80  \\
NGC4710     & S0-a & $-0.9\pm0.8$ & 90    & 16.8 & 2.14  & 4.39 & 3.88 & 0.766 & 160        & 2.00 & 8.47  & 0.77   \\
NGC5102$^m$ & E-S0 & $-2.8\pm0.8$ & 90    & 3.18 & 4.33  & 9.71 & 2.62 & 0.638 & 90         & 0.17 &$<$0.08& 2.54   \\
NGC5170$^n$ & Sc   & $4.8\pm0.6$  & 90    & 22.5 & 6.95  & 7.96 & 5.65 & 0.705 & 245        & 0.20 & -     & -      \\
NGC5253$^o$ & S?   & $7.6\pm4.6$  & 66.7  & 4.07 & 3.87  & 5.01 & 2.36 & 0.300 & $38\pm2$   & 0.18 & -     & 2.34   \\
NGC5422     & S0-a & $-1.5\pm0.9$ & 90    & 30.9 & 1.16  & 2.81 & 4.79 & 0.900 & -          & 0.36 & -     & 34.87  \\
NGC5746$^n$ & SABb & $3.0\pm0.3$  & 83.9  & 24.7 & 3.27  & 7.24 & 6.73 & 0.769 & $319\pm10$ & 0.83 & -     & -      \\
NGC5775$^p$ & SBc  & $5.2\pm0.7$  & 83.22 & 26.7 & 3.48  & 3.70 & 4.38 & 0.660 & $190\pm7$  & 0.67 & 67.59 & 128.79 \\
NGC5866$^q$ & S0-a & $-1.3\pm0.7$ & 90    & 15.3 & 1.46  & 6.31 & 2.32 & 0.785 & -          & 0.24 & 7.71  &$<$3.00 \\
NGC6503$^a$ & Sc   & $5.9\pm0.5$  & 73.5  & 5.27 & 4.09  & 5.94 & 3.00 & 0.559 & $67\pm2$   & 0.08 & 2.84  & 13.91  \\
NGC6764$^s$ & SBbc & $3.6\pm0.6$  & 63.86 & 26.2 & 6.07  & 2.49 & 2.04 & 0.571 & $140\pm4$  & 0.08 & -     & -      \\
NGC7090$^t$ & SBc  & $5.0\pm0.5$  & 90    & 6.28 & 2.76  & 8.15 & 5.00 & 0.451 & 102        & 0.13 & -     & -      \\
NGC7457     & E-S0 & $-2.6\pm0.6$ & 73.05 & 13.2 & 5.56  & 3.95 & 1.75 & 0.815 & -          & 0.13 & 0.05  &$<$0.73 \\
NGC7582$^u$ & SBab & $2.1\pm0.5$  & 68.17 & 23.0 & 1.93  & 6.95 & 2.19 & 0.661 & $195\pm3$  & 0.39 & -     & -      \\
NGC7814     & Sab  & $2.0\pm0.2$  & 70.59 & 18.1 & 3.71  & 4.37 & 2.32 & 0.872 & $231\pm8$  & 0.12 &$<$14.56& 17.10
\enddata
\tablecomments{\scriptsize Listed basic parameters of the galaxies: (1) morphological type, (2) type code \citep{DeVaucouleurs91}, and (3) galactic disk inclination angle, obtained from the \emph{HyperLeda} database (\emph{http://leda.univ-lyon1.fr/}); (4) distances measured using various redshift independent methods, obtained from the \emph{NED} database (\emph{http://nedwww.ipac.caltech.edu/}; see Table~\ref{table:distance} for details); (5) weighted average foreground \ion{H}{1} column densities, obtained from the \ion{H}{1} survey of \citet{Dickey90} (from the HEASARC web tools: \emph{http://heasarc.gsfc.nasa.gov/docs/tools.html}); (6) B-band diameter of the projected major axis at the isophotal level $25\rm~mag~arcsec^{-2}$, (7) axis ratio of the isophote $25\rm~mag~arcsec^{-2}$, (8) B-V color corrected for galactic extinction, inclination and redshift effect, and (9) maximum rotation velocity corrected for inclination (*: for M82, from \citet{Westmoquette09}, see text for details), obtained from \emph{
HyperLeda}; (10) densities of galaxies brighter than -16~mag in the vicinity, obtained from the Nearby Galaxies Catalog (\citealt{Tully88}); molecular (11) and atomic (12) gas masses, obtained from \citet{Bettoni03}.\\
References for the existing studies of diffuse X-ray emission of some individual galaxies (as marked on the galaxy names):
a- \citet{Strickland04a}; b- \citet{Read05}; c- \citet{Boroson11}; d- \citet{David06}; e- \citet{JimenexBailon05}; f- \citet{Li11}; g- \citet{Wang03}; h- \citet{Wang05}; i- \citet{Iwasawa03}; j- \citet{Machacek04}; k- \citet{LiZ11}; l- \citet{Wang01}; m- \citet{Kraft05}; n- \citet{Rasmussen09}; o- \citet{Grimes05}; p- \citet{Li08}; q- \citet{Li09}; r- \citet{Bogdan12a}; s- \citet{Croston08}; t- \citet{Mineo12}; u- \citet{Stefano07}
}\label{table:SampleSelectionPara}
\end{deluxetable}

\begin{deluxetable}{lcllcllcl}
\centering
\tiny %\tiny\scriptsize\footnotesize\small\normalsize\large\Large\LARGE\huge\Huge
%\ptlandscape
  \tabletypesize{\tiny}
  \tablecaption{Methods Used for Distance Measurements}
  \tablewidth{0pt}
  \tablehead{
 \colhead{Name} & \colhead{$m_d$} & \colhead{Method} & \colhead{Name} & \colhead{$m_d$} & \colhead{Method} & \colhead{Name} & \colhead{$m_d$} & \colhead{Method}
}
\startdata
IC2560$^a$  & $32.33\pm0.38$ & Tully-Fisher & NGC3521$^a$ & $30.25\pm0.35$ & Tully-Fisher & NGC4594$^k$ & $29.95\pm0.18$ & SBF          \\
M82$^b$     & $27.74\pm0.04$ & TRGB         & NGC3556$^a$ & $30.14\pm0.35$ & Tully-Fisher & NGC4631$^l$ & $29.41\pm0.04$ & TRGB         \\
NGC24$^a$   & $29.79\pm0.35$ & Tully-Fisher & NGC3628$^j$ & $30.59\pm0.30$ & Tully-Fisher & NGC4666$^a$ & $30.98\pm0.36$ & Tully-Fisher \\
NGC520$^c$  & $32.22\pm0.40$ & Tully-Fisher & NGC3877$^a$ & $30.75\pm0.35$ & Tully-Fisher & NGC4710$^c$ & $31.13\pm0.80$ & Tully est    \\
NGC660$^d$  & $30.83\pm0.33$ & Tully-Fisher & NGC3955$^c$ & $31.57\pm0.80$ & Tully est    & NGC5102$^q$ & $27.51\pm0.16$ & TRGB         \\
NGC891$^e$  & $29.99\pm0.13$ & PNLF         & NGC3957$^c$ & $32.20\pm0.80$ & Tully est    & NGC5170$^o$ & $31.76\pm0.23$ & GCLF         \\
NGC1023$^f$ & $30.32\pm0.16$ & SBF          & NGC4013$^a$ & $31.38\pm0.35$ & Tully-Fisher & NGC5253$^i$ & $28.05\pm0.27$ & Cepheids     \\
NGC1380$^g$ & $31.65\pm0.07$ & SNIa         & NGC4111$^k$ & $30.88\pm0.23$ & SBF          & NGC5422$^c$ & $32.45\pm0.80$ & Tully est    \\
NGC1386$^h$ & $30.93\pm0.25$ & SBF          & NGC4217$^a$ & $31.45\pm0.35$ & Tully-Fisher & NGC5746$^a$ & $31.96\pm0.35$ & Tully-Fisher \\
NGC1482$^c$ & $31.46\pm0.80$ & Tully est    & NGC4244$^l$ & $28.20\pm0.03$ & TRGB         & NGC5775$^c$ & $32.13\pm0.80$ & Tully est    \\
NGC1808$^a$ & $30.45\pm0.36$ & Tully-Fisher & NGC4251$^k$ & $31.46\pm0.20$ & SBF          & NGC5866$^k$ & $30.93\pm0.12$ & SBF          \\
NGC2787$^c$ & $30.58\pm0.80$ & Tully est    & NGC4342$^c$ & $31.13\pm0.80$ & Tully est    & NGC6503$^r$ & $28.61\pm0.23$ & TRGB         \\
NGC2841$^i$ & $30.75\pm0.06$ & Cepheids     & NGC4388$^m$ & $31.16\pm0.47$ & Tully-Fisher & NGC6764$^a$ & $32.09\pm0.38$ & Tully-Fisher \\
NGC3079$^a$ & $31.09\pm0.35$ & Tully-Fisher & NGC4438$^j$ & $30.80\pm0.30$ & Tully-Fisher & NGC7090$^a$ & $28.99\pm0.35$ & Tully-Fisher \\
NGC3115$^f$ & $29.95\pm0.09$ & SBF          & NGC4501$^n$ & $30.99\pm0.08$ & SNIa         & NGC7457$^k$ & $30.61\pm0.21$ & SBF          \\
NGC3198$^i$ & $30.80\pm0.08$ & Cepheids     & NGC4526$^g$ & $31.19\pm0.07$ & SNIa         & NGC7582$^d$ & $31.81\pm0.40$ & Tully-Fisher \\
NGC3384$^f$ & $30.36\pm0.14$ & SBF          & NGC4565$^o$ & $30.23\pm0.20$ & GCLF         & NGC7814$^a$ & $31.29\pm0.36$ & Tully-Fisher \\
NGC3412$^f$ & $30.31\pm0.14$ & SBF          & NGC4569$^p$ & $29.97\pm0.05$ & Tully-Fisher &             &                &
\enddata
\tablecomments{\scriptsize $m_d$ is the distance modulus measured with various redshift independent methods: Tully-Fisher- Tully-Fisher relation \citep{Tully77}; TRGB- Red giant branch star \citep{Lee93}; PNLF- Planetary nebula luminosity function \citep{Ciardullo02}; SNIa- Typa~Ia SN \citep{Branch92}; SBF- Surface brightness fluctuation \citep{Tonry01}; Tully est- Estimation in the Nearby Galaxies Catalog \citep{Tully88}; Cepheids- Cepheid variable \citep{Freedman01}; GCLF- Globular cluster luminosity function \citep{Ferrarese00}.\\
References for the redshift-independent distance measurements of individual galaxies (as marked on the galaxy names):
a- \citet{Tully09}; b- \citet{Dalcanton09}; c- \citet{Tully88}; d- \citet{Springob09}; e- \citet{Ciardullo02}; f- \citet{Blakeslee01}; g- \citet{Jha07}; h- \citet{Jensen03}; i- \citet{Saha06}; j- \citet{Ekholm00}; k- \citet{Tonry01}; l- \citet{Seth05}; m- \citet{Willick97}; n- \citet{Mandel09}; o- \citet{Ferrarese00}; p- \citet{Cortes08}; q- \citet{Davidge08}; r- \citet{Karachentsev03}
}\label{table:distance}
\end{deluxetable}

\begin{deluxetable}{lcccclcccclcccc}
\centering
\tiny %\tiny\scriptsize\footnotesize\small\normalsize\large\Large\LARGE\huge\Huge
%\ptlandscape
  \tabletypesize{\tiny}
  \tablecaption{\emph{Chandra} Observations}
  \tablewidth{0pt}
  \tablehead{
 \colhead{Name} & \colhead{ObsID} & \colhead{Instr} & \colhead{$t_{exp}$} & \colhead{$t_{eff}$} & \colhead{Name} & \colhead{ObsID} & \colhead{Instr} & \colhead{$t_{exp}$} & \colhead{$t_{eff}$} & \colhead{Name} & \colhead{ObsID} & \colhead{Instr} & \colhead{$t_{exp}$} & \colhead{$t_{eff}$}
}
\startdata
IC2560  &  1592 & ACIS-S  &   9.96 &   9.63 & NGC3384 &  4692 & ACIS-S  &  10.03 &   9.47 & NGC4569 &  5911 & ACIS-S  &  39.65 &  29.63 \\
--      &  4908 & ACIS-S  &  55.67 &  54.24 & --      & 11782 & ACIS-S  &  29.04 &  28.67 & NGC4594 &  1586 & ACIS-S  &  18.75 &  18.12 \\
M82     &   361 & ACIS-I  &  33.68 &  32.14 & NGC3412 &  4693 & ACIS-S  &  10.09 &   9.96 & --      &  9532 & ACIS-I  &  86.04 &  83.34 \\
--      &  1302 & ACIS-I  &  15.71 &  15.36 & NGC3521 &  4694 & ACIS-S  &  10.07 &   6.59 & --      &  9533 & ACIS-I  &  90.15 &  87.00 \\
--      &  2933 & ACIS-S  &  18.25 &  17.83 & --      &  9552 & ACIS-I  &  72.41 &  71.39 & NGC4631 &   797 & ACIS-S  &  59.97 &  56.45 \\
NGC24   &  9547 & ACIS-S  &  43.78 &  41.99 & NGC3556 &  2025 & ACIS-S  &  60.12 &  57.98 & NGC4666 &  4018 & ACIS-S  &   5.00 &   4.74 \\
NGC520  &  2924 & ACIS-S  &  49.96 &  32.19 & NGC3628 &  2039 & ACIS-S  &  58.70 &  53.90 & NGC4710 &  9512 & ACIS-S  &  30.16 &  29.57 \\
NGC660  &  1633 & ACIS-S  &   1.94 &   1.91 & --      &  2918 & ACIS-S  &  22.26 &  21.08 & NGC5102 &  2949 & ACIS-S  &  34.66 &  32.77 \\
--      &  4010 & ACIS-S  &   5.13 &   4.87 & --      &  2919 & ACIS-S  &  22.76 &  22.28 & NGC5170 &  3928 & ACIS-I  &  33.45 &  32.58 \\
NGC891  &   794 & ACIS-S  &  51.56 &  29.03 & NGC3877 &   767 & ACIS-S  &  19.16 &  18.71 & NGC5253 &  2032 & ACIS-S  &  57.36 &  49.72 \\
--      &  4613 & ACIS-S  & 120.40 & 100.32 & --      &   768 & ACIS-S  &  23.75 &  22.45 & --      &  7153 & ACIS-S  &  69.12 &  67.66 \\
NGC1023 &  4696 & ACIS-S  &  10.37 &   9.52 & --      &   952 & ACIS-S  &  20.02 &  19.57 & --      &  7154 & ACIS-S  &  67.49 &  62.09 \\
--      &  8197 & ACIS-S  &  48.88 &  47.04 & --      &  1971 & ACIS-S  &  29.55 &  26.82 & NGC5422 &  9511 & ACIS-S  &  18.15 &  17.92 \\
--      &  8198 & ACIS-S  &  50.38 &  49.13 & --      &  1972 & ACIS-S  &  29.09 &  13.82 & --      &  9772 & ACIS-S  &  18.64 &  18.20 \\
--      &  8464 & ACIS-S  &  48.18 &  47.16 & NGC3955 &  2955 & ACIS-S  &  19.96 &  18.73 & NGC5746 &  3929 & ACIS-I  &  37.29 &  35.82 \\
--      &  8465 & ACIS-S  &  45.76 &  44.17 & NGC3957 &  9513 & ACIS-S  &  38.03 &  36.55 & NGC5775 &  2940 & ACIS-S  &  58.96 &  58.02 \\
NGC1380 &  9526 & ACIS-S  &  42.18 &  40.05 & NGC4013 &  4739 & ACIS-S  &  80.11 &  75.86 & NGC5866 &  2879 & ACIS-S  &  34.18 &  30.58 \\
NGC1386 &  4076 & ACIS-S  &  19.89 &  19.64 & NGC4111 &  1578 & ACIS-S  &  15.00 &  14.76 & NGC6503 &   872 & ACIS-S  &  13.36 &  11.27 \\
NGC1482 &  2932 & ACIS-S  &  28.56 &  23.78 & NGC4217 &  4738 & ACIS-S  &  73.66 &  69.37 & NGC6764 &  9269 & ACIS-S  &  20.18 &  19.72 \\
NGC1808 &  3012 & ACIS-S  &  43.43 &  39.52 & NGC4244 &   942 & ACIS-S  &  49.78 &  46.21 & NGC7090 &  7060 & ACIS-S  &  26.41 &  24.53 \\
NGC2787 &  4689 & ACIS-S  &  31.24 &  29.22 & NGC4251 &  4695 & ACIS-S  &  10.18 &   9.85 & --      &  7252 & ACIS-S  &  31.02 &  25.47 \\
NGC2841 &  6096 & ACIS-S  &  28.58 &  26.05 & NGC4342 &  4687 & ACIS-S  &  38.75 &  31.49 & NGC7457 &  4697 & ACIS-S  &   9.11 &   5.63 \\
NGC3079 &  2038 & ACIS-S  &  26.92 &  25.79 & NGC4388 &  1619 & ACIS-S  &  20.23 &  19.75 & --      & 11786 & ACIS-S  &  29.04 &  28.08 \\
NGC3115 &  2040 & ACIS-S  &  37.45 &  34.75 & NGC4438 &  2883 & ACIS-S  &  25.40 &  24.68 & NGC7582 &   436 & ACIS-S  &  13.62 &   6.93 \\
--      & 11268 & ACIS-S  &  41.10 &  40.19 & NGC4501 &  2922 & ACIS-S  &  18.10 &  17.28 & NGC7814 & 11309 & ACIS-S  &  59.11 &  57.74 \\
--      & 12095 & ACIS-S  &  76.65 &  74.50 & NGC4526 &  3925 & ACIS-S  &  44.11 &  38.00 &       &   &     &        &      \\
NGC3198 &  9551 & ACIS-S  &  62.42 &  60.64 & NGC4565 &  3950 & ACIS-S  &  59.95 &  50.31 &   &   &     &        &
\enddata
\tablecomments{\scriptsize $t_{exp}$ and $t_{eff}$ are the original and flare-removed (effective) exposures (in unit of ks) of individual observations. Some exposures of M82 and snapshot observations of other galaxies are not used.}\label{table:ChandraData}
\end{deluxetable}

\begin{deluxetable}{lccccccccc}
\centering
\tiny %\tiny\scriptsize\footnotesize\small\normalsize\large\Large\LARGE\huge\Huge
%\ptlandscape
  \tabletypesize{\tiny}
  \tablecaption{Parameters of the Sample Galaxies (II)}
  \tablewidth{0pt}
  \tablehead{
 \colhead{Name} & \colhead{$m_K$} & \colhead{$M_\star$} & \colhead{$f_{12}$} & \colhead{$f_{25}$} & \colhead{$f_{60}$} & \colhead{$f_{100}$} & \colhead{$L_{IR}$} & \colhead{$SFR_{IR}$} & \colhead{$S_{1.4GHz}$} \\
   & (mag) & ($10^{10}\rm M_\odot$) & (Jy) & (Jy) & (Jy) & (Jy) & ($10^{44}\rm ergs/s$) & ($\rm M_\odot/yr$) & (mJy) \\
   & (1) & (2) & (3) & (4) & (5) & (6) & (7) & (8) & (9)
}
\startdata
IC2560  &$8.85\pm0.03$& $1.08\pm0.03$     & $0.38\pm0.10$   & $0.87\pm0.09$   & $3.54\pm0.50$   & $5.78\pm0.81$   & $0.46\pm0.07$              & $2.1\pm0.3$     & $31\pm2$ \\
M82     &$4.69\pm0.02$& $1.99\pm0.03$     & $53.2\pm3.2$    & $274\pm16$      & $1170\pm70$     & $1140\pm69$     & $1.7\pm0.1$                & $7.7\pm0.5$     & $6433\pm209$ \\
NGC24   &$9.22\pm0.02$& $0.153\pm0.003$   & $<0.26$         & $<0.56$         & $1.26\pm0.13$   & $3.71\pm0.52$   & $0.024_{-0.01}^{+0.002}$   & $0.11_{-0.06}^{+0.007}$ & -- \\
NGC520  &$8.54\pm0.02$& $3.70\pm0.06$     & $0.78\pm0.11$   & $2.83\pm0.28$   & $31.5\pm4.4$    & $48.4\pm6.8$    & $2.6\pm0.4$                & $12\pm2$        & $176\pm5$ \\
NGC660  &$7.40\pm0.02$& $2.98\pm0.04$     & $2.01\pm0.36$   & $7.09\pm0.71$   & $65.4\pm9.2$    & $104\pm10$      & $1.6\pm0.2$                & $7.1\pm0.9$     & $373\pm14$ \\
NGC891  &$5.99\pm0.02$& $4.94\pm0.07$     & $0.92\pm0.23$   & $0.73\pm0.18$   & $34.4\pm6.2$    & $148\pm15$      & $0.55\pm0.08$              & $2.5\pm0.3$     & $239\pm8$ \\
NGC1023 &$6.33\pm0.02$& $6.76\pm0.10$     & --              & --              & --              & --              & --                         & --              & -- \\
NGC1380 &$6.96\pm0.02$& $12.3\pm0.2$      & $<0.25$         & $<0.25$         & $1.05\pm0.06$   & $2.96\pm0.30$   & $0.10_{-0.05}^{+0.004}$    & $0.46_{-0.2}^{+0.02}$ & -- \\
NGC1386 &$8.14\pm0.02$& $1.83\pm0.03$     & $0.50\pm0.03$   & $1.44\pm0.09$   & $5.89\pm0.59$   & $9.92\pm0.99$   & $0.20\pm0.02$              & $0.90\pm0.08$   & $37\pm2$ \\
NGC1482 &$8.56\pm0.02$& $2.29\pm0.04$     & $1.60\pm0.10$   & $4.63\pm0.28$   & $31.5\pm4.4$    & $46.5\pm6.5$    & $1.5\pm0.2$                & $6.5\pm0.8$     & $238\pm8$ \\
NGC1808 &$6.73\pm0.02$& $3.78\pm0.06$     & $4.12\pm0.25$   & $15.9\pm0.95$   & $97.1\pm9.7$    & $136\pm8$       & $1.7\pm0.1$                & $7.8\pm0.6$     & $528\pm20$ \\
NGC2787 &$7.33\pm0.01$& $3.33\pm0.03$     & $<0.63$         & $<0.64$         & $0.66\pm0.07$   & $<2.0$          & $0.057_{-0.05}^{+0.001}$   & $0.26_{-0.2}^{+0.003}$ &$10.9\pm0.5$\\
NGC2841 &$6.16\pm0.02$& $9.86\pm0.10$     & $<0.25$         & $<0.25$         & $2.43\pm0.24$   & $14.0\pm2.0$    & $0.11_{-0.03}^{+0.01}$     & $0.49_{-0.1}^{+0.05}$ & $32\pm3$\\
NGC3079 &$7.35\pm0.02$& $2.97\pm0.04$     & $1.24\pm0.07$   & $2.03\pm0.12$   & $42.9\pm4.3$    & $88.9\pm8.9$    & $1.3\pm0.1$                & $6.1\pm0.6$ & $769\pm27$ \\
NGC3115 &$5.94\pm0.02$& $6.79\pm0.10$     & --              & --              & --              & --              & --                         & --              & -- \\
NGC3198 &$8.04\pm0.02$& $1.04\pm0.02$     & $<0.25$         & $0.46\pm0.05$   & $3.70\pm0.37$   & $14.8\pm1.5$    & $0.14_{-0.03}^{+0.01}$     & $0.62_{-0.1}^{+0.06}$ & $23\pm3$\\
NGC3384 &$6.85\pm0.02$& $4.11\pm0.06$     & --              & --              & --              & --              & --                         & --              & -- \\
NGC3412 &$7.75\pm0.01$& $1.66\pm0.02$     & --              & --              & --              & --              & --                         & --              & -- \\
NGC3521 &$5.87\pm0.02$& $7.05\pm0.10$     & $<0.97$         & $0.91\pm0.23$   & $27.3\pm3.8$    & $85.1\pm8.5$    & $0.47_{-0.09}^{+0.05}$     & $2.1_{-0.4}^{+0.2}$ & $375\pm9$ \\
NGC3556 &$7.15\pm0.02$& $1.61\pm0.03$     & $0.61\pm0.06$   & $1.78\pm0.11$   & $23.5\pm2.4$    & $61.3\pm6.1$    & $0.35\pm0.03$              & $1.6\pm0.2$     & $217\pm7$ \\
NGC3628 &$6.22\pm0.02$& $6.73\pm0.10$     & $2.60\pm0.16$   & $4.64\pm0.46$   & $48.4\pm6.8$    & $103\pm10$      & $1.1\pm0.1$                & $4.8\pm0.5$     & $291\pm9$ \\
NGC3877 &$7.81\pm0.02$& $1.75\pm0.03$     & $0.21\pm0.05$   & $0.35\pm0.05$   & $4.84\pm0.48$   & $19.5\pm1.9$    & $0.16\pm0.02$              & $0.71\pm0.08$   & $41\pm3$  \\
NGC3955 &$8.75\pm0.02$& $1.29\pm0.03$     & $0.56\pm0.06$   & $0.77\pm0.08$   & $8.0\pm1.1$     & $17.6\pm2.5$    & $0.46\pm0.06$              & $2.1\pm0.3$     & $49\pm2$  \\
NGC3957 &$8.69\pm0.02$& $1.11\pm0.02$     & $<0.25$         & $<0.48$         & $0.56\pm0.10$   & $1.77\pm0.18$   & $0.15_{-0.1}^{+0.007}$     & $0.67_{-0.5}^{+0.03}$ & -- \\
NGC4013 &$7.70\pm0.02$& $4.49\pm0.07$     & $<0.25$         & $0.24\pm0.06$   & $5.08\pm0.71$   & $21.5\pm3.0$    & $0.31_{-0.07}^{+0.04}$     & $1.4_{-0.3}^{+0.2}$ & $38\pm2$ \\
NGC4111 &$7.61\pm0.02$& $3.02\pm0.04$     & --              & --              & --              & --              & --                         & -- & $9.4\pm0.5$ \\
NGC4217 &$7.70\pm0.02$& $4.25\pm0.07$     & --              & --              & --              & --              & --                         & -- & $120\pm4$ \\
NGC4244 &$8.11\pm0.02$& $0.088\pm0.002$   & $<0.36$         & $<0.25$         & $<0.45$         & $3.07\pm0.31$   & $0.0043_{-0.003}^{+0.0001}$& $0.019_{-0.01}^{+0.001}$ & -- \\
NGC4251 &$7.80\pm0.01$& $4.23\pm0.04$     & --              & --              & --              & --              & --                         & -- & -- \\
NGC4342 &$9.05\pm0.02$& $1.26\pm0.02$     & --              & --              & --              & --              & --                         & -- & -- \\
NGC4388 &$8.17\pm0.02$& $1.63\pm0.03$     & $1.0\pm0.1$     & $3.55\pm0.21$   & $10.9\pm1.1$    & $17.8\pm2.5$    & $0.50\pm0.05$              & $2.2\pm0.2$ & $119\pm5$ \\
NGC4438 &$7.38\pm0.02$& $3.16\pm0.05$     & $<0.25$         & $<0.26$         & $4.08\pm0.41$   & $10.8\pm1.5$    & $0.12_{-0.03}^{+0.01}$     & $0.53_{-0.1}^{+0.05}$ & $63\pm3$ \\
NGC4501 &$6.33\pm0.02$& $8.11\pm0.10$     & $0.70\pm0.07$   & $0.93\pm0.13$   & $14.2\pm2.0$    & $55.4\pm7.8$    & $0.57\pm0.08$              & $2.6\pm0.4$ & $277\pm9$ \\
NGC4526 &$6.55\pm0.02$& $11.9\pm0.2$      & $<0.33$         & $0.53\pm0.13$   & $5.93\pm0.83$   & $16.0\pm2.2$    & $0.25_{-0.06}^{+0.03}$     & $1.1_{-0.3}^{+0.1}$ & $12.0\pm0.5$ \\
NGC4565 &$6.13\pm0.02$& $5.27\pm0.07$     & $<0.89$         & $0.47\pm0.12$   & $6.04\pm0.85$   & $25.4\pm6.3$    & $0.15_{-0.06}^{+0.02}$     & $0.67_{-0.3}^{+0.1}$ & $55\pm3$ \\
NGC4569 &$6.76\pm0.02$& $2.08\pm0.03$     & $<0.34$         & $0.88\pm0.09$   & $7.2\pm1.0$     & $23.4\pm3.3$    & $0.11_{-0.02}^{+0.01}$     & $0.49_{-0.1}^{+0.06}$ & $71\pm3$ \\
NGC4594 &$5.01\pm0.02$& $15.5\pm0.2$      & $<0.57$         & $0.37\pm0.05$   & $3.18\pm0.45$   & $12.1\pm1.7$    & $0.062_{-0.02}^{+0.006}$   & $0.28_{-0.1}^{+0.03}$ & $93\pm3$ \\
NGC4631 &$6.62\pm0.02$& $1.02\pm0.02$     & $1.81\pm0.18$   & $2.99\pm0.42$   & $51.5\pm7.2$    & $120\pm12$      & $0.37\pm0.04$              & $1.7\pm0.2$ & $446\pm14$ \\
NGC4666 &$7.10\pm0.02$& $4.07\pm0.06$     & $1.09\pm0.07$   & $1.63\pm0.16$   & $25.8\pm3.6$    & $77.1\pm7.7$    & $0.90\pm0.1$               & $4.0\pm0.5$ & $434\pm14$ \\
NGC4710 &$7.67\pm0.02$& $3.35\pm0.05$     & $<0.31$         & $<0.50$         & $6.00\pm0.84$   & $13.3\pm1.9$    & $0.22_{-0.07}^{+0.02}$     & $0.98_{-0.3}^{+0.1}$ & $19\pm1$ \\
NGC5102 &$7.20\pm0.02$& $0.151\pm0.003$   & $<0.25$         & $<0.25$         & $0.83\pm0.08$   & $2.65\pm0.27$   & $0.0021_{-0.001}^{+0.0001}$& $0.0094_{-0.005}^{+0.0005}$ & $3.2\pm0.7$ \\
NGC5170 &$7.82\pm0.02$& $4.73\pm0.08$     & $<0.25$         & $<0.31$         & $1.04\pm0.10$   & $3.72\pm0.52$   & $0.13_{-0.06}^{+0.009}$    & $0.56_{-0.3}^{+0.04}$ & $7\pm2$ \\
NGC5253 &$8.42\pm0.03$& $0.049\pm0.001$   & $2.57\pm0.15$   & $12.1\pm0.73$   & $31.2\pm3.1$    & $29.8\pm3.0$    & $0.075\pm0.006$            & $0.34\pm0.03$ & $85\pm3$ \\
NGC5422 &$8.85\pm0.02$& $4.64\pm0.08$     & --              & --              & --              & --              & --                         & -- & -- \\
NGC5746 &$6.93\pm0.02$& $14.3\pm0.2$      & $<0.25$         & $<0.26$         & $1.07\pm0.11$   & $8.52\pm0.85$   & $0.21_{-0.08}^{+0.02}$     & $0.96_{-0.3}^{+0.07}$ & $15\pm3$ \\
NGC5775 &$7.76\pm0.01$& $6.57\pm0.08$     & $0.71\pm0.04$   & $0.86\pm0.12$   & $15.4\pm1.5$    & $45.2\pm4.5$    & $1.5\pm0.2$                & $6.9\pm0.7$ & $280\pm9$ \\
NGC5866 &$6.95\pm0.02$& $5.53\pm0.08$     & $0.32\pm0.03$   & $0.20\pm0.04$   & $5.13\pm0.31$   & $17.1\pm1.7$    & $0.18\pm0.02$              & $0.82\pm0.07$ & $22\pm1$ \\
NGC6503 &$7.38\pm0.02$& $0.313\pm0.005$   & $0.60\pm0.15$   & $0.45\pm0.11$   & $7.16\pm0.43$   & $25.4\pm1.5$    & $0.033\pm0.003$            & $0.15\pm0.01$ & $37\pm3$ \\
NGC6764 &$9.63\pm0.04$& $0.99\pm0.03$     & $0.38\pm0.04$   & $1.33\pm0.08$   & $6.48\pm0.65$   & $11.9\pm1.2$    & $0.61\pm0.06$              & $2.7\pm0.3$ & $110\pm4$ \\
NGC7090 &$8.40\pm0.02$& $0.148\pm0.003$   & $<0.25$         & $0.29\pm0.04$   & $5.64\pm0.56$   & $17.2\pm1.7$    & $0.031_{-0.006}^{+0.003}$  & $0.14_{-0.03}^{+0.01}$ & -- \\
NGC7457 &$8.33\pm0.03$& $1.22\pm0.03$     & --              & --              & --              & --              & --                         & -- & -- \\
NGC7582 &$7.41\pm0.02$& $6.77\pm0.10$     & $1.35\pm0.14$   & $6.33\pm0.63$   & $48.0\pm6.7$    & $73\pm10$       & $2.9\pm0.4$                & $13\pm2$ & -- \\
NGC7814 &$7.20\pm0.02$& $7.02\pm0.10$     & --              & --              & --              & --              & --                         & -- & --
\enddata
\tablecomments{\scriptsize Listed galaxy parameters: (1) 2MASS K-band apparent magnitude within fiducial ellipse at isophote $K=20\rm~mag~arcsec^{-2}$, obtained from the 2MASS extended sources catalog (\citealt{Skrutskie06}); (2) stellar mass, estimated from the K-band apparent magnitude, distance (d), and extinction corrected B-V color in Table~\ref{table:SampleSelectionPara}, as well as a color-dependent stellar mass-to-light ratio (\citealt{Bell01}); (3-6) \emph{IRAS} fluxes at $12\rm~\mu m$, $25\rm~\mu m$, $60\rm~\mu m$, and $100\rm~\mu m$, obtained from \citet{Fullmer89}; (7) total IR luminosity, defined as $L_{IR}=5.67\times10^5d^2(13.48f_{12}+5.16f_{25}+2.58f_{60}+f_{100})\rm L_\odot$ \citep{Rice88}; (8) SFR estimated from $L_{IR}$ using Kennicutt-relation \citep{Kennicutt98}; (9) 1.4~GHz radio continuum fluxes, obtained from \citet{Condon98}.}\label{table:basicparaII}
\end{deluxetable}

\begin{figure}[!h]
\begin{center}
\epsfig{figure=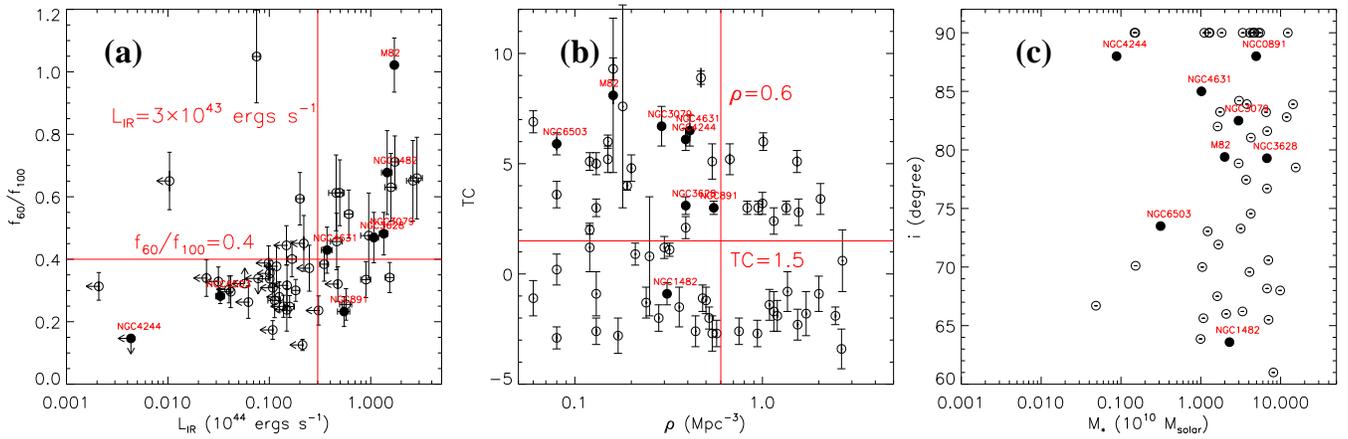,width=1.0\textwidth,angle=0, clip=}
\caption{Galaxy parameter (Tables.~\ref{table:SampleSelectionPara} and \ref{table:basicparaII}) coverage of the present sample: (a) IR luminosity ($L_{IR}$) and far-IR warmth ($f_{60}/f_{100}$); (b) local galaxy number density ($\rho$) and morphological type code (TC); (c) stellar mass ($M_*$) and disk inclination angle ($i$). The filled circles and the names are for galaxies studied by \citet{Strickland04a,Strickland04b}. In (a), the two lines mark the positions of $f_{60}/f_{100}=0.4$ and $L_{IR}=3\times10^{43}\rm~ergs~s^{-1}$, separating starburst (within the upper right portion of the plot) and non-starburst galaxies (the rest). In (b), the horizontal line marks the position of $TC=1.5$, separating bulge- and disk-dominated (or early- and late-type) disk galaxies. The vertical line marks the position of $\rho=0.6$, separating clustered and field galaxies.}\label{fig:sample}
\end{center}
\end{figure}

In total, 53 galaxies are selected for the present study, which form the largest \emph{Chandra} sample of nearby disk galaxies. Key parameters of the sample galaxies and the \emph{Chandra} observations are summarized in Tables~\ref{table:SampleSelectionPara}-\ref{table:basicparaII}. References for those galaxies which have been studied previously for diffuse X-ray emission with the \emph{Chandra} data are given in Table~\ref{table:SampleSelectionPara}; 20 galaxies are presented for the first time. The sample coverages of the IR luminosity and warmth ($f_{60}/f_{100}$), morphological type, local galaxy number density, and stellar mass are shown in Fig.~\ref{fig:sample}.

In this paper, we utilize the Spearman's rank order coefficient ($r_s$; by definition, $-1<r_s<1$) to describe the goodness of a correlation. This simple goodness description works for any pair of variables that show a monotonic correlation, insensitive to its exact form. We consider $|r_s|\gtrsim0.6$ or $0.3\lesssim|r_s|\lesssim0.6$ as a tight or weak correlation, and $|r_s|\lesssim0.3$ as no correlation. We further characterize important tight correlations with simple expressions, together with the corresponding dispersion measurements [the root mean square ($rms$) around the fitted relations]. All errors are quoted at the $1~\sigma$ confidence level.

\subsection{Multi-Wavelength Galaxy Properties}\label{PaperIsubsec:MultiBandProperties}

\subsubsection{SF Tracers}\label{PaperIsubsubsec:SFtracer}

There are several commonly-used SF tracers (for reviews, e.g., see \citealt{Kennicutt98,Calzetti08}). For example, the ultraviolet (UV, $\lambda\sim912-3000\rm~{\AA}$) emission and the hydrogen recombination lines (H$\alpha$, H$\beta$, P$\alpha$, etc.) are the products of massive stars and are often adopted as direct SF indicators. However, such relatively short wavelength radiation is often highly attenuated by dust in a highly inclined galactic disk. Here, we estimate the SFR of an individual galaxy using its total IR luminosity (Table~\ref{table:basicparaII}), which is mainly from the dust-reprocessed young stellar emission and is little affected by the extinction.

\begin{figure}[!h]
\begin{center}
\epsfig{figure=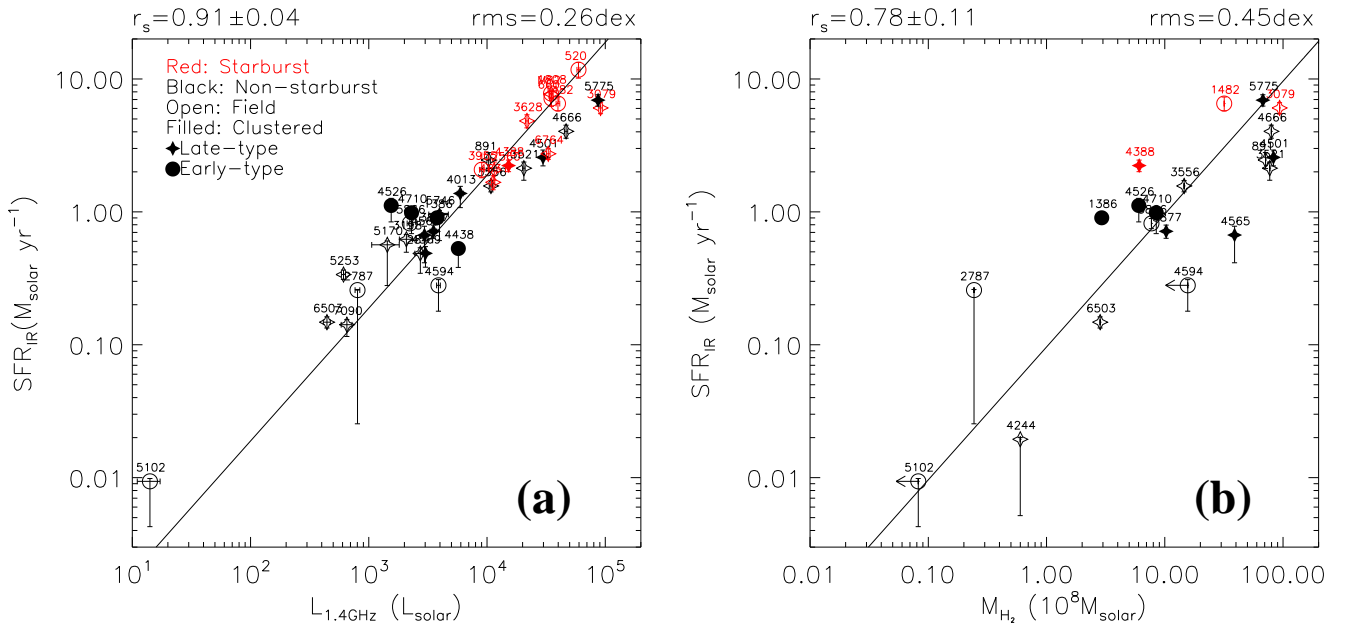,width=1.0\textwidth,angle=0, clip=}
\caption{SFR calculated with the total IR luminosity ($SFR_{IR}$) plotted against other SF tracers: (a) The 1.4~GHz radio continuum luminosity ($L_{1.4GHz}$); (b) The molecular gas mass ($M_{H_2}$; data from \citealt{Bettoni03}). The symbols are noted in (a). NGC names of the galaxies are marked next to the symbols. The solid lines represents the linear fits which are detailed in the text (Eqs.~\ref{equi:SFRFIRL14} and \ref{equi:SFRFIRMH2}). We also mark the Spearman's rank order coefficient ($r_s$) and the root mean square around the linear fits ($rms$) on top of each panel.}\label{fig:SFtracer}
\end{center}
\end{figure}

We further check if such a simple estimation of the SFR is consistent with other extinction-free SF tracers, such as the radio continuum emission (e.g., \citealt{Yun01}) and the molecular gas mass (e.g., \citealt{Gao04}). Here we use the mass of the molecular gas (instead of that of the total cold gas including atomic gas, as sometimes considered in similar analysis, e.g., \citealt{Strickland04a}), because it shows a better correlation with the SF activity (e.g., \citealt{Bigiel08,Genzel10}). The relations between different SF tracers are shown in Fig.~\ref{fig:SFtracer}, and can be expressed with the following equations:
\begin{equation}\label{equi:SFRFIRL14}
{\rm SFR_{IR}}=(1.89\pm0.02)\times10^{-4}~(L_{1.4GHz}/L_\odot),
\end{equation}
\begin{equation}\label{equi:SFRFIRMH2}
{\rm SFR_{IR}}=(0.096\pm0.002)~(M_{H_2}/10^8M_\odot).
\end{equation}
Despite the tight correlations, all these SF tracers have their own uncertainties. For example, IR emission can be produced by circum-stellar dust in early-type galaxies \citep{Temi07,Temi09}; radio continuum emission can be produced by AGN (e.g., \citealt{David06,Giodini10}); SF inactive galaxies may still host significant amount of molecular gas \citep{Welch03,Welch10}. Furthermore, even if the current SFR is accurately estimated, the different SF history of different types of galaxies also mean that their stellar mass built-up and so feedback history may be different (e.g., for S0 galaxies, \citealt{vandenBergh09}). It is still not clear if this difference in SF history can significantly affect the observed coronal properties, or their relations to other galaxy properties (e.g., the $L_X-{\rm SFR}$ and $L_X-L_K$ relations; \citealt{Li11}). The above uncertainties, either in the measurement of SFR or the different SF histories, may result in additional scatters in the following correlation analysis (Paper~II).

In addition to the global SFR, another key property of SF is its specific intensity (SFR per unit area or unit mass), which is often traced by the IR warmth ($f_{60}/f_{100}$, a measurement of dust temperature) \citep{Strickland04a}. It is not surprising that the SFR correlates well with the IR warmth (Fig.~\ref{fig:sample}a), since a galaxy with a higher SFR also tends to have a higher SF intensity. However, this correlation has a large dispersion, which suggests that other factors such as the SF mode (e.g., nuclear starburst vs. disk-wide star forming) may affect the observed IR properties. We will thus use both the IR luminosity and warmth to define the starburst subsample (\S\ref{PaperIsubsubsec:subsample}).

\subsubsection{Galaxy Mass}\label{PaperIsubsubsec:GalaxyMass}

Galaxy mass is often described in two ways: the stellar mass (responsible for old stellar feedback) and the gravitational mass (related to the gravitational confinement), which are closely correlated with each other (e.g., \citealt{Bell01,Crain10a}). We use the \emph{2MASS} K-band luminosity \citep{Skrutskie06} and the color-dependent mass-to-light ratio \citep{Bell01} to estimate the stellar mass ($M_*$, Table~\ref{table:basicparaII}). The application of this method is somewhat inaccurate for edge-on galaxies, especially starburst ones, because the extinction can be strong, even in near-IR. The uncertainty in the mass estimate can lead to scatters in the subsequent statistical analysis.

\begin{figure}[!h]
\begin{center}
\epsfig{figure=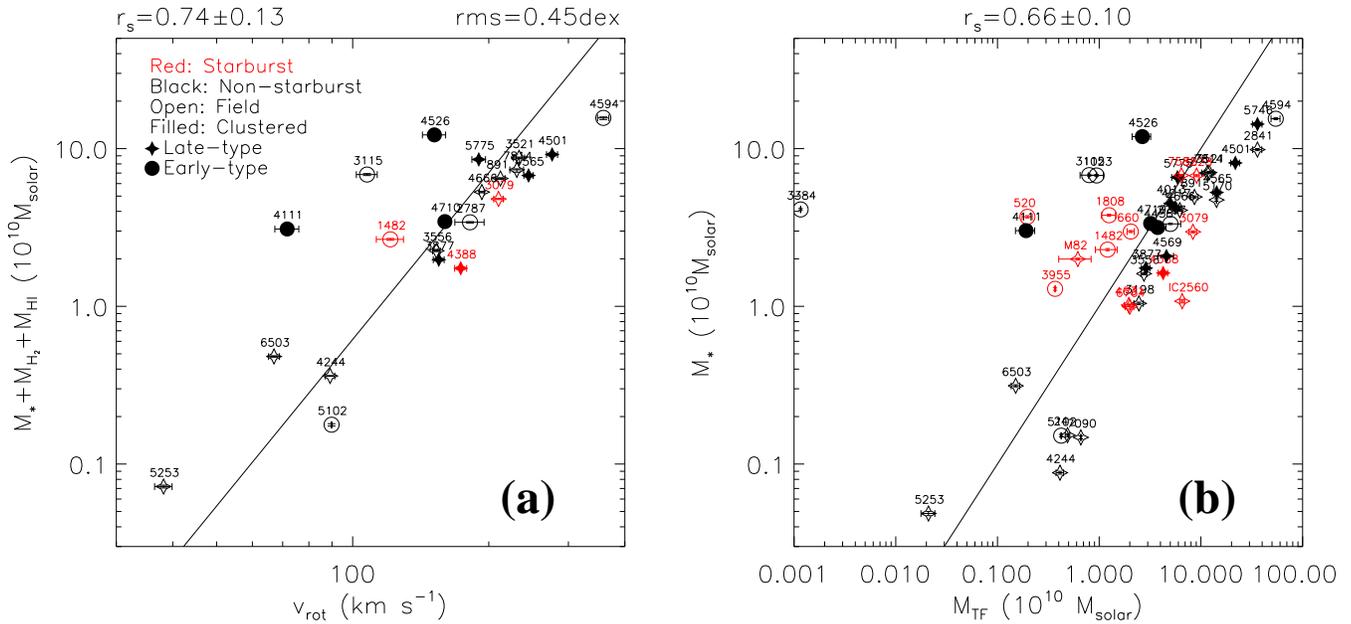,width=1.0\textwidth,angle=0, clip=}
\caption{(a) Tully-Fisher relation for the sample galaxies. The total baryonic mass is defined as the total mass of stars and cold gas (molecular and atomic gas). The solid line is the Tully-Fisher relation for the total baryonic mass obtained from \citet{Bell01}, which is used to calculate $M_{TF}$ in the present work. $rms$ around this relation is marked on the top right corner. (b) Comparison between the stellar mass ($M_*$) and the baryonic mass ($M_{TF}$). The solid line marks where the two masses equal to each other. The symbols are noted in (a). NGC names of the galaxies are marked next to the symbols. We also mark the Spearman's rank order coefficient ($r_s$) on top of each panel.}\label{fig:GalaxyMass}
\end{center}
\end{figure}

The baryon mass ($M_{TF}$) is estimated from the rotation velocity ($v_{rot}$ in Table~\ref{table:SampleSelectionPara}) together with the baryon Tully-Fisher relation, which represents a tight correlation between the total baryon mass (though only contributions from stars and cool ISM are included; Fig.~\ref{fig:GalaxyMass}) and the rotation velocity of spiral galaxies (\citealt{Bell01}). The relation implies an approximate constancy of the baryon to gravitational mass ratio in such galaxies. Our inferred $M_{TF}$ of a galaxy is thus expected to be proportional to its gravitational mass. The rotation velocity is obtained from the \emph{HyperLeda} database and is estimated from the 21~cm line widths and/or the rotation curves (generally from H$\alpha$ emission), except for M82 (see \S\ref{PaperIsec:Individual} for the reason). For data homogenization, the epidemic method as detailed in \citet{Paturel03} is used. We emphasize that the accuracy of the mass obtained this way may be affected by other factors: (1) The galaxy inclination correction of $v_{rot}$ in \emph{HyperLeda} is conducted using the inclination angle defined in optical ($i$ in Table~\ref{table:SampleSelectionPara}), which may be different from those of \ion{H}{1} or H$\alpha$ gases. (2) Some galaxies in our sample are apparently undergoing tidal interactions with their companions. The gas in such a galaxy may be strongly disturbed. Consequently, their $v_{rot}$ and mass may not follow the Tully-Fisher relation. (3) S0 galaxies may not be well described by the Tully-Fisher relation, which is defined only for spiral galaxies. We have double-checked the published rotation curves of some galaxies in our sample (e.g., \citealt{Sofue97}) and found that their flattening velocities are consistent with $v_{rot}$ as given in \emph{HyperLeda}. We thus consider that the estimated $M_{TF}$ values are generally reliable, except for a few extreme individuals (see later discussion in \S\ref{PaperIsec:Individual}), and sufficiently good for a
statistical comparison as is intended in Paper~II.

\subsubsection{Definition of Subsamples}\label{PaperIsubsubsec:subsample}

We further define some subsamples of galaxies for later in-depth comparisons. \citet{Strickland04a,Strickland04b} separate starburst and non-starburst galaxies by $f_{60}/f_{100}=0.4$, based on the assumption that galaxies with more intense and/or compact SF tend to have higher dust temperature. From Fig.~\ref{fig:sample}a, we see that some non-starburst galaxies with low IR luminosities can also have high $f_{60}/f_{100}$ ratios. In the present work, we define starburst galaxies as those with both $f_{60}/f_{100}>0.4$ and $L_{IR}>3\times10^{43}\rm~ergs~s^{-1}$. Other galaxies are defined as non-starburst galaxies. According to this definition, a couple of dwarf galaxies with high $f_{60}/f_{100}$ ratio and low $L_{IR}$ are not considered as starbursts; the $f_{60}/f_{100}$ ratios of these low $L_{IR}$ galaxies tend to be contaminated by other non-SF-related heating processes (e.g., due to heating by old stars and/or weak AGNs; \citealt{Temi07,Temi09}).

We use the local galaxy number density ($\rho$) to characterize the galaxy environment (Fig.~\ref{fig:sample}b; Table~\ref{table:SampleSelectionPara}). Galaxies with $\rho\leq0.6$ are defined as being in the field, while those with $\rho>0.6$ as being clustered.

We further separate galaxies with different morphological types. Early- and late-type disk galaxies are separated by $TC=1.5$ or roughly the ``Sa'' type (Fig.~\ref{fig:sample}b).

\section{Data Reduction and Analysis}\label{PaperIsec:DateReduction}

\subsection{Chandra Data Reduction}\label{PaperIsubsec:DateReduction}

\subsubsection{Basic Calibration}\label{PaperIsubsubsec:Calibration}

The \emph{Chandra} data used in the present work are summarized in Table~\ref{table:ChandraData}. The data are comprised of 72 ACIS-S and 7 ACIS-I observations of the 53 galaxies. We reprocess the data using CIAO (\emph{Chandra} Interactive Analysis of Observations) v.4.2 and the corresponding \emph{Chandra} Calibration Database (CALDB). We reduce all the data in a \emph{uniform} manner, which allows for comparison among them with minimal calibration biases.

We start our calibration with the level=1 raw data. We remove afterglow, representing residual charges from the interaction of cosmic rays in a CCD using the CIAO tool \emph{``dmtcalc''}. We flagged out hot pixels with \emph{``acis\_run\_hotpix''}. We then generate a new level=2 event file, using \emph{``acis\_process\_events''}. The \emph{``eventdef''} and \emph{``check\_vf\_pha''} parameters are set according to the observational modes, while the TGAIN (ACIS Time-dependent Gain) and CTI (Charge Transfer Inefficiency) calibrations are conducted following the official recommendations of the \emph{Chandra} data analysis guide to account for the time-dependent degradation of the CCD quantum efficiency. The remaining events are filtered for good grade, status, and time, and the steak events are also removed. Such basic calibrations result in a new level=2 event file for further analysis.

\subsubsection{Background Flare Removal}\label{PaperIsubsubsec:Flare}

We clean the light curve of each observation in the 2.5-7~keV band, which is sensitive to background flares (short periods of significantly enhanced background count rate; \citealt{Markevitch03,Hickox06}). The light curve is extracted from regions located in the same CCD chips as those used for the later scientific analysis, with bright point sources excluded. We initially identify significant background flares using a $2.5~\sigma$ clip, check the light curves individually, and change the threshold slightly if necessary. We then filter the data, removing the time intervals with these background flares, resulting in the effective exposure time of the good time intervals ($t_{eff}$) as listed in Table~\ref{table:ChandraData}.

\subsubsection{Merging Multiple Observations}\label{PaperIsubsubsec:Merge}

For some galaxies, there are multiple observations. Many \emph{Chandra} observations have taken on M82, but we only use the three observations analyzed by \citet{Strickland04a}, which are sufficient for our measurements of the global coronal properties. We merge multiple observations together to maximize counting statistics for imaging and spectral analysis. For the present study, we find that the default astrometric accuracy is good enough (typically $\lesssim0\farcs5$).

For spatial analysis, we use the CIAO tool \emph{``reproject\_events''} and \emph{``dmmerge''} to combine multiple event files into a single one. While for spectral analysis, we use \emph{``specextract''} to extract spectra from individual observations and stack them.

\subsubsection{Non-X-ray Background Measurements}\label{PaperIsubsubsec:Background}

We utilize ACIS stowed data to estimate the quiescent instrument background \citep{Markevitch03,Hickox06}. We find the stowed background data set which best matches an individual observation, according to its observing date. The exposure of the stowed data is adjusted so that the resultant counts rate in the 10-12/10-14~keV band [for the front/back-illuminated (FI/BI) chips] to be the same as that of the target observation.

\subsubsection{Exposure Correction}\label{PaperIsubsubsec:Exposure}

We generate exposure maps for spatial analysis. An exposure map accounts for the variation of the effective area across the detector field of view (FoV) and the telescope dithering. Unlike a spectral response file, an exposure map must be averaged in a relatively broad band. We produce an exposure map using a power law spectral weighting (power law index=1.7), which is a reasonable characterization for the cosmic background and the typical point-like sources. For an accurate count flux calculation, for example, we make the exposure corrections in up to four relatively narrow bands (see \S\ref{PaperIsubsubsec:PointSource}). For image construction and rough spatial analysis, we may use corrections in broader bands. For quantitative calculations, more accurate spectral response files are adopted (\S\ref{PaperIsubsec:spec}).

\subsubsection{Point Source Detection and Removal}\label{PaperIsubsubsec:PointSource}

Most of the discrete X-ray sources are detected with very limited counting statistics in a typical \emph{Chandra} observation. Furthermore, the detection threshold can strongly depend on the point-spread function (PSF), the local background, and the effective exposure, which all vary significantly across the FoV. Our procedure (refer to \citealt{Wang04} for details) includes the map detection and maximum likelihood analyses of individual sources, as well as the conventional wavelet detection, and gives a better iterative estimation of the local background (than using the usual wavelet source detection tools alone), which is critical in detecting faint point sources, especially in a crowded field. Using this procedure, we perform source detection in the broad (B, $0.3-7\rm~keV$), soft (S, $0.3-1.5\rm~keV$) and hard (H, $1.5-7\rm~keV$) bands. This multi-band detection may help us to reduce false detection and to roughly identify some sources with peculiar hardness ratio. The detected sources will be used for the subsequent image construction (\S\ref{PaperIsubsubsec:Image}), spatial (\S\ref{PaperIsubsec:spatial}), and spectral analyses (\S\ref{PaperIsubsec:spec}).

\subsubsection{Image Construction}\label{PaperIsubsubsec:Image}

To map out the diffuse X-ray emission, we remove the detected discrete sources from the images (count, background, and exposure). A circular region of twice the 90\% energy enclosed radius (EER) is excluded around each source of a count rate $(CR)\lesssim 0.01\rm~cts~s^{-1}$. For a brighter source, the removal radius is further multiplied by a factor of $1+log(CR/0.01)$. Generally about 96\% of the source counts are excluded in such a removal \citep{Wang04}. The images, with or without point sources subtracted, are further smoothed with the CIAO tool \emph{``csmooth''} (with a Gaussian kernel), adopting adaptive smoothing scales to match the preset range of the signal-to-noise ratio. The stowed-background and exposure maps are smoothed with the smoothing scale (kernel size) map adaptively calculated for the count image in the broad band. Such an image construction is performed with the same scale map for different bands and is shown as a tri-color image of each galaxy in Fig.~\ref{fig:imagea} to highlight the hardness variation. We emphasize that such smoothed images are only used to show the outstanding diffuse X-ray emission features, while all the quantitative measurements are based on the data without any smoothing.

\begin{figure}[!h]
\begin{center}
\epsfig{figure=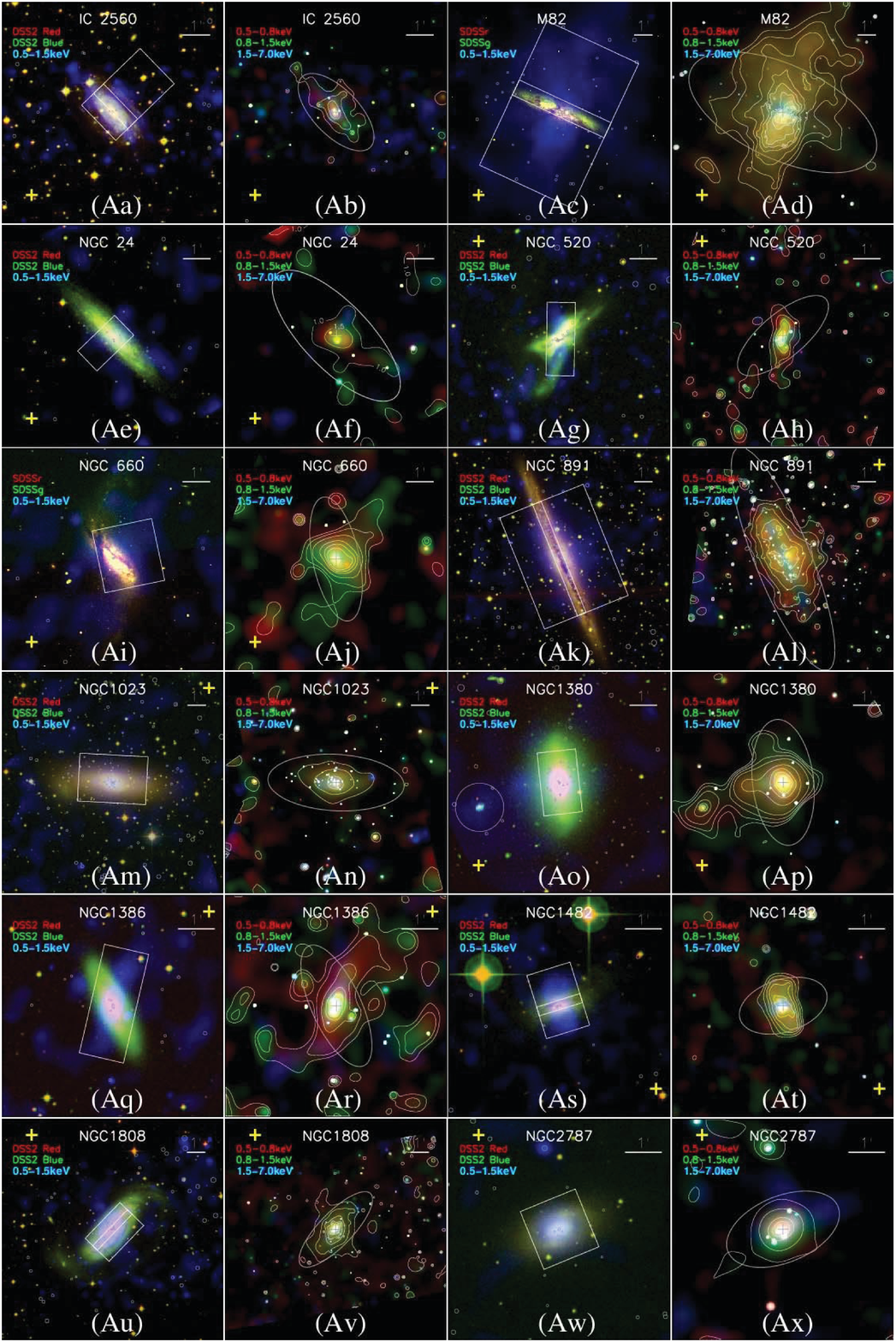,width=0.85\textwidth,angle=0, clip=}
\caption{Images of the sample galaxies. Two for each galaxy. The left panel shows the color-coded images: \emph{Red:} DSS red or SDSS r-band image; \emph{Green:} DSS blue or SDSS g-band image; and \emph{Blue:} the point-source-removed, non-X-ray-background-subtracted, exposure-corrected, and adaptively smoothed \emph{Chandra} 0.5-1.5~keV image. The small circles mark the removed X-ray point sources. The boxes show the spectral analysis and/or luminosity calculation regions; an inner one, if presented, shows the galactic disk that is filtered out. Some additional regions, such as the large circular region of NGC~1380 and NGC~3384, as well as the box region to the south of NGC~4565, are not used for spectral extraction of the coronae, but are filtered-out regions in sky background analysis or for the analysis of some regions of particular interest (see \S\ref{PaperIsec:Individual} for details). See \S\ref{PaperIsubsec:spatial} and \ref{PaperIsubsec:spec} for the definition of these regions. The yellow plus marks the (far) side of the galaxy, which is less obscured by the dusty cool gas in the disk. The right panel of each galaxy presents the \emph{Chandra} images: \emph{Red:} 0.5-0.8~keV image; \emph{Green:} 0.8-1.5~keV; and \emph{Blue:} 1.5-7~keV. While point sources are not removed in these images, the contours represent the source-removed 0.5-1.5~keV intensity distribution as in the left panel. The ellipse marks the stellar light extension defined by $D_{25}$ and $i$ in Table~\ref{table:SampleSelectionPara}.}\label{fig:imagea}
\end{center}
\end{figure}
\addtocounter{figure}{-1}
\begin{figure}[!h]
\begin{center}
\epsfig{figure=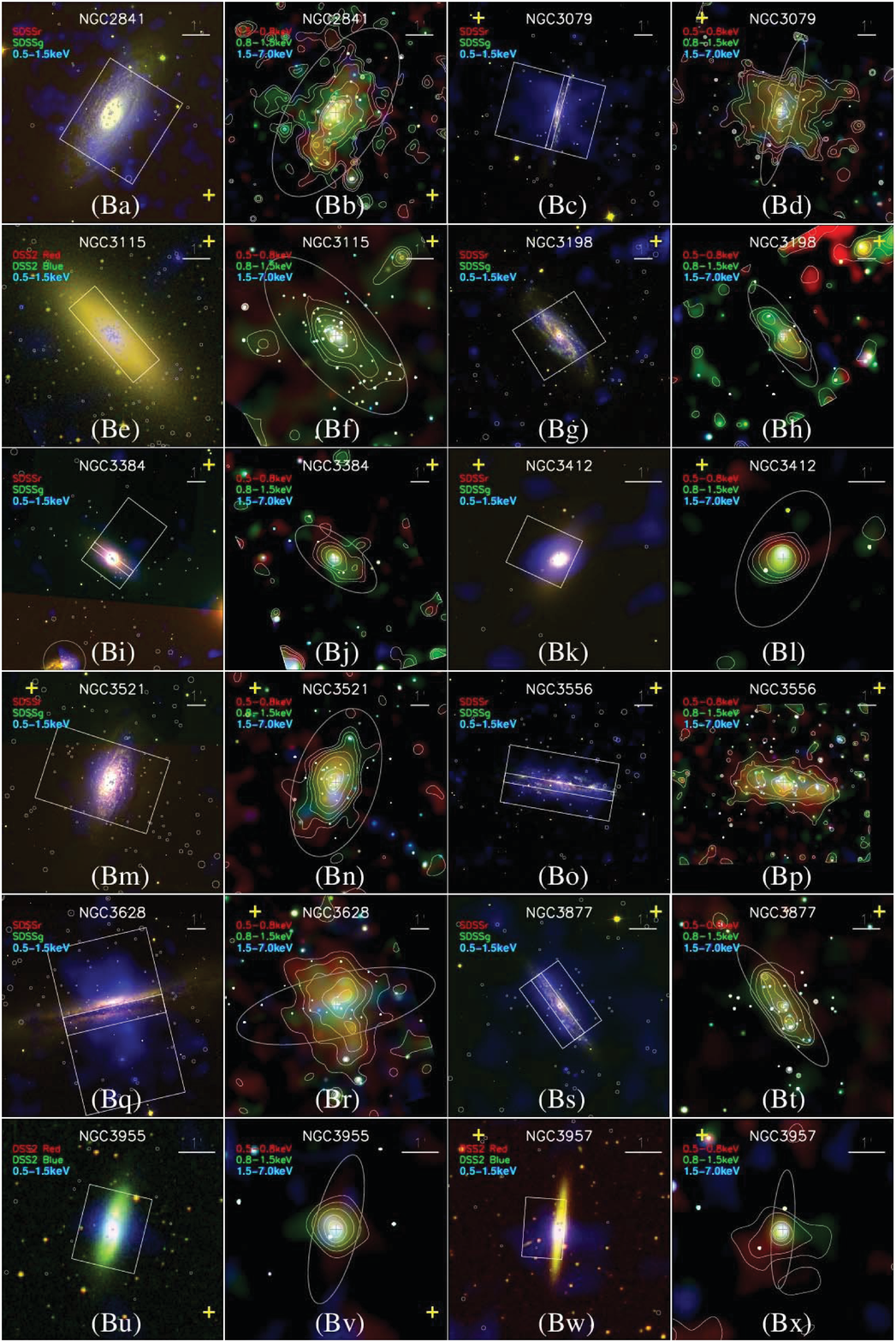,width=0.85\textwidth,angle=0, clip=}
\caption{continued.}%\label{fig:imageb}
\end{center}
\end{figure}
\addtocounter{figure}{-1}
\begin{figure}[!h]
\begin{center}
\epsfig{figure=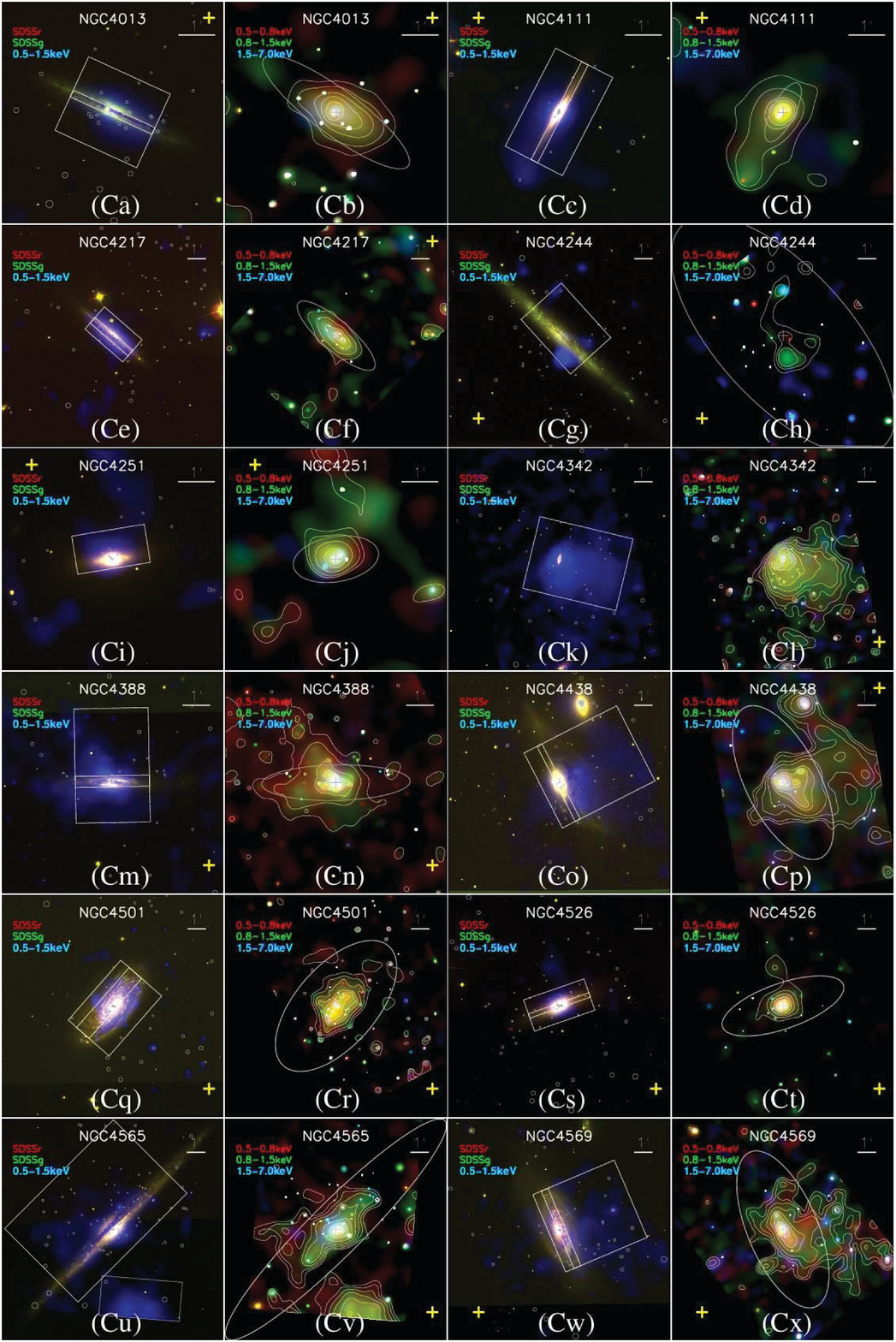,width=0.85\textwidth,angle=0, clip=}
\caption{continued.}%\label{fig:imagec}
\end{center}
\end{figure}
\addtocounter{figure}{-1}
\begin{figure}[!h]
\begin{center}
\epsfig{figure=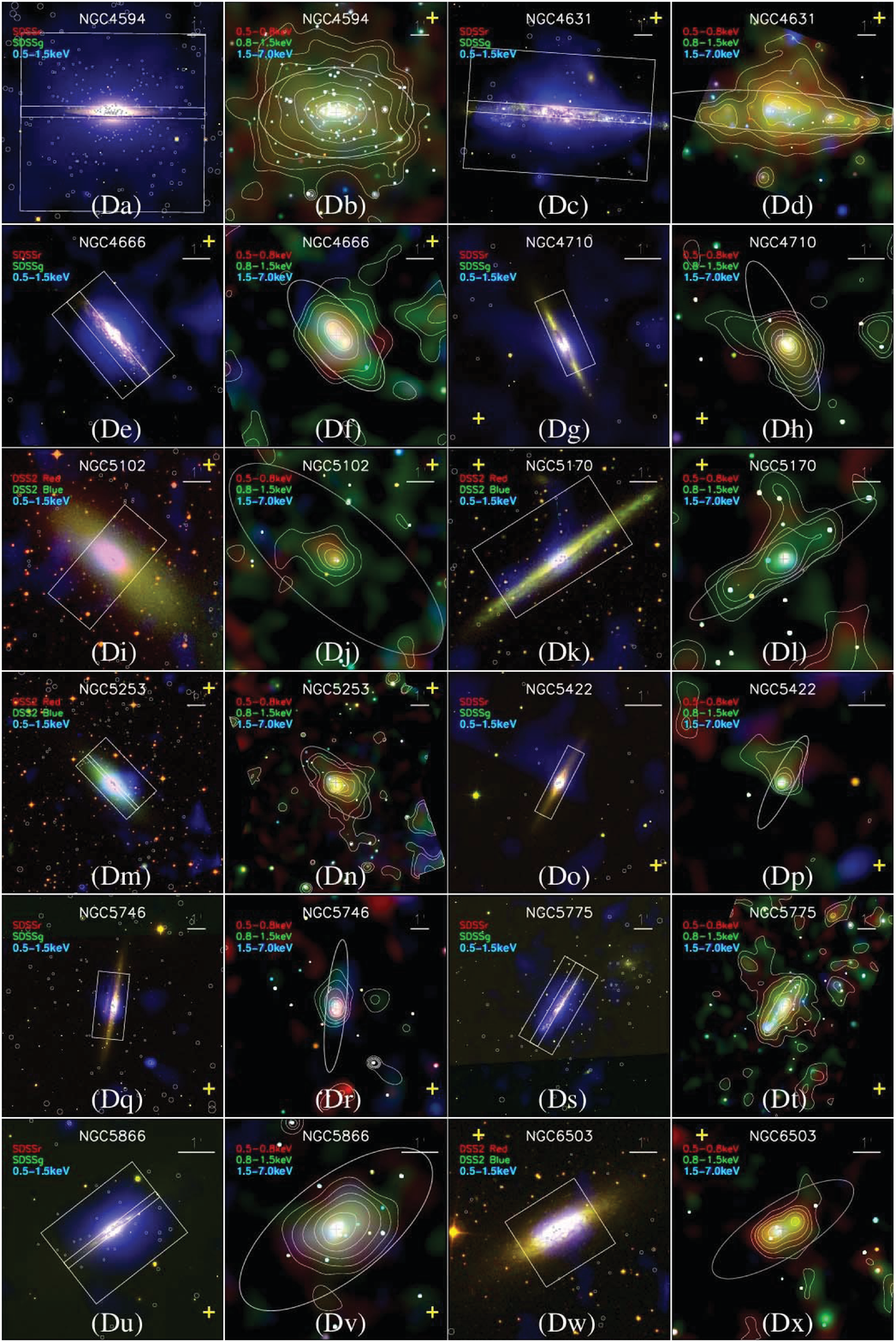,width=0.85\textwidth,angle=0, clip=}
\caption{continued.}%\label{fig:imaged}
\end{center}
\end{figure}
\addtocounter{figure}{-1}
\begin{figure}[!h]
\begin{center}
\epsfig{figure=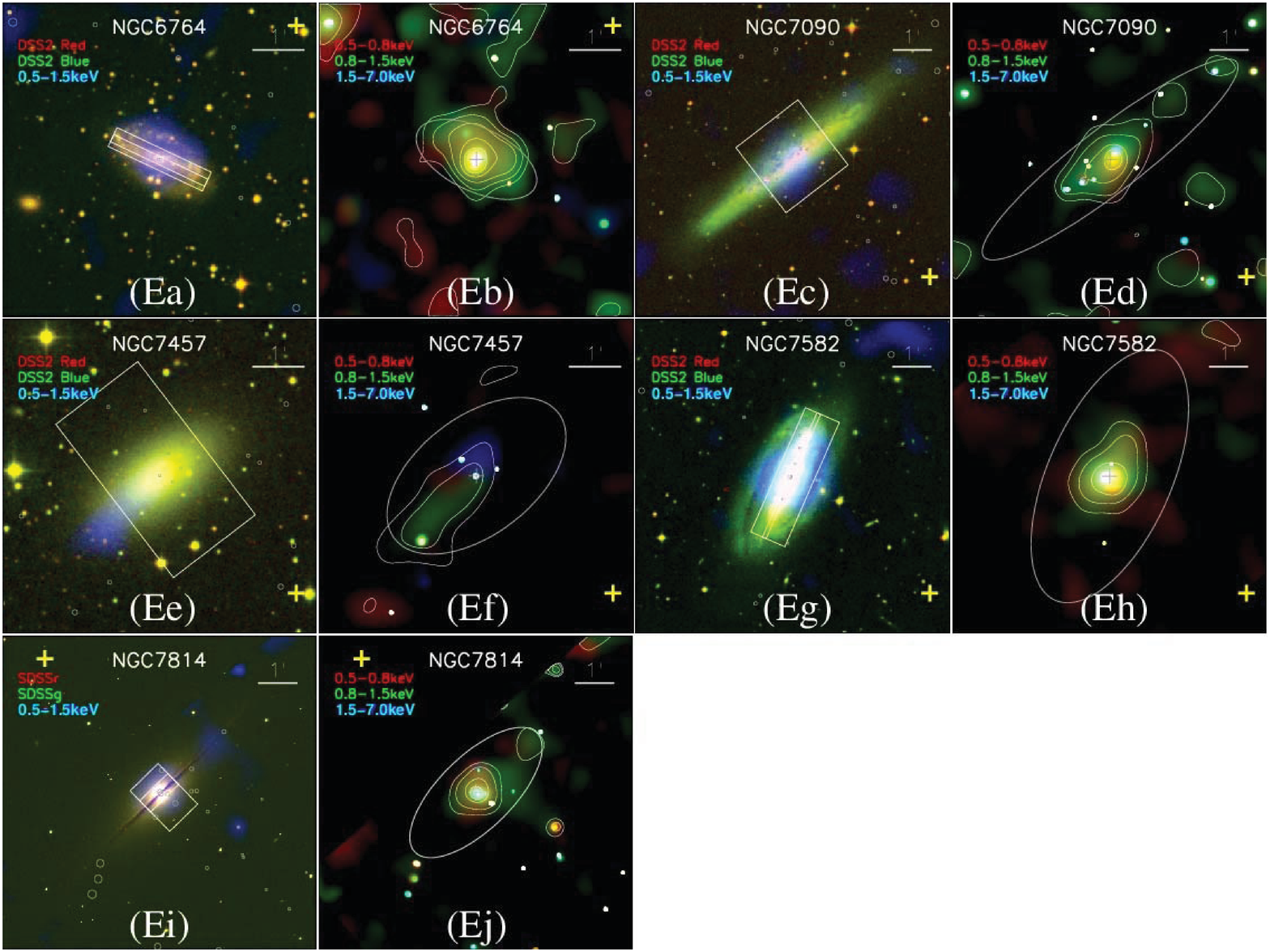,width=0.85\textwidth,angle=0, clip=}
\caption{continued.}%\label{fig:imagee}
\end{center}
\end{figure}

\subsection{Spatial Analysis}\label{PaperIsubsec:spatial}

We define the vertical range that contains most of the significant coronal emission. This range can be used uniformly for spectral extraction and for obtaining the overall luminosities of the coronae. We extract the surface brightness profile in the 0.5-1.5~keV band for each galaxy along its minor (vertical) axis. The horizontal range used to extract such a profile is mostly $|r|<D_{25}/4$, which typically contains bulk of the coronal emission (Fig.~\ref{fig:imagea}), and is adjusted for only a few galaxies with exceptionally large or small diffuse X-ray extensions (e.g., NGC~4342, Fig.~\ref{fig:imagea}Ck,Cl).

We fit each vertical profile with an exponential function plus a linear function with a slope to characterize the background. To minimize effects due to the absorption and/or enhanced emission in the galactic disk, our fit excludes an interval that corresponds to a significant valley and/or peak (if there is any) in the profile at the galactic major axis (shown as the inner box in Fig.~\ref{fig:imagea} for some galaxies, if presented). The best fits to the profiles of all the galaxies are shown in Fig.~\ref{fig:profilesa}, while the fitted scale heights ($h_{exp,+,-}$) are listed in Table~\ref{table:scaleheight}. Such a simple vertical profile extraction and fitting are \emph{not} aimed to specify the spatial distribution of the coronae (in fact, the unresolved stellar contribution can be important in some bulge-dominated galaxies, \S\ref{PaperIsubsec:spec}), but instead, to roughly characterize their extensions, which will chiefly be used for a uniform definition of the spectral analysis regions.

\begin{figure}[!h]
\begin{center}
\epsfig{figure=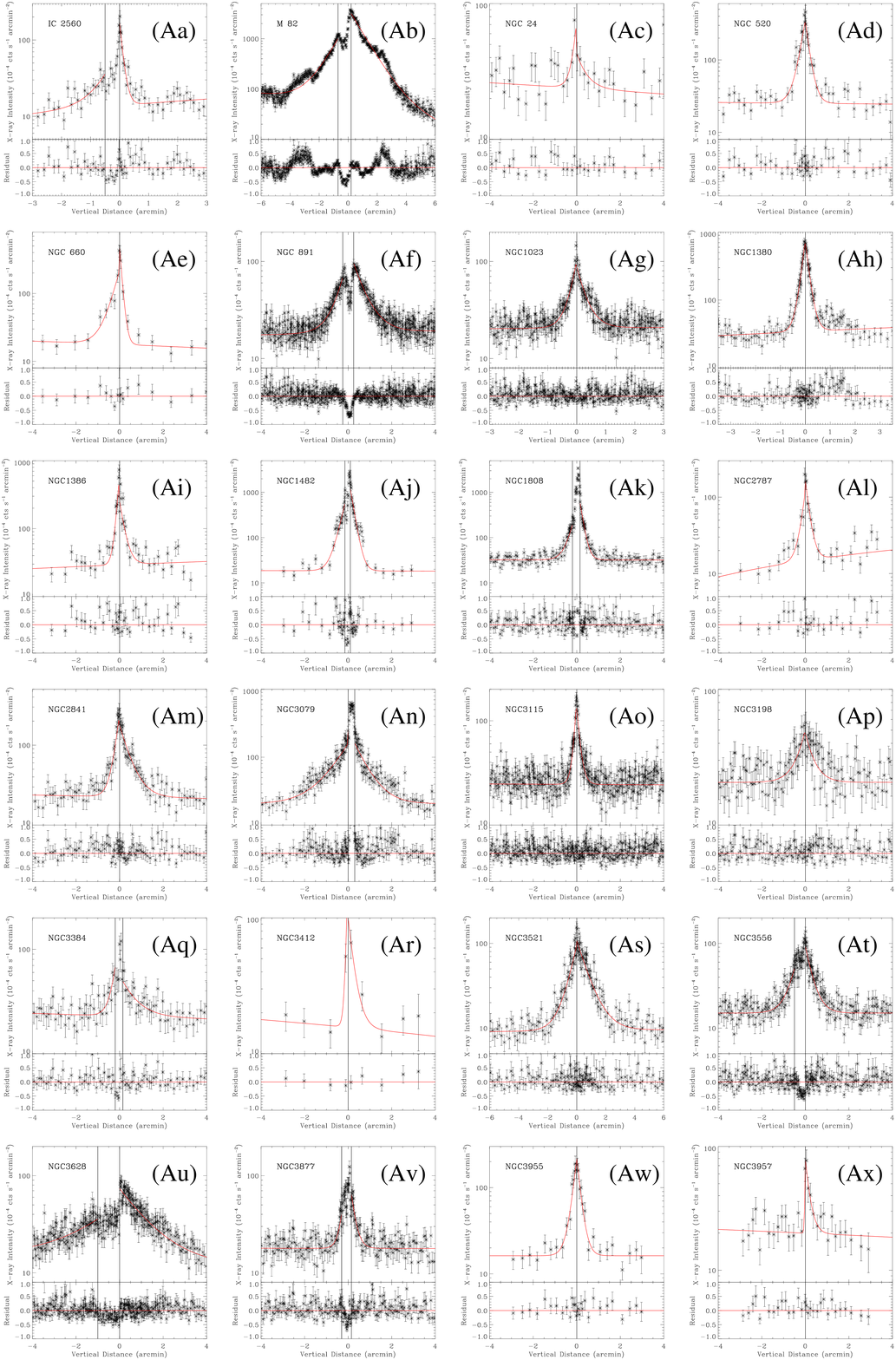,width=0.85\textwidth,angle=0, clip=}
\caption{Point-source-removed 0.5-1.5~keV vertical brightness distribution of the sample galaxies (see \S\ref{PaperIsubsec:spatial} for details). For each galaxy, the positive (less obscured) side and the galactic plane-parallel range used to extract the profile are shown in the corresponding panel of Fig.~\ref{fig:imagea}. The black vertical line marks the range filtered out for the galactic disk obscuration. The red solid curve represents a fit to the profile with an exponential plus a constant background. The lower sub-panel shows the residuals defined as $(data-model)/model$.}\label{fig:profilesa}
\end{center}
\end{figure}
\addtocounter{figure}{-1}
\begin{figure}[!h]
\begin{center}
\epsfig{figure=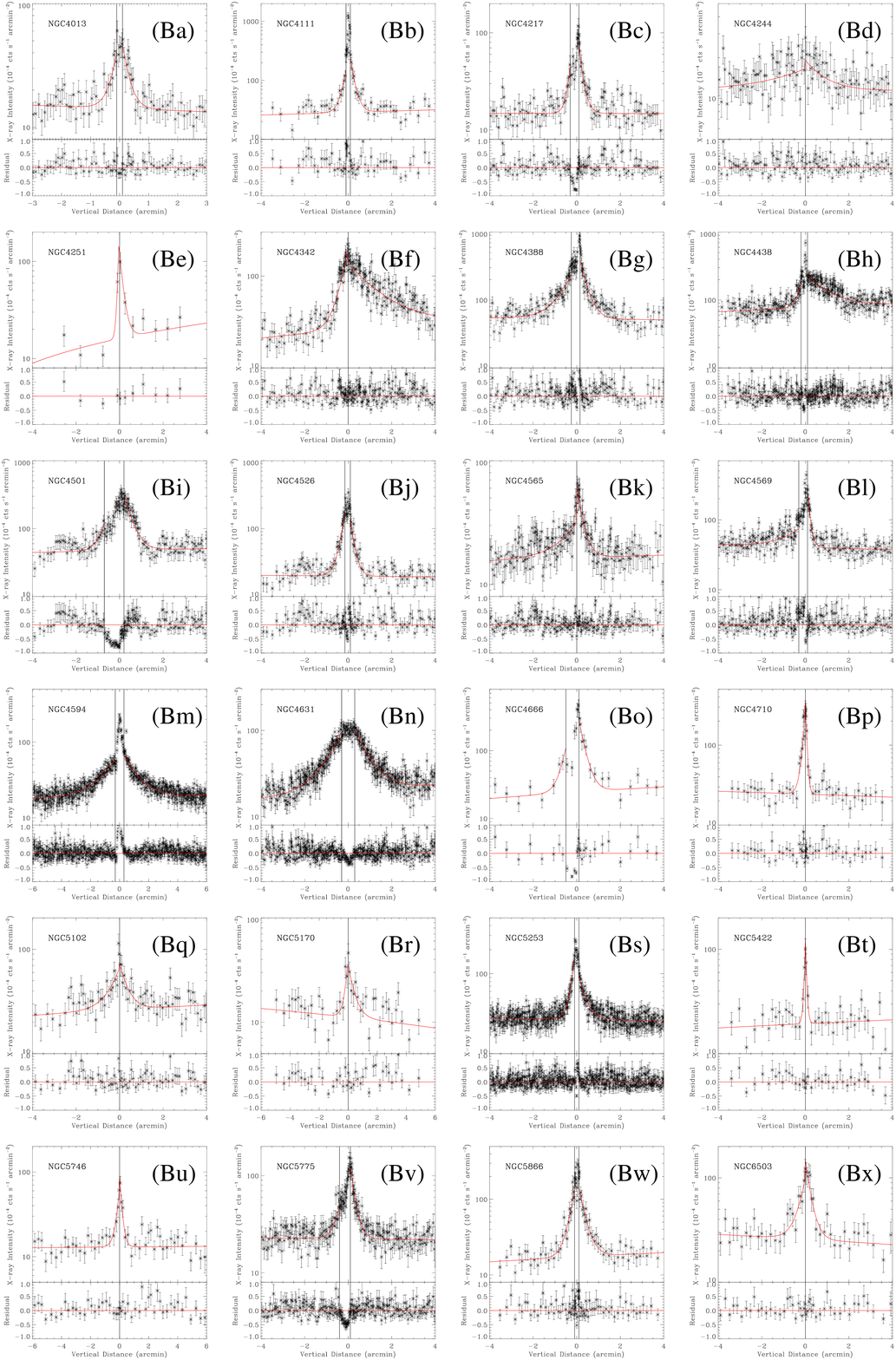,width=0.85\textwidth,angle=0, clip=}
\caption{continued.}%\label{fig:profilesb}
\end{center}
\end{figure}
\addtocounter{figure}{-1}
\begin{figure}[!h]
\begin{center}
\epsfig{figure=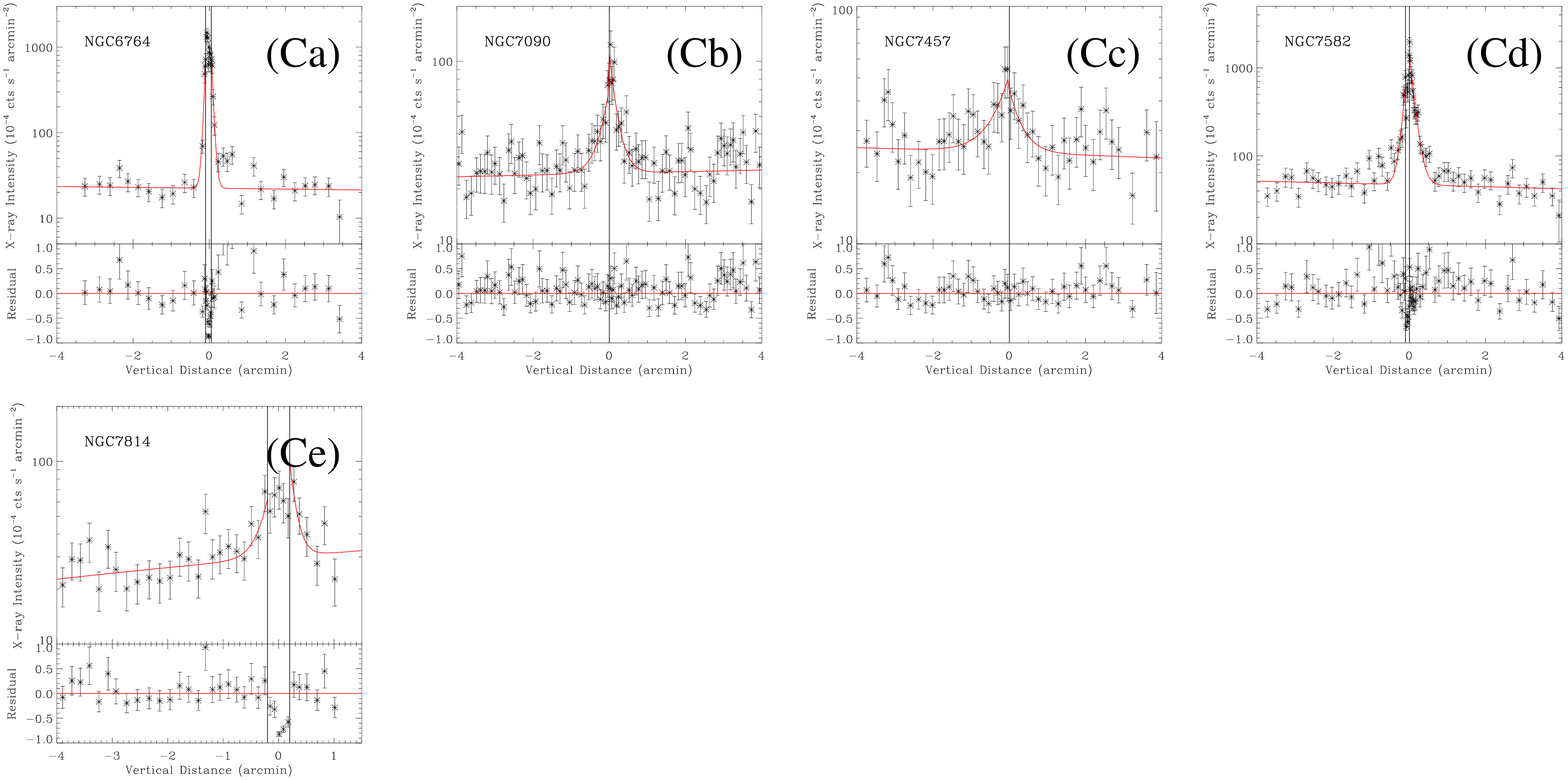,width=0.85\textwidth,angle=0, clip=}
\caption{continued.}%\label{fig:profilesc}
\end{center}
\end{figure}

%\clearpage
\begin{deluxetable}{lcc}
\centering
\tiny %\tiny\scriptsize\footnotesize\small\normalsize\large\Large\LARGE\huge\Huge
%\ptlandscape
  \tabletypesize{\tiny}
  \tablecaption{Diffuse X-ray Scale Height}
  \tablewidth{0pt}
  \tablehead{
 \colhead{Name} & \colhead{$h_{exp,+}$} & \colhead{$h_{exp,-}$}
}
\startdata
IC2560 & $0.11\pm0.02$ & $0.48\pm0.09$ \\
M82 & $0.89\pm0.004$ & $0.84\pm0.01$ \\
NGC24 & $0.44\pm0.31$ & $0.18\pm0.16$ \\
NGC520 & $0.18\pm0.03$ & $0.22\pm0.02$ \\
NGC660 & $0.10\pm0.05$ & $0.34\pm0.09$ \\
NGC891 & $0.56\pm0.01$ & $0.42\pm0.01$ \\
NGC1023 & $0.30\pm0.01$ & $0.22\pm0.01$ \\
NGC1380 & $0.14\pm0.006$ & $0.14\pm0.005$ \\
NGC1386 & $0.20\pm0.27$ & $0.10\pm0.04$ \\
NGC1482 & $0.15\pm0.24$ & $0.21\pm0.41$ \\
NGC1808 & $0.14\pm0.007$ & $0.22\pm0.02$ \\
NGC2787 & $0.17\pm0.02$ & $0.17\pm0.10$ \\
NGC2841 & $0.43\pm0.03$ & $0.19\pm0.02$ \\
NGC3079 & $0.58\pm0.05$ & $0.76\pm0.05$ \\
NGC3115 & $0.14\pm0.01$ & $0.11\pm0.006$ \\
NGC3198 & $0.39\pm0.05$ & $0.40\pm0.04$ \\
NGC3384 & $0.64\pm0.22$ & $0.21\pm0.08$ \\
NGC3412 & $0.26\pm0.27$ & $0.07\pm0.08$ \\
NGC3521 & $0.75\pm0.03$ & $0.52\pm0.03$ \\
NGC3556 & $0.31\pm0.02$ & $0.32\pm0.03$ \\
NGC3628 & $1.24\pm0.08$ & $1.82\pm0.55$ \\
NGC3877 & $0.19\pm0.02$ & $0.16\pm0.02$ \\
NGC3955 & $0.17\pm0.11$ & $0.18\pm0.09$ \\
NGC3957 & $0.20\pm0.14$ & $0.02\pm0.04$ \\
NGC4013 & $0.22\pm0.04$ & $0.22\pm0.05$ \\
NGC4111 & $0.18\pm0.03$ & $0.17\pm0.06$ \\
NGC4217 & $0.18\pm0.02$ & $0.14\pm0.06$ \\
NGC4244 & $0.90\pm0.31$ & $2.05\pm0.90$ \\
NGC4251 & $0.16\pm0.15$ & $0.05\pm0.04$ \\
NGC4342 & $1.40\pm0.13$ & $0.31\pm0.01$ \\
NGC4388 & $0.29\pm0.04$ & $0.54\pm0.05$ \\
NGC4438 & $0.94\pm0.05$ & $0.19\pm0.03$ \\
NGC4501 & $0.33\pm0.01$ & $0.26\pm0.10$ \\
NGC4526 & $0.17\pm0.01$ & $0.19\pm0.01$ \\
NGC4565 & $0.28\pm0.03$ & $0.85\pm0.33$ \\
NGC4569 & $0.11\pm0.009$ & $0.85\pm0.19$ \\
NGC4594 & $0.86\pm0.04$ & $1.08\pm0.04$ \\
NGC4631 & $0.64\pm0.02$ & $0.68\pm0.02$ \\
NGC4666 & $0.25\pm0.04$ & $0.24\pm0.07$ \\
NGC4710 & $0.05\pm0.01$ & $0.12\pm0.02$ \\
NGC5102 & $0.35\pm0.16$ & $0.50\pm0.25$ \\
NGC5170 & $0.40\pm0.10$ & $0.19\pm0.06$ \\
NGC5253 & $0.29\pm0.02$ & $0.16\pm0.007$ \\
NGC5422 & $0.05\pm0.03$ & $0.05\pm0.05$ \\
NGC5746 & $0.15\pm0.03$ & $0.19\pm0.12$ \\
NGC5775 & $0.20\pm0.01$ & $0.22\pm0.03$ \\
NGC5866 & $0.27\pm0.02$ & $0.22\pm0.02$ \\
NGC6503 & $0.25\pm0.03$ & $0.31\pm0.07$ \\
NGC6764 & $0.04\pm0.01$ & $0.04\pm0.02$ \\
NGC7090 & $0.19\pm0.03$ & $0.23\pm0.05$ \\
NGC7457 & $0.34\pm0.23$ & $0.40\pm0.20$ \\
NGC7582 & $0.11\pm0.01$ & $0.08\pm0.03$ \\
NGC7814 & $0.11\pm0.10$ & $0.18\pm0.18$
\enddata
\tablecomments{\scriptsize Scale heights of the exponential model from the fits to the 0.5-1.5~keV vertical brightness profiles in the positive and negative sides (``+'' and ``-'') of each galaxy. Errors are measured by resampling the data points within their statistical errors. Unresolved stellar contributions cannot be easily modeled and are not subtracted in these model fits.}\label{table:scaleheight}
\end{deluxetable}

\subsection{Spectral Analysis}\label{PaperIsubsec:spec}

\subsubsection{Spectra Extraction and Background Subtraction}\label{PaperIsubsubsec:SpectraExtraction}

We extract the spectrum of a galaxy from the region with the same horizontal extension and disk exclusion as those used for extracting its vertical surface brightness profile (\S\ref{PaperIsubsec:spatial}), except for several individuals with distorted diffuse X-ray morphology (for these galaxies, the regions are selected to cover the bulk of the diffuse X-ray emission, e.g., NGC~520 and NGC~1386; Fig.~\ref{fig:imagea}). The outermost vertical extension is taken to be $5~h_{exp}$, except for NGC~4342 ($\lesssim3~h_{exp}$ to limit the region on the S3 chip) and NGC~4244 ($h_{exp}$ is not well constrained due to the low counting statistics; the region is taken to cover the bulk of the diffuse X-ray emission). CCD edges and gaps are filtered out to minimize the uncertainty in the instrument spectral response, which may be affected by the dithering of the observations. We limit the on-source and background regions to the S3 chip for ACIS-S observations and to the I0-3 chips for ACIS-I observations. Fig.~\ref{fig:imagea} includes the outlines of the adopted spectral extraction regions, although they may be cut short by filtered CCD edges.

If different spectral regions are adopted, we correct for the derived luminosities by assuming the soft X-ray intensity profile can be characterized by the exponential model described in \S\ref{PaperIsubsec:spatial}. The regions typically include the most luminous part of the coronae, except for the residual of some disk absorption or emission features (Fig.~\ref{fig:profilesa}) and some very extended extraplanar features (e.g., NGC~2841; Fig.~\ref{fig:imagea}Ba, Bb). These residual features typically only account for a small fraction of the coronal emission and will be individually noted in \S\ref{PaperIsec:Individual}. Generally, they cannot seriously affect our statistical comparisons of the global coronal properties within such a large sample.

We extract the on-corona (source) and background spectra, as well as the corresponding response files using the CIAO tool \emph{``specextract''}, which is officially recommended for extended sources. The \emph{``specextract''} script automatically runs a set of CIAO tools. In particular, it has the ability to determine when \emph{``mkacisrmf''} should be used in place of \emph{``mkrmf''} in building the RMF (the response matrix file), based on the gain and CTI correction adopted for the event file, as well as the type and focal plane temperature of the CCD. The RMF and ARF (Ancillary Response Files) files are weighted using a detector weight map (WMAP) in 0.3-2~keV, created for the source and background files, respectively.

The sky background can be complicated in some fields (e.g., \citealt{Li08}). Therefore, we prefer a local background subtraction in most cases. Most of our sample galaxies are small enough to be covered by a single CCD chip, allowing for background regions selected within the same CCD, but still beyond $5~h_{exp}$. For a few galaxies with very extended coronae (filling the entire CCD, e.g., M82; Fig.~\ref{fig:imagea}Ac,Ad), the FoV of the \emph{Chandra} observations is too small to allow for a local background subtraction. In such a case, we use the blank-sky background files, which are reprojected and exposure adjusted to match the 10-12/10-14~keV (for FI/BI CCDs) count rate in the source spectrum. The coronal emission in such cases are strong. The uncertainty in the background estimate should not cause significant problem.

For each galaxy with good counting statistics, we rebin the background subtracted spectrum to achieve a minimum signal-to-noise ratio ($\gtrsim 3$) for spectral fitting with $\chi^2$ statistic.

\subsubsection{Spectral Modeling}\label{PaperIsubsubsec:SpectraAnalysis}

To model the spectra of the galactic coronae, we need to subtract the residual contributions from discrete sources. In \emph{X-ray faint} galaxies, after removing the brightest X-ray point sources, the unresolved stellar contribution may still be non-negligible, or may even dominate the hot gas emission. \citet{Revnivtsev07,Revnivtsev09} have successfully calibrated this individually faint stellar source component in M32, which is not massive enough to hold a galactic corona, or by directly resolving them in the Galactic ridge (also see \citealt{LiZ07b,Revnivtsev08,Bogdan11,Mineo11}). This component is mainly comprised of cataclysmic variables (CVs) and coronal active binaries (ABs), which individually have a luminosity of $\lesssim10^{34}\rm~ergs~s^{-1}$ (hereafter the CV+AB component). The CV+AB contribution can be characterized with a $kT=0.5\rm~keV$ thermal plasma plus a $\Gamma=1.9$ power law \citep{Revnivtsev08}. The characterization of this CV+AB component (or its X-ray spectral contribution) scaled with the stellar mass enables us to quantitatively isolate the truly diffuse hot gas emission in X-ray faint, especially bulge-dominated galaxies (e.g., \citealt{Li09,Li11,LiZ11,Boroson11,Wong11}). As a result, we can just simply include a fixed accumulated CV+AB component in the spectral modeling of an \emph{early-type} disk galaxy. Exceptions are made for two individuals, NGC~520 and NGC~660, which show highly distorted morphologies with warped disks \citep{vanDriel95}; the old stellar population and hence the CV+AB contribution are too uncertain to quantify. In fact, the diffuse X-ray emission is strongly elongated along the galaxies' minor axes, indicating that the stellar contribution is not important and has a significantly different distribution from their stellar light (Fig.~\ref{fig:imagea}Ag-Aj). Similarly, we also neglect the stellar contribution for late-type galaxies, which is expected to be small and to be largely absorbed by the gas-rich galactic disks. In addition, we also include a residual contribution from the removed sources due to counts spilled out of the removal circles and from X-ray binaries below our detection limit for each galaxy. This contribution is expected to be small because of the flat luminosity function of X-ray binaries at low luminosities (e.g., \citealt{Kim10}). As is usually done, the spectral shape of this component is assumed to be a simple power law, which may also approximately account for the possible residual instrumental and/or sky background. This component is typically less important in soft X-ray, e.g., usually more than an order of magnitude lower than the hot gas emission in the 0.5-1.5~keV band (Figs.~\ref{fig:speca} and \ref{fig:spec2T}).

We fit a coronal spectrum with a 1-T thermal plasma model (XSPEC model MEKAL or VMEKAL; \citealt{Mewe85}) (Tables~\ref{table:specmodel}-\ref{table:specpara1Tmetal}). Elements are divided into two groups: O-like (C, N, O, Ne, Mg, Si, S, Ar, Ca) and Fe-like (the other elements); the respective abundances of O-like elements are fixed to be solar. For 14 spectra with relatively high counting statistics, we fit the Fe/O abundance ratio (see \S\ref{PaperIsubsubsec:ThermalChemicalStates} for details). For other spectra, we simply set the ratio to be 0.3 for late-type galaxies \citep{Strickland04a} and 1 for early-type ones \citep{Kim11}. All the spectral components are subject to the Galactic foreground absorption (WABS; the $N_H$ values are listed in Table~\ref{table:SampleSelectionPara}). Although strong variation in the absorption column on scales of several tens of pc may be expected in superwinds, such an uncertainty in $N_H$ has very limited effects on the modeling of the accumulated spectra of the galactic coronae \citep{Strickland00b}. For galaxies with particularly thick cool gas disks (e.g., NGC~5775; \citealt{Irwin94}), there is also uncertainty in the residual X-ray absorption from the emission right behind the disks. For most of our sample galaxies, the spectral quality is not sufficiently high to allow for directly constraining $N_H$. For galaxies with good counting statistics, we fit the $N_H$ value and found it is typically not significantly different from the Galactic foreground value. We thus fix $N_H$ at the Galactic value, assuming that the intrinsic absorption of the extraplanar emission is not significant, although it may be important for galaxies with other orientations \citep{Mineo12}.

\begin{figure}[!h]
\begin{center}
\epsfig{figure=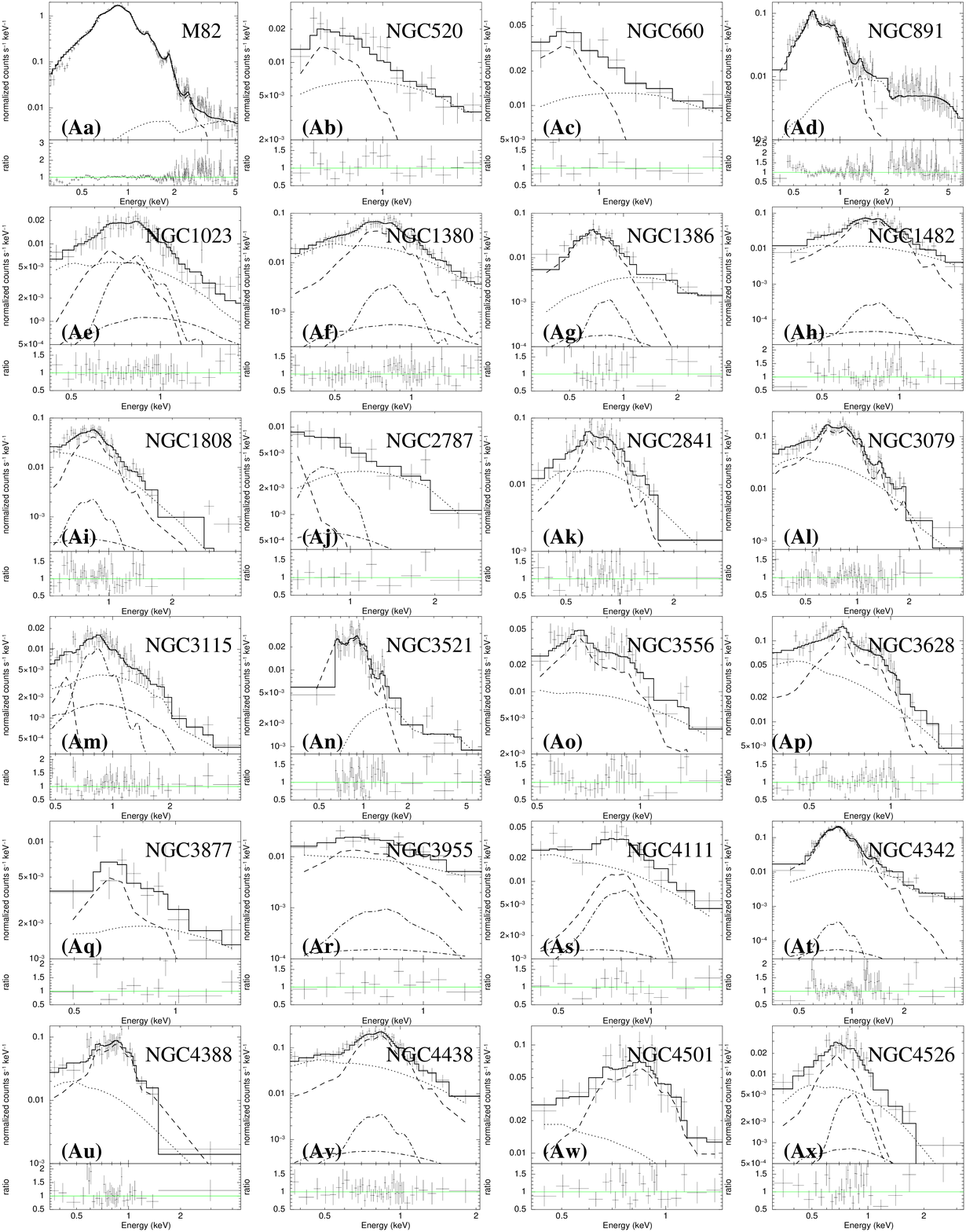,width=0.85\textwidth,angle=0, clip=}
\caption{Spectra of the apparently diffuse X-ray emission from the sample galaxies. Each spectrum is fitted with an 1-T thermal plasma model of a fixed abundance ratio  (dashed curve), plus other components: CV+AB (dash-dotted) and point source residuals (dotted). The CV+AB component, if significant, is modeled with a thermal plasma and a power law, while the source residuals with a power law (see \S\ref{PaperIsubsec:spec} for details). The solid curve represents the total model spectrum.}\label{fig:speca}
\end{center}
\end{figure}
\addtocounter{figure}{-1}
\begin{figure}[!h]
\begin{center}
\epsfig{figure=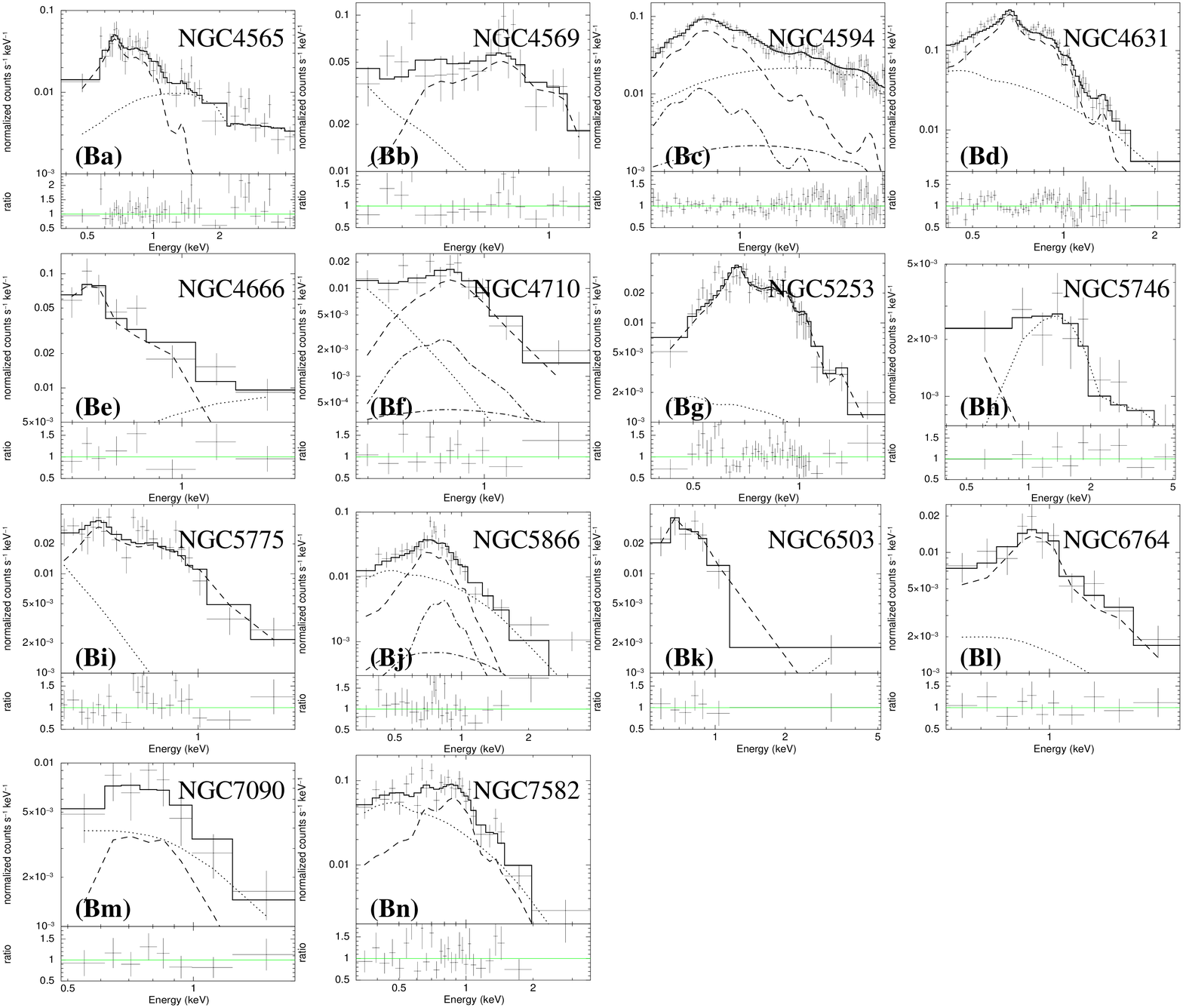,width=0.85\textwidth,angle=0, clip=}
\caption{continued.}%\label{fig:specb}
\end{center}
\end{figure}

\begin{figure}[!h]
\begin{center}
\epsfig{figure=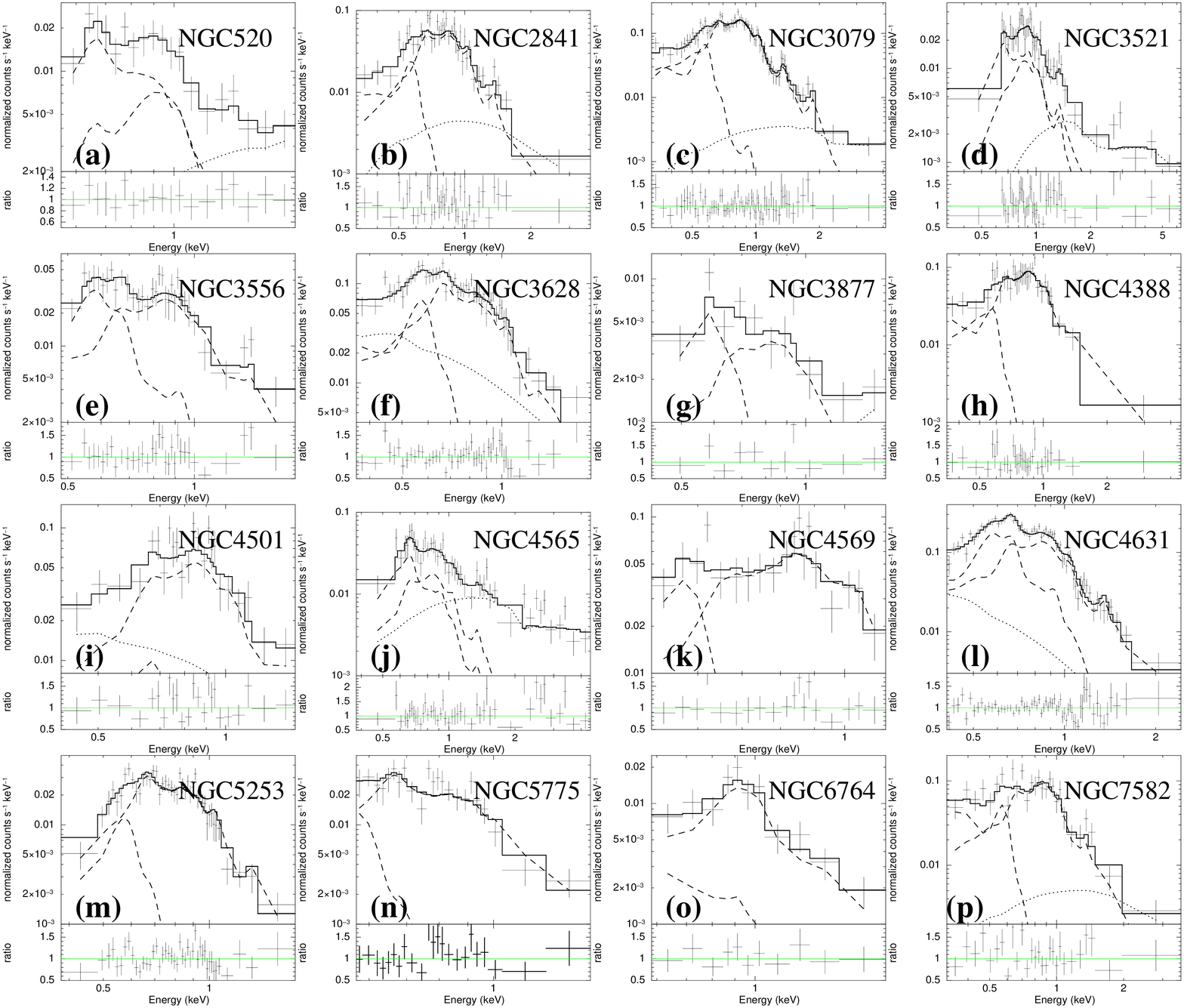,width=0.85\textwidth,angle=0, clip=}
\caption{2-T plasma model fit of the apparently diffuse X-ray spectra of the sample galaxies. Symbols are the same as those in Fig.~\ref{fig:speca}.}\label{fig:spec2T}
\end{center}
\end{figure}

We also test a 2-T model for several \emph{late-type} galaxies to characterize the temperature range of the coronae (\S\ref{PaperIsec:Introduction}), while the abundances are fixed for simplicity. For 16 galaxies, such a modeling shows evidence for the presence of distinct different temperature components. We list the parameters of the two thermal components in Table~\ref{table:specpara2T} and show the best-fit spectra in Fig.~\ref{fig:spec2T}.

We emphasize that with the limited quality, the spectral data themselves are not sufficient to justify a specific decomposition of various components, which is selected to represent the physical plausibility (e.g., the presence of optically-thin plasma) and the best calibration (e.g., the CV+AB contribution). We also fix as many parameters as possible (e.g., the CV+AB component, and in some cases the abundances). But when a component (e.g., the power law component used to characterize the residual background and stellar contributions) is hard to quantify with reasonable certainty, we choose to fit it so that its uncertainty can be propagated into the estimation of other spectral parameters. However, in almost all cases, we find that the power law component is less important (than the thermal and/or the CV+AB component) in soft X-ray and thus has very limited effect on the modeling of the galactic coronae. Typically, each spectral model fit contains a total of only 4-6 free model parameters (i.e., the power law index and normalization, one or two temperatures plus the corresponding normalizations of the thermal plasma), which can be usefully constrained with reasonably good counting statistics of a spectrum (Table~\ref{table:specmodel}).

For a corona with inadequate counting statistics for the above spectral fit, we estimate the luminosity simply from the background-subtracted 0.5-1.5~keV count rate detected in the same region as would be used for a spectral extraction. This conversion assumes a thermal plasma of a typical 0.3~keV temperature. The other parameters (abundances and the CV+AB component) are adopted in the same way as those used in the spectral fit cases. But the power law component is omitted, because its contribution is typically negligible, compared to the uncertainty in the luminosity estimate.

We compare the results (the directly measured hot gas properties, such as luminosity, temperature, abundance ratio, and emission measure; Table~\ref{table:1Tspec}) to those from previous analyses of the same \emph{Chandra} data (33 galaxies; Table~\ref{table:SampleSelectionPara}). Most of the results are consistent with each other within the quoted errors or with suitable corrections (e.g., for the differences in the spectral extraction regions and/or the subtraction of stellar sources). Several galaxies with significantly inconsistent results are individually noted in \S\ref{PaperIsec:Individual}.

The errors of the hot gas parameters as listed in Tables~\ref{table:specmodel}-\ref{table:specpara2T} are statistical only (i.e., from the spectral fits). Systematic uncertainties (e.g., in the assumed spectral components and models), which are difficult to quantify, will be discussed in \S\ref{PaperIsec:Discussion}, and will be reflected in the dispersions of subsequent correlation relations of the parameters (Paper~II).

\clearpage
%\begin{deluxetable}{lcccccccc}
\begin{longtable}{ccccccccc}
\centering
\tiny %\tiny\scriptsize\footnotesize\small\normalsize\large\Large\LARGE\huge\Huge
  \tablecaption{Spectral Fit Details}
 \tablehead{
\colhead{Name} & \colhead{Model} & \colhead{Para} & \colhead{Total/Net Cts} & \colhead{$\chi^2/\rm DoF$} & \colhead{$\Gamma$} & \colhead{$N_{power}$} & \colhead{$N_{CV,mekal}$} \\
 (1) & (2) & (3) & (4) & (5) & (6) & (7) & (8)
}
%\startdata
%IC2560  & Fixed  & 1 & 53/35      & -          & -                      & -                      & -    \\
M82     & MEKAL  & 4 & 70448/64174& 571.21/180 & $-0.47\pm0.05$         & $0.72_{-0.23}^{+0.29}$ & -    \\
        & VMEKAL & 5 & 70448/64174& 467.98/179 & $0.40_{-0.07}^{+0.06}$ & $2.50_{-0.93}^{+0.63}$ & -    \\
%NGC24   & Fixed  & 1 & 53/35      & -          & -                      & -                      & -    \\
NGC520  & MEKAL  & 4 & 614/395    & 12.29/15   & $1.54_{-0.24}^{+0.41}$ & $1.33_{-0.31}^{+0.45}$ & -    \\
        & 2MEKAL & 6 & 614/395    & 3.70/13    & $-0.59_{-0.95}^{+0.86}$& $0.36_{-0.26}^{+0.41}$ & -    \\
NGC660  & MEKAL  & 4 & 223/179    & 5.76/7     & $1.04_{-0.40}^{+0.45}$ & $2.74_{-1.27}^{+1.57}$ & -    \\
NGC891  & MEKAL  & 4 & 21944/9074 & 232.12/129 & $0.19\pm0.04$          & $1.61\pm0.17$          & -    \\
        & VMEKAL & 5 & 21944/9074 & 222.52/128 & $0.29_{-0.04}^{+0.06}$ & $1.80_{-0.16}^{+0.20}$ & -    \\
NGC1023 & MEKAL  & 4 & 6398/2358  & 30.58/45   & $3.11_{-0.28}^{+0.33}$ & $1.03\pm0.15$          & 0.46 \\
NGC1380 & MEKAL  & 4 & 1971/1592  & 44.53/54   & $3.07\pm0.17$          & $2.93_{-0.37}^{+0.26}$ & 0.19 \\
NGC1386 & MEKAL  & 4 & 815/472    & 25.48/18   & $0.55_{-0.29}^{+0.28}$ & $0.67_{-0.29}^{+0.42}$ & 0.06 \\
NGC1482 & MEKAL  & 4 & 977/874    & 53.96/39   & $2.37_{-0.23}^{+0.22}$ & $1.77_{-0.26}^{+0.25}$ & 0.015\\
        & VMEKAL & 5 & 977/874    & 44.24/38   & $2.96_{-0.51}^{+0.67}$ & $0.70_{-0.53}^{+0.32}$ & 0.015\\
NGC1808 & MEKAL  & 4 & 1782/1160  & 48.41/40   & $3.37_{-0.34}^{+0.28}$ & $1.87\pm0.28$          & 0.11 \\
        & VMEKAL & 5 & 1782/1160  & 39.05/39   & $-0.94_{-1.48}^{+0.35}$& $0.05_{-0.05}^{+0.08}$ & 0.11 \\
NGC2787 & MEKAL  & 4 & 394/216    & 4.11/7     & $1.01_{-0.73}^{+0.45}$ & $0.64_{-0.24}^{+0.19}$ & 0.22 \\
NGC2841 & MEKAL  & 4 & 2019/1049  & 44.05/35   & $2.46_{-0.35}^{+0.32}$ & $2.51_{-0.66}^{+0.30}$ & -    \\
        & VMEKAL & 5 & 2019/1049  & 42.44/34   & $2.41\pm0.26$          & $2.65_{-0.50}^{+0.46}$ & -    \\
        & 2MEKAL & 6 & 2019/1049  & 38.13/33   & $1.24_{-0.45}^{+0.90}$ & $0.89_{-0.60}^{+0.77}$ & -    \\
NGC3079 & MEKAL  & 4 & 4620/2765  & 69.30/63   & $2.46_{-0.23}^{+0.16}$ & $3.45_{-0.79}^{+0.36}$ & -    \\
        & VMEKAL & 5 & 4620/2765  & 69.07/62   & $2.46_{-0.21}^{+0.20}$ & $3.31_{-0.67}^{+0.65}$ & -    \\
        & 2MEKAL & 6 & 4620/2765  & 46.70/61   & $0.16_{-0.41}^{+0.38}$ & $0.55_{-0.34}^{+0.48}$ & -    \\
NGC3115 & MEKAL  & 4 & 4816/1868  & 50.68/50   & $1.80\pm0.22$          & $0.87\pm0.08$          & 0.59 \\
%NGC3198 & Fixed  & 1 & 53/35      & -          & -                      & -                      & -    \\
%NGC3384 & Fixed  & 1 &
%NGC3412 & Fixed  & 1 &
NGC3521 & MEKAL  & 4 & 7292/1815  & 41.64/38   & $0.51_{-0.11}^{+0.22}$ & $0.79_{-0.27}^{+0.37}$ & -    \\
        & 2MEKAL & 6 & 7292/1815  & 41.00/36   & $0.31_{-0.43}^{+0.39}$ & $0.61_{-0.27}^{+0.37}$ & -    \\
NGC3556 & MEKAL  & 4 & 2564/1222  & 34.97/29   & $2.14_{-0.52}^{+0.60}$ & $1.30_{-0.41}^{+0.35}$ & -    \\
        & VMEKAL & 5 & 2564/1222  & 34.97/28   & $2.13_{-0.57}^{+0.56}$ & $1.31_{-0.41}^{+0.35}$ & -    \\
        & 2MEKAL & 6 & 2564/1222  & 22.51/27   & $<4.45$                & $0.09_{-0.05}^{+0.18}$ & -    \\
NGC3628 & MEKAL  & 4 & 7419/3665  & 66.20/45   & $3.20_{-0.16}^{+0.20}$ & $3.04_{-0.59}^{+0.33}$ & -    \\
        & VMEKAL & 5 & 7419/3665  & 64.27/44   & $2.88_{-0.18}^{+0.16}$ & $3.74_{-0.41}^{+0.46}$ & -    \\
        & 2MEKAL & 6 & 7419/3665  & 50.73/43   & $3.32_{-0.40}^{+0.50}$ & $1.55_{-0.72}^{+0.71}$ & -    \\
NGC3877 & MEKAL  & 4 & 1056/372   & 12.49/7    & $1.33_{-0.64}^{+0.76}$ & $0.31_{-0.12}^{+0.10}$ & -    \\
        & 2MEKAL & 6 & 1056/372   & 10.29/5    & $<-0.05$               & $0.07_{-0.02}^{+0.16}$ & -    \\
NGC3955 & MEKAL  & 4 & 301/220    & 6.21/8     & $2.97_{-1.08}^{+0.84}$ & $1.28_{-0.42}^{+0.39}$ & 0.05 \\
%NGC3957 & Fixed  & 1 &
%NGC4013 & Fixed  & 1 &
NGC4111 & MEKAL  & 4 & 518/341    & 14.14/13   & $2.69_{-0.34}^{+0.36}$ & $1.75_{-0.38}^{+0.35}$ & 0.34 \\
%NGC4217 & Fixed  & 1 &
%NGC4244 & Fixed  & 1 &
%NGC4251 & Fixed  & 1 &
NGC4342 & MEKAL  & 4 & 5131/2932  & 81.58/57   & $1.27_{-0.14}^{+0.17}$ & $2.33_{-0.46}^{+0.48}$ & 0.02 \\
        & VMEKAL & 5 & 5131/2932  & 79.99/56   & $0.99_{-0.25}^{+0.22}$ & $1.68_{-0.65}^{+0.67}$ & 0.02 \\
NGC4388 & MEKAL  & 4 & 2837/1103  & 23.71/28   & $3.00_{-0.45}^{+0.30}$ & $1.49_{-1.17}^{+0.88}$ & -    \\
        & 2MEKAL & 6 & 2837/1103  & 18.16/26   & $<1.54$                & $0.09_{-0.09}^{+0.47}$ & -    \\
NGC4438 & MEKAL  & 4 & 6421/2653  & 19.96/38   & $2.82_{-0.28}^{+0.25}$ & $5.42_{-0.96}^{+1.35}$ & 0.17 \\
NGC4501 & MEKAL  & 4 & 1269/678   & 23.43/20   & $3.44_{-0.58}^{+2.08}$ & $1.37_{-1.22}^{+0.97}$ & -    \\
        & 2MEKAL & 6 & 1269/678   & 23.14/18   & $3.31_{-0.63}^{+4.64}$ & $1.30_{-1.26}^{+1.05}$ & -    \\
NGC4526 & MEKAL  & 4 & 1311/659   & 30.24/23   & $2.64_{-0.46}^{+0.45}$ & $0.86_{-0.19}^{+0.18}$ & 0.26 \\
        & VMEKAL & 5 & 1311/659   & 30.14/22   & $2.60_{-0.46}^{+0.51}$ & $0.88_{-0.20}^{+0.18}$ & 0.26 \\
NGC4565 & MEKAL  & 4 & 7998/1979  & 40.79/41   & $0.48_{-0.12}^{+0.14}$ & $1.75_{-0.34}^{+0.40}$ & -    \\
        & 2MEKAL & 6 & 7998/1979  & 40.11/39   & $0.39\pm0.23$          & $1.58_{-0.30}^{+0.42}$ & -    \\
NGC4569 & MEKAL  & 4 & 1822/805   & 19.99/15   & $6.34_{-2.47}^{+1.92}$ & $0.60_{-0.38}^{+2.27}$ & -    \\
        & 2MEKAL & 6 & 1822/805   & 15.52/13   & $<7.82$                & $<1.01$                & -    \\
NGC4594 & MEKAL  & 4 & 19456/9949 & 110.17/85  & $1.11\pm0.24$          & $6.35_{-0.79}^{+0.88}$ & 1.32 \\
        & VMEKAL & 5 & 19456/9949 & 109.65/84  & $1.03_{-0.31}^{+0.21}$ & $5.91_{-0.95}^{+0.82}$ & 1.32 \\
NGC4631 & MEKAL  & 4 & 14145/7322 & 104.40/58  & $2.59_{-0.22}^{+0.20}$ & $3.68_{-0.69}^{+0.32}$ & -    \\
        & VMEKAL & 5 & 14145/7322 & 101.83/57  & $2.51_{-0.19}^{+0.18}$ & $3.74_{-0.44}^{+0.57}$ & -    \\
        & 2MEKAL & 6 & 14145/7322 & 67.38/56   & $3.84_{-0.48}^{+1.83}$ & $0.68_{-0.56}^{+0.60}$ & -    \\
NGC4666 & MEKAL  & 4 & 187/132    & 4.79/4     & $-0.32_{-0.76}^{+1.51}$& $1.12_{-0.80}^{+2.43}$ & -    \\
NGC4710 & MEKAL  & 4 & 305/221    & 7.49/8     & $8.77_{-0.72}^{+0.59}$ & $0.07_{-0.04}^{+0.11}$ & 0.15 \\
%NGC5102 & Fixed  & 1 &
%NGC5170 & Fixed  & 1 &
NGC5253 & MEKAL  & 4 & 6862/2721  & 64.02/40   & $3.21_{-1.35}^{+0.72}$ & $0.18_{-0.16}^{+0.17}$ & -    \\
        & VMEKAL & 5 & 6862/2721  & 59.12/39   & $2.51_{-0.76}^{+0.68}$ & $0.34_{-0.16}^{+0.19}$ & -    \\
        & 2MEKAL & 6 & 6862/2721  & 50.58/38   & -                      & -                      & -    \\
%NGC5422 & Fixed  & 1 &
NGC5746 & MEKAL  & 4 & 484/233    & 4.40/6     & $0.87_{-0.28}^{+0.14}$ & $0.72\pm0.17$          & -    \\
NGC5775 & MEKAL  & 4 & 1306/679   & 20.45/19   & $<9.50$                & $0.04_{-0.02}^{0.08}$  & -    \\
        & 2MEKAL & 6 & 1306/679   & 19.31/19   & -                      & -                      & -    \\
NGC5866 & MEKAL  & 4 & 1365/724   & 32.99/24   & $2.89_{-0.42}^{+0.36}$ & $1.17_{-0.28}^{+0.26}$ & 0.20 \\
        & VMEKAL & 5 & 1365/724   & 20.44/23   & $1.70_{-0.50}^{+0.59}$ & $1.13\pm0.29$          & 0.20 \\
NGC6503 & MEKAL  & 4 & 934/229    & 1.20/3     & $-2.10_{-0.42}^{+1.75}$& $0.03_{-0.02}^{+0.19}$ & -    \\
NGC6764 & MEKAL  & 4 & 188/167    & 4.34/7     & $3.17\pm2.20$          & $<0.91$                & -    \\
        & 2MEKAL & 6 & 188/167    & 3.94/5     & $<0.03$                & $<0.80$                & -    \\
NGC7090 & MEKAL  & 4 & 683/240    & 2.61/4     & $2.89_{-0.69}^{+1.04}$ & $0.57_{-0.37}^{+0.19}$ & -    \\
%NGC7457 & Fixed  & 1 &
NGC7582 & MEKAL  & 4 & 768/540    & 27.59/24   & $2.71_{-0.21}^{+0.25}$ & $3.90_{-1.16}^{+0.83}$ & -    \\
        & 2MEKAL & 6 & 768/540    & 25.89/22   & $0.37_{-2.55}^{+0.51}$ & $0.83_{-0.78}^{+3.09}$ & -
%NGC7814 & Fixed  & 1 &
%\enddata
\label{table:specmodel}
\end{longtable}
{\scriptsize \tablecomments{Listed items: (1) Galaxy name; (2) XSPEC spectral emission model used to characterize the galactic corona, where MEKAL represents an optically-thin thermal plasma with all the metal abundances fixed, VMEKAL is the same plasma, but with the Fe/O ratio set free, and 2MEKAL contains two plasma components of different temperatures and with fixed metal abundances (see \S\ref{PaperIsubsec:spec} for details); (3) The total number of fitted parameters; (4) Total/background-subtracted net number of counts; (5) $\chi^2$/degrees-of-freedom of the spectral fitting; (6) Photon index of the power law component; (7) Normalization of the power law component in unit of $10^{-5}\rm~photons~keV^{-1}~cm^{-2}~s^{-1}$; (8) Normalization of the thermal component of the CV+AB contribution (scaled with the stellar mass), with a fixed ratio to the normalization of the power law component (of the CV+AB contribution), as detailed in \S\ref{PaperIsubsec:spec}. Fitted and derived parameters of the coronae are summarized in Tables~\ref{table:1Tspec}-\ref{table:specpara2T}.}}
%\end{deluxetable}
%\clearpage

%\clearpage
\begin{deluxetable}{lccccccc}
\centering
\tiny %\tiny\scriptsize\footnotesize\small\normalsize\large\Large\LARGE\huge\Huge
%\ptlandscape
  \tabletypesize{\tiny}
  \tablecaption{Hot Gas Properties from the 1-T Model Fits}
  \tablewidth{0pt}
  \tablehead{
 \colhead{Name} & \colhead{$L_{hot}$} & \colhead{$T_{hot}$} & \colhead{$EM$} & \colhead{$n_e$} & \colhead{$M_{hot}$} & \colhead{$E_{hot}$} & \colhead{$t_{cool}$} \\
    & ($10^{38}\rm ergs/s$) & (keV) & ($10^{-2}\rm cm^{-6}kpc^3$) & ($f^{-1/2}10^{-3}\rm cm^{-3}$) & ($f^{1/2}10^8\rm M_\odot$) & ($f^{1/2}10^{55}\rm ergs$) & ($f^{1/2}\rm Gyr$) \\
   & (1) & (2) & (3) & (4) & (5) & (6) & (7)
}
\startdata
IC2560 & $66.34_{-4.01}^{+4.01}$ & 0.3 & $23.54_{-1.42}^{+1.42}$ & $7.96_{-0.24}^{+0.24}$ & $7.31_{-0.22}^{+0.22}$ & $41.72_{-13.96}^{+13.96}$ & $2.00_{-0.68}^{+0.68}$ \\
M82 & $68.76_{-0.30}^{+0.29}$ & $0.61_{-0.003}^{+0.003}$ & $17.92_{-0.08}^{+0.08}$ & $29.99_{-0.06}^{+0.06}$ & $1.48_{-0.003}^{+0.003}$ & $17.17_{-0.10}^{+0.09}$ & $0.79_{-0.01}^{+0.01}$ \\
NGC0024 & $1.70_{-0.33}^{+0.33}$ & 0.3 & $0.49_{-0.09}^{+0.09}$ & $13.43_{-1.29}^{+1.29}$ & $0.09_{-0.01}^{+0.01}$ & $0.51_{-0.18}^{+0.18}$ & $0.95_{-0.38}^{+0.38}$ \\
NGC0520 & $19.06_{-6.22}^{+3.71}$ & $0.29_{-0.03}^{+0.05}$ & $3.55_{-1.15}^{+0.96}$ & $5.56_{-0.90}^{+0.75}$ & $1.58_{-0.26}^{+0.21}$ & $8.79_{-1.65}^{+1.82}$ & $1.46_{-0.55}^{+0.42}$ \\
NGC0660 & $11.73_{-2.03}^{+3.05}$ & $0.52_{-0.13}^{+0.11}$ & $1.50_{-0.76}^{+0.52}$ & $4.60_{-1.16}^{+0.79}$ & $0.80_{-0.20}^{+0.14}$ & $7.94_{-2.85}^{+2.18}$ & $2.15_{-0.86}^{+0.81}$ \\
NGC0891 & $22.26_{-0.51}^{+0.50}$ & $0.34_{-0.01}^{+0.01}$ & $5.67_{-0.14}^{+0.14}$ & $3.99_{-0.05}^{+0.05}$ & $3.51_{-0.04}^{+0.04}$ & $22.47_{-0.46}^{+0.46}$ & $3.21_{-0.10}^{+0.10}$ \\
NGC1023 & $2.81_{-0.70}^{+0.59}$ & $0.26_{-0.02}^{+0.03}$ & $0.55_{-0.11}^{+0.12}$ & $2.26_{-0.23}^{+0.24}$ & $0.60_{-0.06}^{+0.06}$ & $2.98_{-0.39}^{+0.45}$ & $3.37_{-0.95}^{+0.87}$ \\
NGC1380 & $38.95_{-3.64}^{+3.35}$ & $0.33_{-0.02}^{+0.02}$ & $6.48_{-0.62}^{+0.85}$ & $6.91_{-0.33}^{+0.46}$ & $2.31_{-0.11}^{+0.15}$ & $14.34_{-1.03}^{+1.28}$ & $1.17_{-0.14}^{+0.14}$ \\
NGC1386 & $14.86_{-1.97}^{+1.17}$ & $0.26_{-0.01}^{+0.01}$ & $2.88_{-0.35}^{+0.31}$ & $9.49_{-0.57}^{+0.51}$ & $0.75_{-0.04}^{+0.04}$ & $3.73_{-0.29}^{+0.28}$ & $0.80_{-0.12}^{+0.09}$ \\
NGC1482 & $37.21_{-3.31}^{+3.05}$ & $0.38_{-0.03}^{+0.03}$ & $5.72_{-0.52}^{+0.53}$ & $12.93_{-0.58}^{+0.60}$ & $1.09_{-0.05}^{+0.05}$ & $7.91_{-0.71}^{+0.78}$ & $0.68_{-0.08}^{+0.09}$ \\
NGC1808 & $9.09_{-1.02}^{+0.92}$ & $0.58_{-0.06}^{+0.03}$ & $1.21_{-0.13}^{+0.12}$ & $5.61_{-0.29}^{+0.29}$ & $0.53_{-0.03}^{+0.03}$ & $5.83_{-0.68}^{+0.45}$ & $2.04_{-0.33}^{+0.26}$ \\
NGC2787 & $1.77_{-1.32}^{+1.85}$ & $0.18_{-0.18}^{+0.11}$ & $1.22_{-0.56}^{+32.15}$ & $8.03_{-1.85}^{+105.80}$ & $0.38_{-0.09}^{+4.95}$ & $1.31_{-1.35}^{+17.29}$ & $2.35_{-2.99}^{+31.12}$ \\
NGC2841 & $19.37_{-3.17}^{+2.77}$ & $0.41_{-0.04}^{+0.07}$ & $4.70_{-0.43}^{+0.87}$ & $4.85_{-0.22}^{+0.45}$ & $2.40_{-0.11}^{+0.22}$ & $18.90_{-2.03}^{+3.75}$ & $3.10_{-0.61}^{+0.76}$ \\
NGC3079 & $65.90_{-3.77}^{+3.61}$ & $0.51_{-0.02}^{+0.02}$ & $16.42_{-0.67}^{+1.37}$ & $4.82_{-0.10}^{+0.20}$ & $8.43_{-0.17}^{+0.35}$ & $81.73_{-3.90}^{+5.14}$ & $3.94_{-0.29}^{+0.33}$ \\
NGC3115 & $0.41_{-0.19}^{+0.13}$ & $0.08_{-0.08}^{+0.04}$ & $2.31_{-1.37}^{+1.34}$ & $9.01_{-2.67}^{+2.61}$ & $0.63_{-0.19}^{+0.18}$ & $0.97_{-1.01}^{+0.54}$ & $7.45_{-8.50}^{+4.71}$ \\
NGC3198 & $15.62_{-1.69}^{+1.69}$ & 0.3 & $4.18_{-0.45}^{+0.45}$ & $4.15_{-0.22}^{+0.22}$ & $2.49_{-0.14}^{+0.14}$ & $14.21_{-4.80}^{+4.80}$ & $2.89_{-1.02}^{+1.02}$ \\
NGC3384 & $7.18_{-1.26}^{+1.26}$ & 0.3 & $1.46_{-0.26}^{+0.26}$ & $4.14_{-0.36}^{+0.36}$ & $0.87_{-0.08}^{+0.08}$ & $4.97_{-1.71}^{+1.71}$ & $2.20_{-0.85}^{+0.85}$ \\
NGC3412 & $9.82_{-1.07}^{+1.07}$ & 0.3 & $2.63_{-0.29}^{+0.29}$ & $16.61_{-0.90}^{+0.90}$ & $0.39_{-0.02}^{+0.02}$ & $2.23_{-0.75}^{+0.75}$ & $0.72_{-0.26}^{+0.26}$ \\
NGC3521 & $20.24_{-1.37}^{+1.35}$ & $0.36_{-0.02}^{+0.03}$ & $5.21_{-0.45}^{+0.39}$ & $4.18_{-0.18}^{+0.16}$ & $3.08_{-0.13}^{+0.11}$ & $20.87_{-1.37}^{+1.87}$ & $3.27_{-0.31}^{+0.37}$ \\
NGC3556 & $5.73_{-1.16}^{+1.01}$ & $0.33_{-0.02}^{+0.02}$ & $1.46_{-0.28}^{+0.30}$ & $2.54_{-0.25}^{+0.26}$ & $1.42_{-0.14}^{+0.15}$ & $8.92_{-1.03}^{+1.08}$ & $4.94_{-1.15}^{+1.06}$ \\
NGC3628 & $24.45_{-2.06}^{+1.86}$ & $0.32_{-0.01}^{+0.01}$ & $6.09_{-0.40}^{+0.64}$ & $2.37_{-0.08}^{+0.12}$ & $6.35_{-0.21}^{+0.33}$ & $39.22_{-2.08}^{+2.60}$ & $5.09_{-0.51}^{+0.51}$ \\
NGC3877 & $1.21_{-0.68}^{+0.61}$ & $0.30_{-0.06}^{+0.05}$ & $0.30_{-0.13}^{+0.17}$ & $2.45_{-0.52}^{+0.69}$ & $0.31_{-0.07}^{+0.09}$ & $1.76_{-0.52}^{+0.59}$ & $4.63_{-2.93}^{+2.80}$ \\
NGC3955 & $10.77_{-6.27}^{+3.67}$ & $0.31_{-0.05}^{+0.29}$ & $1.84_{-1.36}^{+1.09}$ & $3.80_{-1.40}^{+1.13}$ & $1.20_{-0.44}^{+0.36}$ & $6.98_{-2.81}^{+6.88}$ & $2.06_{-1.45}^{+2.14}$ \\
NGC3957 & $19.68_{-2.95}^{+2.95}$ & 0.3 & $4.18_{-0.63}^{+0.63}$ & $5.99_{-0.45}^{+0.45}$ & $1.72_{-0.13}^{+0.13}$ & $9.84_{-3.36}^{+3.36}$ & $1.59_{-0.59}^{+0.59}$ \\
NGC4013 & $14.47_{-1.39}^{+1.39}$ & 0.3 & $3.95_{-0.38}^{+0.38}$ & $5.03_{-0.24}^{+0.24}$ & $1.94_{-0.09}^{+0.09}$ & $11.07_{-3.73}^{+3.73}$ & $2.43_{-0.85}^{+0.85}$ \\
NGC4111 & $3.74_{-2.13}^{+1.36}$ & $0.44_{-0.10}^{+0.12}$ & $0.52_{-0.23}^{+0.27}$ & $2.40_{-0.52}^{+0.62}$ & $0.54_{-0.12}^{+0.14}$ & $4.57_{-1.45}^{+1.74}$ & $3.88_{-2.52}^{+2.04}$ \\
NGC4217 & $20.35_{-1.59}^{+1.59}$ & 0.3 & $5.50_{-0.43}^{+0.43}$ & $6.58_{-0.26}^{+0.26}$ & $2.07_{-0.08}^{+0.08}$ & $11.80_{-3.96}^{+3.96}$ & $1.84_{-0.63}^{+0.63}$ \\
NGC4244 & $1.08_{-0.14}^{+0.14}$ & 0.3 & $0.30_{-0.04}^{+0.04}$ & $6.25_{-0.39}^{+0.39}$ & $0.12_{-0.01}^{+0.01}$ & $0.67_{-0.23}^{+0.23}$ & $1.98_{-0.72}^{+0.72}$ \\
NGC4251 & $23.05_{-4.52}^{+4.52}$ & 0.3 & $4.49_{-0.88}^{+0.88}$ & $8.56_{-0.84}^{+0.84}$ & $1.30_{-0.13}^{+0.13}$ & $7.39_{-2.57}^{+2.57}$ & $1.02_{-0.41}^{+0.41}$ \\
NGC4342 & $85.89_{-2.92}^{+2.89}$ & $0.53_{-0.02}^{+0.02}$ & $11.50_{-0.42}^{+0.40}$ & $3.87_{-0.07}^{+0.07}$ & $7.34_{-0.13}^{+0.13}$ & $74.14_{-2.56}^{+2.55}$ & $2.74_{-0.13}^{+0.13}$ \\
NGC4388 & $44.31_{-5.55}^{+7.00}$ & $0.61_{-0.05}^{+0.04}$ & $11.71_{-1.63}^{+1.99}$ & $6.67_{-0.46}^{+0.57}$ & $4.34_{-0.30}^{+0.37}$ & $49.98_{-5.66}^{+5.62}$ & $3.58_{-0.60}^{+0.69}$ \\
NGC4438 & $52.76_{-3.95}^{+3.65}$ & $0.52_{-0.03}^{+0.03}$ & $7.23_{-0.79}^{+0.52}$ & $3.34_{-0.18}^{+0.12}$ & $5.35_{-0.29}^{+0.19}$ & $53.29_{-4.30}^{+3.97}$ & $3.21_{-0.35}^{+0.33}$ \\
NGC4501 & $32.12_{-6.13}^{+5.89}$ & $0.56_{-0.07}^{+0.05}$ & $8.20_{-1.40}^{+1.40}$ & $5.34_{-0.46}^{+0.45}$ & $3.79_{-0.32}^{+0.32}$ & $40.45_{-5.98}^{+5.23}$ & $4.00_{-0.97}^{+0.90}$ \\
NGC4526 & $8.84_{-2.05}^{+1.83}$ & $0.27_{-0.02}^{+0.04}$ & $1.69_{-0.36}^{+0.37}$ & $3.03_{-0.32}^{+0.33}$ & $1.38_{-0.15}^{+0.15}$ & $7.02_{-0.94}^{+1.25}$ & $2.52_{-0.67}^{+0.69}$ \\
NGC4565 & $8.87_{-0.84}^{+0.87}$ & $0.36_{-0.02}^{+0.04}$ & $2.28_{-0.23}^{+0.25}$ & $1.47_{-0.07}^{+0.08}$ & $3.82_{-0.19}^{+0.21}$ & $26.24_{-2.16}^{+2.95}$ & $9.39_{-1.18}^{+1.40}$ \\
NGC4569 & $11.43_{-4.45}^{+1.97}$ & $0.56_{-0.04}^{+0.04}$ & $2.99_{-0.91}^{+0.45}$ & $4.20_{-0.64}^{+0.31}$ & $1.76_{-0.27}^{+0.13}$ & $18.83_{-3.16}^{+2.05}$ & $5.23_{-2.21}^{+1.07}$ \\
NGC4594 & $20.66_{-1.29}^{+1.08}$ & $0.60_{-0.01}^{+0.01}$ & $2.77_{-0.19}^{+0.16}$ & $1.30_{-0.04}^{+0.04}$ & $5.27_{-0.18}^{+0.15}$ & $60.50_{-2.46}^{+2.16}$ & $9.30_{-0.69}^{+0.59}$ \\
NGC4631 & $18.55_{-0.69}^{+0.66}$ & $0.35_{-0.01}^{+0.01}$ & $4.67_{-0.12}^{+0.26}$ & $3.01_{-0.04}^{+0.08}$ & $3.83_{-0.05}^{+0.11}$ & $25.15_{-0.58}^{+0.83}$ & $4.30_{-0.19}^{+0.21}$ \\
NGC4666 & $27.01_{-11.97}^{+4.41}$ & $0.27_{-0.05}^{+0.04}$ & $6.91_{-2.91}^{+1.16}$ & $5.58_{-1.17}^{+0.47}$ & $3.06_{-0.64}^{+0.26}$ & $15.74_{-4.51}^{+2.59}$ & $1.85_{-0.98}^{+0.43}$ \\
NGC4710 & $6.00_{-3.51}^{+0.80}$ & $0.63_{-0.06}^{+0.10}$ & $0.80_{-0.38}^{+0.13}$ & $4.62_{-1.08}^{+0.36}$ & $0.43_{-0.10}^{+0.03}$ & $5.11_{-1.29}^{+0.87}$ & $2.70_{-1.72}^{+0.59}$ \\
NGC5102 & $0.58_{-0.11}^{+0.11}$ & 0.3 & $0.13_{-0.02}^{+0.02}$ & $10.95_{-1.08}^{+1.08}$ & $0.03_{-0.003}^{+0.003}$ & $0.16_{-0.06}^{+0.06}$ & $0.90_{-0.36}^{+0.36}$ \\
NGC5170 & $24.85_{-6.30}^{+6.30}$ & 0.3 & $9.01_{-2.28}^{+2.28}$ & $2.35_{-0.30}^{+0.30}$ & $9.46_{-1.20}^{+1.20}$ & $54.00_{-19.26}^{+19.26}$ & $6.90_{-3.02}^{+3.02}$ \\
NGC5253 & $1.09_{-0.07}^{+0.07}$ & $0.35_{-0.01}^{+0.02}$ & $0.28_{-0.02}^{+0.02}$ & $8.07_{-0.27}^{+0.24}$ & $0.09_{-0.003}^{+0.003}$ & $0.57_{-0.03}^{+0.04}$ & $1.65_{-0.13}^{+0.15}$ \\
NGC5422 & $18.36_{-3.36}^{+3.36}$ & 0.3 & $3.46_{-0.63}^{+0.63}$ & $5.50_{-0.50}^{+0.50}$ & $1.55_{-0.14}^{+0.14}$ & $8.87_{-3.07}^{+3.07}$ & $1.53_{-0.60}^{+0.60}$ \\
NGC5746 & $12.87_{-7.50}^{+4.79}$ & $0.16_{-0.16}^{+0.12}$ & $4.60_{-1.41}^{+95.71}$ & $2.66_{-0.41}^{+27.68}$ & $4.27_{-0.66}^{+44.49}$ & $13.23_{-13.38}^{+138.08}$ & $3.26_{-3.81}^{+34.07}$ \\
NGC5775 & $36.35_{-4.25}^{+3.64}$ & $0.38_{-0.04}^{+0.05}$ & $9.16_{-1.91}^{+1.07}$ & $2.49_{-0.26}^{+0.15}$ & $9.08_{-0.95}^{+0.53}$ & $66.00_{-9.41}^{+8.80}$ & $5.76_{-1.06}^{+0.96}$ \\
NGC5866 & $8.96_{-1.56}^{+1.34}$ & $0.31_{-0.03}^{+0.04}$ & $1.55_{-0.30}^{+0.33}$ & $3.16_{-0.31}^{+0.34}$ & $1.21_{-0.12}^{+0.13}$ & $7.26_{-0.96}^{+1.13}$ & $2.57_{-0.56}^{+0.56}$ \\
NGC6503 & $1.54_{-0.21}^{+0.18}$ & $0.42_{-0.06}^{+0.09}$ & $0.39_{-0.05}^{+0.05}$ & $7.47_{-0.47}^{+0.47}$ & $0.13_{-0.01}^{+0.01}$ & $1.02_{-0.16}^{+0.23}$ & $2.10_{-0.43}^{+0.54}$ \\
NGC6764 & $23.48_{-9.84}^{+6.76}$ & $0.75_{-0.11}^{+0.13}$ & $6.95_{-2.57}^{+1.72}$ & $14.13_{-2.61}^{+1.74}$ & $1.22_{-0.22}^{+0.15}$ & $17.41_{-4.15}^{+3.70}$ & $2.35_{-1.13}^{+0.84}$ \\
NGC7090 & $0.44_{-0.38}^{+0.32}$ & $0.44_{-0.14}^{+0.13}$ & $0.09_{-0.05}^{+0.11}$ & $4.82_{-1.31}^{+2.80}$ & $0.05_{-0.01}^{+0.03}$ & $0.40_{-0.16}^{+0.26}$ & $2.87_{-2.76}^{+2.81}$ \\
NGC7457 & $4.97_{-1.26}^{+1.26}$ & 0.3 & $1.15_{-0.29}^{+0.29}$ & $4.24_{-0.54}^{+0.54}$ & $0.67_{-0.09}^{+0.09}$ & $3.85_{-1.37}^{+1.37}$ & $2.46_{-1.08}^{+1.08}$ \\
NGC7582 & $62.94_{-11.88}^{+10.51}$ & $0.67_{-0.07}^{+0.08}$ & $16.36_{-2.69}^{+3.80}$ & $8.23_{-0.68}^{+0.96}$ & $4.91_{-0.40}^{+0.57}$ & $62.43_{-8.57}^{+10.21}$ & $3.15_{-0.73}^{+0.74}$ \\
NGC7814 & $5.89_{-0.97}^{+0.97}$ & 0.3 & $1.81_{-0.30}^{+0.30}$ & $12.27_{-1.01}^{+1.01}$ & $0.37_{-0.03}^{+0.03}$ & $2.09_{-0.72}^{+0.72}$ & $1.12_{-0.43}^{+0.43}$
\enddata
\tablecomments{\scriptsize Hot gas properties: (1) absorption corrected 0.5-2~keV luminosity; (2) temperature; (3) volume emission measure; (4) electron number density; (5) mass; (6) thermal energy; (7) radiative cooling timescale. All the parameters are estimated from the 1-T fit with fixed abundance ratio. The temperature is fixed at $0.3\rm~keV$ for some galaxies with low counting statistic (those without errors). See \S\ref{PaperIsubsec:spec} for details.}\label{table:1Tspec}
\end{deluxetable}

%\clearpage
\begin{deluxetable}{lcccccccc}
\centering
\tiny %\tiny\scriptsize\footnotesize\small\normalsize\large\Large\LARGE\huge\Huge
%\ptlandscape
  \tabletypesize{\tiny}
  \tablecaption{Hot Gas Properties from 1-T Model Fit with Free Abundance Ratio}
  \tablewidth{0pt}
  \tablehead{
 \colhead{Name} & \colhead{$L_X$} & \colhead{$T_X$} & \colhead{$EM$} & \colhead{$n_e$} & \colhead{$M_{hot}$} & \colhead{$E_{hot}$} & \colhead{$t_{cool}$} & \colhead{$Fe/\alpha$}\\
    & ($10^{38}\rm ergs/s$) & (keV) & ($10^{-2}\rm cm^{-6}kpc^3$) & ($f^{-1/2}10^{-3}\rm cm^{-3}$) & ($f^{1/2}10^8\rm M_\odot$) & ($f^{1/2}10^{55}\rm ergs$) & ($f^{1/2}\rm Gyr$) & \\
   & (1) & (2) & (3) & (4) & (5) & (6) & (7) & (8)
}
\startdata
M82 & $67.68_{-0.38}^{+0.39}$ & $0.62_{-0.003}^{+0.004}$ & $16.29_{-0.12}^{+0.24}$ & $28.60_{-0.11}^{+0.21}$ & $1.41_{-0.01}^{+0.01}$ & $16.51_{-0.11}^{+0.15}$ & $0.77_{-0.01}^{+0.01}$ & $0.36_{-0.01}^{+0.01}$ \\
NGC891 & $21.85_{-0.60}^{+0.48}$ & $0.32_{-0.01}^{+0.01}$ & $5.09_{-0.19}^{+0.15}$ & $3.78_{-0.07}^{+0.05}$ & $3.33_{-0.06}^{+0.05}$ & $20.26_{-0.53}^{+0.50}$ & $2.94_{-0.11}^{+0.10}$ & $0.43_{-0.04}^{+0.06}$ \\
NGC1482 & $48.96_{-2.73}^{+3.57}$ & $0.49_{-0.04}^{+0.03}$ & $11.22_{-1.88}^{+2.06}$ & $18.12_{-1.52}^{+1.66}$ & $1.53_{-0.13}^{+0.14}$ & $14.41_{-1.65}^{+1.58}$ & $0.93_{-0.12}^{+0.12}$ & $0.36_{-0.06}^{+0.11}$ \\
NGC1808 & $16.60_{-0.79}^{+0.73}$ & $0.51_{-0.03}^{+0.03}$ & $4.00_{-0.39}^{+0.41}$ & $10.22_{-0.50}^{+0.52}$ & $0.97_{-0.05}^{+0.05}$ & $9.31_{-0.70}^{+0.69}$ & $1.78_{-0.16}^{+0.15}$ & $0.34_{-0.05}^{+0.06}$ \\
NGC2841 & $17.75_{-2.83}^{+2.48}$ & $0.38_{-0.05}^{+0.04}$ & $3.88_{-0.41}^{+0.76}$ & $4.40_{-0.23}^{+0.43}$ & $2.18_{-0.12}^{+0.21}$ & $15.67_{-2.29}^{+2.34}$ & $2.80_{-0.61}^{+0.57}$ & $0.48_{-0.12}^{+0.19}$ \\
NGC3079 & $65.29_{-4.96}^{+4.99}$ & $0.51_{-0.03}^{+0.03}$ & $16.48_{-1.38}^{+1.41}$ & $4.83_{-0.20}^{+0.21}$ & $8.44_{-0.35}^{+0.36}$ & $81.59_{-5.84}^{+5.92}$ & $3.97_{-0.41}^{+0.42}$ & $0.31_{-0.05}^{+0.06}$ \\
NGC3556 & $5.73_{-1.18}^{+1.03}$ & $0.33_{-0.03}^{+0.06}$ & $1.45_{-0.24}^{+0.31}$ & $2.53_{-0.21}^{+0.27}$ & $1.41_{-0.12}^{+0.15}$ & $8.86_{-1.14}^{+1.80}$ & $4.91_{-1.19}^{+1.33}$ & $0.30_{-0.12}^{+0.12}$ \\
NGC3628 & $23.32_{-2.08}^{+1.08}$ & $0.30_{-0.02}^{+0.01}$ & $4.73_{-0.42}^{+0.50}$ & $2.09_{-0.09}^{+0.11}$ & $5.60_{-0.25}^{+0.30}$ & $32.18_{-2.49}^{+2.11}$ & $4.38_{-0.52}^{+0.35}$ & $0.45_{-0.08}^{+0.48}$ \\
NGC4342 & $91.18_{-3.34}^{+3.53}$ & $0.54_{-0.02}^{+0.02}$ & $14.03_{-1.97}^{+2.14}$ & $4.28_{-0.30}^{+0.33}$ & $8.10_{-0.57}^{+0.62}$ & $83.24_{-6.45}^{+6.93}$ & $2.90_{-0.25}^{+0.27}$ & $0.79_{-0.12}^{+0.16}$ \\
NGC4526 & $8.77_{-1.95}^{+1.77}$ & $0.26_{-0.04}^{+0.07}$ & $1.59_{-0.38}^{+0.44}$ & $2.94_{-0.36}^{+0.41}$ & $1.34_{-0.16}^{+0.19}$ & $6.56_{-1.21}^{+1.89}$ & $2.37_{-0.69}^{+0.84}$ & $1.25_{-0.78}^{+1.25}$ \\
NGC4594 & $21.95_{-1.77}^{+1.66}$ & $0.60_{-0.01}^{+0.01}$ & $3.33_{-0.84}^{+0.77}$ & $1.42_{-0.18}^{+0.16}$ & $5.78_{-0.73}^{+0.67}$ & $66.18_{-8.44}^{+7.79}$ & $9.57_{-1.44}^{+1.34}$ & $0.81_{-0.19}^{+0.29}$ \\
NGC4631 & $18.16_{-0.75}^{+0.72}$ & $0.33_{-0.01}^{+0.01}$ & $4.42_{-0.22}^{+0.18}$ & $2.93_{-0.07}^{+0.06}$ & $3.72_{-0.09}^{+0.08}$ & $23.68_{-0.79}^{+0.73}$ & $4.14_{-0.22}^{+0.21}$ & $0.37_{-0.04}^{+0.04}$ \\
NGC5253 & $1.04_{-0.08}^{+0.08}$ & $0.32_{-0.02}^{+0.02}$ & $0.24_{-0.02}^{+0.02}$ & $7.41_{-0.31}^{+0.32}$ & $0.08_{-0.003}^{+0.003}$ & $0.48_{-0.03}^{+0.03}$ & $1.47_{-0.15}^{+0.15}$ & $0.48_{-0.07}^{+0.08}$ \\
NGC5866 & $10.59_{-2.12}^{+1.43}$ & $0.14_{-0.001}^{+0.001}$ & $1.73_{-0.88}^{+0.47}$ & $3.33_{-0.85}^{+0.46}$ & $1.28_{-0.33}^{+0.18}$ & $3.34_{-0.85}^{+0.47}$ & $1.00_{-0.32}^{+0.19}$ & $<401.10$
\enddata
\tablecomments{\scriptsize Hot gas properties of the 1-T fit with the Fe/$\alpha$ ratio set free: (1) extinction corrected 0.5-2~keV luminosity; (2) temperature; (3) volume emission measure; (4) electron number density; (5) mass; (6) thermal energy; (7) radiative cooling timescale; (8) Fe/$\alpha$ ratio. See \S\ref{PaperIsubsec:spec} for details.}\label{table:specpara1Tmetal}
\end{deluxetable}

%\clearpage
\begin{deluxetable}{lcccccccc}
\centering
\tiny %\tiny\scriptsize\footnotesize\small\normalsize\large\Large\LARGE\huge\Huge
%\ptlandscape
  \tabletypesize{\tiny}
  \tablecaption{Hot Gas Properties from the 2-T Plasma Model}
  \tablewidth{0pt}
  \tablehead{
 \colhead{Name} & \colhead{$L_{X,low}$} & \colhead{$T_{X,low}$} & \colhead{$EM_{X,Low}$} & \colhead{$L_{X,high}$} & \colhead{$T_{X,high}$} & \colhead{$EM_{X,high}$} & \colhead{$L_{X,total}$} & \colhead{$T_{X,L_W}$} \\
    & ($10^{38}\rm ergs/s$) & (keV) & ($10^{-2}\rm cm^{-6}kpc^3$) & ($10^{38}\rm ergs/s$) & (keV) & ($10^{-2}\rm cm^{-6}kpc^3$) & ($10^{38}\rm ergs/s$) & (keV) \\
    & (1) & (2) & (3) & (4) & (5) & (6) & (7) & (8)
}
\startdata
NGC520 & $23.08_{-20.57}^{+7.05}$ & $0.35_{-0.09}^{+0.07}$ & $5.86_{-2.80}^{+2.20}$ & $12.94(<22.76)$ & $0.77_{-0.19}^{+0.27}$ & $3.61_{-2.61}^{+4.91}$ & $36.1_{-6.9}^{+4.5}$ & $0.50_{-0.13}^{+0.14}$ \\
NGC2841 & $4.20_{-1.57}^{+1.40}$ & $0.11_{-0.02}^{+0.03}$ & $6.22_{-2.38}^{+7.92}$ & $23.56_{-1.76}^{+2.31}$ & $0.49_{-0.05}^{+0.04}$ & $5.99_{-0.42}^{+0.62}$ & $27.7_{-1.9}^{+3.0}$ & $0.43_{-0.05}^{+0.04}$ \\
NGC3079 & $10.28_{-1.53}^{+1.49}$ & $0.12_{-0.01}^{+0.02}$ & $9.00_{-2.64}^{+2.59}$ & $77.08_{-3.52}^{+2.64}$ & $0.56_{-0.02}^{+0.02}$ & $19.63_{-0.90}^{+0.80}$ & $87.1_{-3.6}^{+2.9}$ & $0.51_{-0.02}^{+0.02}$ \\
NGC3521 & $14.00(<19.75)$ & $0.31_{-0.05}^{+0.05}$ & $3.51_{-1.49}^{+3.77}$ & $7.18(<14.12)$ & $0.56(<4.09)$ & $1.88(<3.28)$ & $21.2_{-2.0}^{+1.7}$ & $0.40_{-0.22}^{+1.20}$ \\
NGC3556 & $3.29_{-0.66}^{+0.57}$ & $0.20_{-0.03}^{+0.03}$ & $0.96_{-0.14}^{+0.14}$ & $5.94_{-0.76}^{+0.58}$ & $0.61_{-0.03}^{+0.05}$ & $1.55_{-0.11}^{+0.17}$ & $9.2_{-0.5}^{+0.5}$ & $0.47_{-0.03}^{+0.05}$ \\
NGC3628 & $7.37_{-1.71}^{+5.42}$ & $0.14_{-0.02}^{+0.03}$ & $4.75_{-1.34}^{+1.66}$ & $25.01_{-6.76}^{+2.23}$ & $0.40_{-0.03}^{+0.19}$ & $6.39_{-0.74}^{+0.55}$ & $34.0_{-3.8}^{+2.5}$ & $0.34_{-0.03}^{+0.15}$ \\
NGC3877 & $0.80(<1.13)$ & $0.18(<0.24)$ & $0.27_{-0.15}^{+2.31}$ & $1.38_{-0.62}^{+0.28}$ & $0.59_{-0.24}^{+0.13}$ & $0.35_{-0.14}^{+0.09}$ & $2.1_{-0.8}^{+0.3}$ & $0.44_{-0.21}^{+0.11}$ \\
NGC4388 & $5.61_{-2.53}^{+1.81}$ & $0.09(<0.12)$ & $15.31_{-8.71}^{+17.55}$ & $51.82_{-3.31}^{+3.09}$ & $0.61_{-0.04}^{+0.04}$ & $13.36_{-1.42}^{+1.03}$ & $57.4_{-7.1}^{+3.1}$ & $0.56_{-0.05}^{+0.04}$ \\
NGC4501 & $3.06(<10.59)$ & $0.29(<0.29)$ & $1.00(<3.68)$ & $30.48(<37.90)$ & $0.58_{-0.08}^{+0.08}$ & $7.55_{-1.89}^{+2.56}$ & $33.6_{-6.9}^{+8.1}$ & $0.55_{-0.10}^{+0.10}$ \\
NGC4565 & $4.98(<8.70)$ & $0.29_{-0.10}^{+0.10}$ & $1.29_{-1.24}^{+1.14}$ & $4.74(<6.62)$ & $0.56(<0.56)$ & $1.17(<1.61)$ & $9.7_{-1.3}^{+1.0}$ & $0.42_{-0.32}^{+0.32}$ \\
NGC4569 & $4.19_{-1.51}^{+0.88}$ & $0.10(<0.15)$ & $24.83_{-19.39}^{+6.32}$ & $13.22_{-1.92}^{+1.13}$ & $0.56_{-0.05}^{+0.02}$ & $3.42_{-0.25}^{+0.25}$ & $17.4_{-2.8}^{+1.0}$ & $0.45_{-0.06}^{+0.03}$ \\
NGC4631 & $9.33_{-1.15}^{+0.97}$ & $0.24_{-0.02}^{+0.01}$ & $2.40_{-0.25}^{+0.23}$ & $13.77_{-1.41}^{+1.53}$ & $0.58_{-0.02}^{+0.02}$ & $3.58_{-0.37}^{+0.37}$ & $23.3_{-1.1}^{+0.9}$ & $0.44_{-0.02}^{+0.02}$ \\
NGC5253 & $0.20_{-0.06}^{+0.08}$ & $0.14_{-0.02}^{+0.08}$ & $0.12_{-0.03}^{+0.06}$ & $1.04_{-0.09}^{+0.07}$ & $0.40_{-0.02}^{+0.03}$ & $0.26_{-0.02}^{+0.01}$ & $1.24_{-0.04}^{+0.04}$ & $0.36_{-0.02}^{+0.04}$ \\
NGC5775 & $6.85_{-3.81}^{+9.90}$ & $0.08(<0.20)$ & $11.90_{-9.94}^{+157.80}$ & $38.22(<40.85)$ & $0.37(<0.61)$ & $9.64_{-7.06}^{+0.83}$ & $45.1_{-3.6}^{+3.1}$ & $0.33(<0.55)$ \\
NGC6764 & $4.59(<10.84)$ & $0.30(<0.30)$ & $1.25(<8.51)$ & $24.36_{-19.57}^{+5.84}$ & $0.75_{-0.10}^{+0.10}$ & $6.87(<8.66)$ & $29.6_{-11.1}^{+9.8}$ & $0.68_{-0.13}^{+0.13}$ \\
NGC7582 & $14.61(<17.97)$ & $0.09(<0.11)$ & $41.45_{-35.11}^{+19.45}$ & $90.19_{-28.83}^{+12.12}$ & $0.64_{-0.05}^{+0.05}$ & $24.25_{-3.16}^{+3.29}$ & $105.0_{-6.8}^{+11.6}$ & $0.56_{-0.05}^{+0.04}$
\enddata
\tablecomments{\scriptsize See \S\ref{PaperIsubsec:spec} for details.}\label{table:specpara2T}
\end{deluxetable}

\section{Notes on Individual Galaxies}\label{PaperIsec:Individual}

We herein describe multi-wavelength properties of individual galaxies, concentrating on those that show distinct properties and/or are not well studied in literatures.

\emph{M82}: As a member of the M81 group, M82 is known to be surrounded by large-scale structures of atomic gas, produced by tidal interaction \citep{Yun94}. The rotation curve of M82 declines outward without a significant flattening part \citep{Sofue97}. The $v_{rot}$ value of $\sim65\rm~km~s^{-1}$ obtained from \emph{Hyperleda} is likely measured at a projected radius of $\sim3^\prime$ and may be strongly affected by the tidal interaction. Indeed, the rotation curve measured with the stellar absorption lines peaks at $\sim1^\prime$ with a peak velocity of $\sim100\rm~km~s^{-1}$ (as adopted in the present work; Table~\ref{table:SampleSelectionPara}), where the tidal effect should be negligible \citep{Westmoquette09}. The surrounding gas may also help to confine the outflow and hence enhance the X-ray emission, as manifested most vividly by the presence of an H$\alpha$ and X-ray ridge, $11^\prime$ or 11.6~kpc north of the galactic disk \citep{Lehnert99}.

\emph{NGC~520}: This system represents an ongoing merger between two galaxies: one gas-rich and the other gas-poor \citep{Read05}. Only the nucleus and its surrounding region of the gas-rich one are bright in X-ray. The diffuse emission, in particular, is elongated in the direction perpendicular to the gas-rich disk \citep{Read05} (also see Fig.~\ref{fig:imagea}Ag,Ah). We thus adjust the spectrum extraction region to cover the bulk of this dominant diffuse X-ray feature (Fig.~\ref{fig:imagea}Ag). Both the number of X-ray bright point sources and the diffuse X-ray emission, relative to its SFR, are smaller than other merging galaxies, such as the Mice and Antennae galaxy systems. These distinct X-ray properties are suggested to be a result of a merger between gas-rich and gas-poor galaxies, instead of between gas-rich and gas-rich ones \citep{Read05}.

\emph{NGC~660}: This galaxy has a severely tidally-disturbed disk surrounded by a gas-rich polar ring. The stellar population of the polar ring is young, with a considerable fraction consisting of blue and red supergiants \citep{Karataeva04}. Although continuous SF is evidenced in the polar ring, it shows no significant soft X-ray enhancement in the short \emph{Chandra} observations (Table~\ref{table:ChandraData}; Fig.~\ref{fig:imagea}Ai). Bright CO emission exhibits a clear central concentration, similar to those seen in nuclear starburst galaxies \citep{Israel09}. Polarized extraplanar optical emission is attributed to the light scattered by dust grains residing up to $\sim2.5\rm~kpc$ from the stellar disk \citep{Alton00}, which are likely entrained in an energetic outflow, as revealed by the corona elongated along the minor axis of the galactic disk (Fig.~\ref{fig:imagea}Ai,Aj).

\emph{NGC~1380}: This is a lenticular galaxy $27.8^\prime$ from the center of the Fornax Cluster. The measured temperature of $\sim0.33\rm~keV$ is significantly lower than that of the surrounding intra-cluster medium (ICM) \citep{Scharf05}. Therefore, the coronal gas likely arises from the galaxy itself. The temperature obtained from the \emph{Chandra} observation is lower than the \emph{ROSAT} value of $\sim0.5\rm~keV$ \citep{Schlegel98}, which is likely due to contamination from point sources or the ICM. The coronal emission is bright and extended toward southeast, with a blob of diffuse X-ray emission located nearby. This very extended X-ray-emitting blob is centered on a small spiral galaxy, which is, however, not detected as a compact X-ray source (a bright point-like X-ray source CXO~J033639.4-345847 is indeed detected $\sim1^\prime$ off the center and is apparently not related; Fig.~\ref{fig:imagea}Ao). If the blob is associated with the spiral galaxy and if its distance is the same as that of NGC~1380, the X-ray to K-band luminosity ratio, $6\times10^{38}{\rm~ergs~s^{-1}}/5\times10^8\rm~L_\odot$, would then be much higher than typical isolated galaxies. In addition, if the diffuse X-ray is dominated by thermal emission from hot gas, the hardness ratio of the blob suggests a temperature of $0.55_{-0.17}^{+0.10}\rm~keV$, likely higher than that of NGC~1380 ($0.33\pm0.02\rm~keV$, Table~\ref{table:1Tspec}). We thus speculate that the blob is more likely to be a background galaxy group, if not a local enhancement of the Fornax cluster happening to coincide with a background galaxy.

\emph{NGC~1386}: This galaxy is also a member of the Fornax cluster. It hosts a low luminosity Seyfert-2 AGN with Fe K-shell lines detected in X-ray \citep{Iyomoto97,Bennert06}. The diffuse soft X-ray emission is faint, extended, and apparently highly distorted, especially in the outer part (Fig.~\ref{fig:imagea}Aq, Ar; but the distorted outermost contour in Fig.~\ref{fig:imagea}Ar is close to the background level, may or may not relate to the galaxy itself, thus is not included in the spectral analysis). Analysis of the diffuse X-ray emission extracted from the box in Fig.~\ref{fig:imagea}Aq shows that the thermal component has a temperature of $\sim0.26\rm~keV$, significantly lower than the ICM value \citep{Scharf05}. Therefore, the bulk of the hot gas in the galaxy vicinity is likely associated with the galaxy itself and reshaped in the cluster environment.

\emph{NGC~2787}: This is a low mass S0 galaxy rich in atomic gas, although the molecular gas mass is relatively low \citep{Welch03,Sage06,Li11}. Most of the atomic gas is located in a ring-like structure with a diameter of $\sim6\farcm4$ \citep{Shostak87}, significantly larger than the extent of the stellar light (Fig.~\ref{fig:imagea}Aw). The unresolved soft X-ray emission shows a round morphology and has a spectrum consistent with primarily stellar in origin (Fig.~\ref{fig:speca}Aj; \citealt{Li11}).

\emph{NGC~2841}: This is a massive spiral galaxy. Some plume-like diffuse X-ray emission features are apparently connected to the disk, some of which may form a limb-brightened ``X-shaped'' structure around the galaxy's nucleus, similar as those often found in nuclear starburst galaxies \citep{Strickland04a}. The coronal luminosity obtained here is significantly lower than the estimation using \emph{ROSAT} observation \citep{Benson00}, which is clearly due to the differences in removing point sources and in separating the disk and halo components.

\emph{NGC~3079}: This galaxy hosts a kpc-scale nuclear superbubble, whose diffuse X-ray emission is closely associated with various H$\alpha$ filaments \citep{Cecil02}. It is still not clear whether the nuclear starburst or the Seyfert~2 AGN \citep{Ho97} is the dominant driving source of the bubble. An \ion{H}{1} observation of the galaxy and its two companions (NGC~3073 and MCG~9-17-9) reveals that NGC~3073 exhibits an elongated \ion{H}{1} tail remarkably aligned with the nucleus of NGC~3079, and is probably shaped by the ram pressure stripping of the outflowing gas from NGC~3079 \citep{Irwin87}. NGC~3079 is thought to have both nuclear-concentrated and disk-wide SFs and have the diffuse X-ray characteristics of both types of galaxies (``X-shaped'' and disk-wide extended) \citep{Strickland04a}.

\emph{NGC~3115}: This galaxy contains primarily an old stellar population, shows little on-going SF, and is extremely poor in cold gas \citep{Li11}. The galaxy hosts a massive nuclear BH \citep{Kormendy96}; its Bondi radius is readily resolved by \emph{Chandra} \citep{Wong11}. Spectral analysis of the unresolved X-ray emission shows that it mainly consists of unresolved stellar sources, with little contribution from hot gas (Fig.~\ref{fig:speca}Am). Previous detection of the $\sim0.5\rm~keV$ hot gas from short \emph{Chandra} observations \citep{David06} is more likely due to the contamination of unresolved point sources, i.e., the CV+AB component (\S\ref{PaperIsubsubsec:SpectraAnalysis}).

\emph{NGC~3384}: This galaxy is a member of the M96 group. Its collision with M96 is thought to be the major triggering mechanism for the very large-scale, ring-like, star forming gaseous structure (the ``Leo Ring'', \citealt{Thilker09,MichelDansac10}). The galaxy has a very low rotation velocity ($\sim17\rm~km~s^{-1}$, Table~\ref{table:SampleSelectionPara}) which may be highly affected by the interaction. Its $M_*$ is significantly higher than $M_{TF}$ (Fig.~\ref{fig:GalaxyMass}b); the latter may \emph{not} be physical.

\emph{NGC~3628}: This galaxy is a member of the Leo triplet, and has a huge optical and \ion{H}{1} tidal tail ($\sim80\rm~kpc$ extension) \citep{Kormendy74,Rots78}, which is the result of the interaction with its companions NGC~3627 and NGC~3623. Clumpy SF regions are detected in this tidal tail, with a SF age of $\sim10^8\rm~yr$ \citep{Chromey98}. Apparently extended X-ray emission around this nuclear starburst galaxy was detected in \emph{Einstein} \citep{Fabbiano90} and \emph{ROSAT} \citep{Dahlem96} observations. In higher resolution \emph{Chandra} images (Fig.~\ref{fig:imagea}Bq, Br), the diffuse X-ray emission is dominated by a vertically elongated component, which is associated with the H$\alpha$ emission on various scales, probably representing a well collimated nuclear outflow \citep{Strickland04a}.

\emph{NGC~4244}: This is a quiescent low-mass late-type galaxy, with very weak SF, no nuclear activity, and a small bulge. Even with a nearly perfect edge-on orientation ($i\sim88^\circ$, Table~\ref{table:SampleSelectionPara}), the galaxy shows little extraplanar diffuse X-ray emission in a deep \emph{Chandra} observation ($t_{eff}\sim46\rm~ks$, Table~\ref{table:ChandraData}; Fig.~\ref{fig:imagea}Cg, Ch; \citealt{Strickland04a}). The detected emission comes from primarily some patches, with a total luminosity of $\sim10^{38}\rm~ergs~s^{-1}$, putting the galaxy into the X-ray faintest group of the present sample.

\emph{NGC~4342}: This is a low-mass galaxy located in the Virgo cluster. As a dwarf galaxy with little SF, it appears to be unusually bright in X-ray and hosts an exceptionally massive nuclear BH \citep{Bogdan12a,Bogdan12b}. It is suggested to reside in a massive dark matter halo. The very extended and lopsided morphology of the diffuse X-ray emission, as well as the sharp northeastern boundary (or a cold front) (Fig.~\ref{fig:imagea}Ck,Cl), suggests that the galaxy is moving supersonically in the ICM \citep{Bogdan12b}.

\emph{NGC~4388}: This galaxy is also located in the Virgo cluster. It hosts a Seyfert~2 nucleus \citep{Shirai08}. The extended diffuse soft X-ray emission is correlated well with the ionization cone found in optical-line emission \citep{Iwasawa03}. The relatively high temperature ($\sim0.6\rm~keV$; Table~\ref{table:1Tspec}) indicates that the gas represents a mixture of the stripped ISM and the surrounding ICM \citep{Wezgowiec11}.

\emph{NGC~4438}: It belongs to an interacting galaxy pair in the Virgo cluster \citep{Kenney08} and is much more X-ray luminous than galaxies with similar SFR or stellar mass. In addition, the diffuse X-ray emission of NGC~4438 seems to connect both members of the pair (Fig.~\ref{fig:imagea}Co,Cp) and is apparently correlated with many cool gas features (\citealt{Machacek04}). The SF and AGN are not particularly active in the galaxy, so the very luminous and asymmetric X-ray emission is likely produced and shaped by other mechanisms, such as cool-hot gas interaction, ram-pressure stripping, and turbulent mixing (e.g., \citealt{Kenney95,Vollmer05,Vollmer09}).

\emph{NGC~4565}: This galaxy is inactive in SF \citep{Rand92} and hosts a weak Seyfert~2 AGN \citep{Chiaberge06}. The diffuse soft X-ray emission consists of three components (Fig.~\ref{fig:imagea}Cu, Cv) with distinct morphologies: 1) a round one apparently associated with the small stellar bulge, 2) an elongated one along the disk, and 3) an lopsided one extending from the disk toward northeast. The very asymmetric, large-scale, diffuse X-ray morphology of this third component is probably due to the ram pressure that the galaxy is subject to against the ambient IGM \citep{Wang05}.

\emph{NGC~4569}: This is one of the largest and most \ion{H}{1} deficient galaxies in the Virgo cluster \citep{Wezgowiec11}. It shows strong evidence for ongoing ICM-ISM interaction. \ion{H}{1} line synthesis observations have showed that the galaxy has lost more than 90\% of its \ion{H}{1} gas \citep{Solanes01}. The diffuse soft X-ray emission extends to the west (Fig.~\ref{fig:imagea}Cw, Cx), coinciding with a giant \ion{H}{1} and H$\alpha$ arm, which likely represents the gas stripped from the disk about 300~Myr ago \citep{Vollmer04}.

\emph{NGC~4594}: This, often known as the ``Sombrero'' galaxy, is a massive, bulge-dominated Sa galaxy. It contains a quiescent nucleus and little SF, but is very luminous in X-ray ($\sim2\times10^{39}\rm~ergs~s^{-1}$, Table~\ref{table:1Tspec}). Quantitative calculations indicate that the observed energy, mass, and metal contents of the corona are much less than expected from stellar feedback \citep{LiZ07a}.

\emph{NGC~4631}: Strong starburst is evidenced in this galaxy, apparently triggered by interactions with neighboring galaxies. Multi-wavelength observations \citep{Wang01,Irwin11} reveal that the diffuse X-ray morphology resembles the radio halo, indicating a close connection between the outflows of hot gas, cosmic ray, and magnetic field from the galactic disk. Enhanced diffuse X-ray emission is also apparently enclosed by numerous H$\alpha$-emitting loops blistered out from the central disk of the galaxy.

\emph{NGC~4666}: This galaxy hosts a Compton-thick AGN and intense SF \citep{Persic04}, which are thought to be caused by gravitational interaction with the neighboring galaxy NGC~4668 \citep{Walter04}. The galaxy is in many respects very similar to NGC~4631, showing a strong galactic superwind as revealed by multi-wavelength observations \citep{Dahlem97}. The \emph{Chandra} observation has a too short exposure ($t_{eff}\sim4.7\rm~ks$, Table~\ref{table:ChandraData}) to resolve the structure of the superwind, but still shows prominent vertical-elongated diffuse X-ray emission (Fig.~\ref{fig:imagea}De, Df).

\emph{NGC~5170}: This is a massive quiescent spiral galaxy, with a small stellar bulge and a perfect edge-on orientation (Table~\ref{table:SampleSelectionPara}), thus good for testing the accretion scenario of the corona formation (e.g., \citealt{White91,Benson00}). \citet{Rasmussen09} analyzed the \emph{Chandra} data, but failed to detect any significant extraplanar diffuse X-ray emission (most of the X-ray emission in Fig.~\ref{fig:imagea}Dk, Dl can be attributed to the residual contribution from point sources). The upper limit is consistent with our measurement, accounting for the different spectral regions adopted in the two analyses.

\emph{NGC~5253}: In our sample, this galaxy has the highest specific SFR (${\rm SFR}/M_*=6.9\rm~M_\odot~yr^{-1}/10^{10}M_\odot$; Table~\ref{table:SampleSelectionPara}) \citep{Calzetti97}, compared to M82, which has ${\rm SFR}/M_*\sim3.9\rm~M_\odot~yr^{-1}/10^{10}M_\odot$. However, NGC~5253 appears to be relatively X-ray faint and has a less vertically extended corona than those of typical starburst galaxies (Fig.~\ref{fig:imagea}Dm,Dn), probably because of its shallow gravitational potential.

\emph{NGC~5746}: This galaxy, similar as NGC~5170, is also a massive quiescent spiral galaxy, but has a slightly smaller inclination angle and a larger stellar bulge (Table~\ref{table:SampleSelectionPara}). Previous analysis of the \emph{Chandra} data also failed to detect the extraplanar hot gas \citep{Rasmussen09}. There is some apparently diffuse X-ray emission in the nuclear region, most of which may be nonthermal in origin (Fig.~\ref{fig:speca}Bh).

\emph{NGC~5775}: This galaxy is interacting with its companion NGC~5774 (and probably also with another small companion IC~1070), which probably triggers the active SF over the disk. \ion{H}{1} observation shows that gas transfer from NGC~5774 to NGC~5775 is ongoing through two ``bridges'' \citep{Irwin94}. The diffuse X-ray emission shows characteristics of both nuclear starburst and disk-wide SF, i.e., with both vertically and horizontally elongated components (Fig.~\ref{fig:imagea}Ds, Dt). A less smoothed X-ray image shows a 10~kpc-diameter shell-like feature in the southern halo, but the current X-ray data (lack of spectral information on the ``shell'') cannot be used to say much about its nature \citep{Li08}.

\emph{NGC~5866}: This galaxy resembles NGC~3115 in stellar mass, morphological type, stellar population, and environment (Tables~\ref{table:SampleSelectionPara}, \ref{table:basicparaII}), but is much richer in cool gas and significantly more X-ray luminous \citep{Li11}. Detailed spectral analysis shows a prominent Fe~L-shell line feature, but no significant oxygen line, suggesting a high Fe/O abundance ratio \citep{Li09}. Optical observation reveals various extraplanar dusty filaments, which are suggested to be cool gas outflows produced by Type~Ia SNe.

\emph{NGC~6764}: This galaxy hosts both AGN and nuclear starburst. The two kpc-scale radio bubbles spatially coincide with the diffuse X-ray emission with a high fitted temperature ($\sim0.75\rm~keV$) \citep{Croston08}. In addition, the vertical profile peaks sharply toward the galactic disk (Fig.~\ref{fig:profilesa}Ca). It is therefore most likely that the hot gas is heated by the AGN jet/ISM interactions.

\emph{NGC~7090}: This is a bulge-less, low-mass galaxy with a patchy dust lane. The SF is weak and scattered over the disk, but the radio continuum emission is very extended and asymmetric \citep{Dahlem01,Dahlem06}. An H$\alpha$ image reveals a faint halo in the form of knots, and morphologically coincident with the extended radio continuum emission \citep{Rossa03a,Rossa03b}. Both the \ion{H}{1} gas distribution and the kinematics are distorted. But there is no companion galaxy that might have caused these distortions \citep{Dahlem05}. The diffuse X-ray emission is weak and concentrated only in the central region. A large fraction of the emission can be attributed to the residual emission from point sources (Figs.~\ref{fig:imagea}Ec, Ed, \ref{fig:speca}Bm; \citealt{Mineo12}).

\emph{NGC~7582}: This galaxy hosts a prominent dust lane being spatially coincident with the excess of X-ray absorption. Two ``hotspots'' with highly ionized oxygen and neon line emissions (stronger \ion{O}{8} and \ion{Ne}{5} emission than \ion{O}{7} and \ion{Ne}{4}) are detected in soft X-ray, and are thought to be ionized by the leaking photons from the AGN instead of from the disk SF regions \citep{Stefano07}.

\section{Discussion}\label{PaperIsec:Discussion}

\subsection{Derived Hot Gas Parameters}\label{PaperIsubsec:DerivedParameters}

In addition to the directly fitted parameters of a galactic corona from the spectral analysis (luminosity, temperature, and emission measure) (\S\ref{PaperIsubsec:spec}), we further derive several potentially useful parameters: the electron number density ($n_e$), total mass ($M_{hot}$), thermal energy ($E_{th}$), and radiative cooling timescale ($t_{cool}$) (Table~\ref{table:1Tspec}). The estimation of these parameters depends on the uncertain spatial distribution of the X-ray-emitting gas. We parameterize this uncertainty into an unknown effective volume filling factor $f$ of the gas (e.g., $f=1$ if the distribution is uniform). In high-resolution optical and X-ray observations of starburst galaxies (e.g., NGC~253, \citealt{Strickland00b}), the soft X-ray emission is often observed to be spatially correlated with H$\alpha$ emitting filaments. The hot gas is thus apparently produced at the interface between SN-driven winds and the cooler ambient ISM instead of from the wind material itself \citep{Strickland00a,Strickland00b}. Therefore, at least in such starburst galaxies, the volume filling factor of the X-ray emitting gas is likely far below unity and can only be poorly constrained in few galaxies with deep and high-resolution X-ray observations (e.g., \citealt{Strickland00b}). The filling factor represents a large uncertainty in the estimation of the derived hot gas parameters.

We obtain the average electron number density and the total hot gas mass from the emission measure, assuming a cylinder with its radius and height as defined for the spectral extraction. Using the mass and the average temperature, we further estimate the total thermal energy contained in the volume. We finally use the extinction-corrected luminosity from the best-fit plasma model and the total thermal energy to infer the radiative cooling timescale of the corona.

A few systematic corrections are needed to obtain uniform estimates of the parameters for the coronae. \emph{First}, the fields excluded from the removal of the sources and CCD edges are sometimes non-negligible. For each of the galaxies, we correct for the missed luminosity by the ratio of the areas before and after the removal. \emph{Second}, we correct for the loss of the coronal luminosity due to the filtered out ``disk'' range, by extrapolating the vertical intensity profile into the disk, using the best-fit exponential model as obtained in \S\ref{PaperIsubsec:spatial}. The correction is then the ratio of the accumulated model fluxes, with and without the extrapolation. These two corrections are merged to obtain a total correction factor for the coronal luminosity ($f_L$). Similarly, we obtain the corresponding correction factors for the derived parameters, based on the fitted intensity profiles. We further extrapolate $n_e$ and $t_{cool}$ into the disk, assuming constant temperature and metal abundances. In this case, the soft X-ray intensity is assumed to be proportional to $n_e^2$. $M_{hot}$ and $E_{th}$ are calculated in the same volume as that for the luminosity calculation. All these correction factors are listed in Table~\ref{table:correctionfactor}.

Except for the systematical corrections discussed above, all the directly measured or derived parameters may be affected by other factors. Therefore, their physical interpretation may not be straightforward. Here we discuss the measurement uncertainties and the physical meanings of the derived parameters ($n_e$, $M_{hot}$, $E_{th}$, $t_{cool}$). Our results will be based primarily on the directly measured parameters $L_X$ and $T_X$, which will be more thoroughly discussed in the next subsection (\S\ref{PaperIsubsec:Contaminations}).

\emph{Density}: For each galaxy, the coronal gas density, $n_e$, is inferred from the emission measure and further interpolated into the galactic disk following the soft X-ray intensity profile. Physically, the emission measure anti-correlates with the absolute abundance which is poorly constrained in low-resolution soft X-ray spectroscopy. Therefore, the absolute value of $n_e$ is meaningful only if both the absolute abundance and the volume filling factor $f$ can be well constrained. Fortunately, except for a few extreme cases (e.g., \citealt{Li09}), most of the galaxies are not observed to have very super- or sub-solar abundances (e.g., \citealt{Martin02,Kim04,Humphrey04,Humphrey06,Ji09}). The assumption of solar metallicity of oxygen (\S\ref{PaperIsubsec:spec}) then only has limited effects on the measured $n_e$. Furthermore, due to the possible existence of higher temperature hot gas \citep{Strickland07}, even if a pressure balance is reached, as often assumed in the estimation of $n_e$ (e.g., \citealt{Li08}), the soft X-ray-emitting gas of a galactic corona is still expected to have a volume filling factor far below unity \citep{Strickland00a,Strickland00b,Strickland02}. Therefore, $n_e$ in Table~\ref{table:1Tspec} could be adopted as an lower limit to the real hot gas density (so we have included $f$ in the table).

\emph{Hot gas mass and thermal energy}: Also due to the unknown volume filling factor, $M_{hot}$ and $E_{th}$ as listed in Table~\ref{table:1Tspec} are only upper limits to the real coronal mass and thermal energy. In general, the coronal gas only takes a small fraction of the total galactic baryonic mass, typically $<5\%M_*$ (Tables~\ref{table:basicparaII} and \ref{table:1Tspec}), even for the most X-ray luminous galaxies. Furthermore, much of the feedback material may be stored in a higher temperature phase. In hydrodynamical simulations of galactic scale SN feedbacks, the majority of the feedback energy from either starburst cores or SF inactive galactic bulges is supposed to be stored in a volume-filling hot ($T>10^7\rm~K$) phase of ISM, which triggers the fast galactic wind and transport large amount of energy (but relatively small amount of mass) out of the host galaxy \citep{Strickland00a,Strickland09}. Deep \emph{Chandra} and \emph{XMM-Newton} observations do have revealed such a high-temperature component through the detection of $\sim$6.7~keV FeK$\alpha$ emission line in the starburst core of the nearby superwind galaxy M82 \citep{Strickland07}. However, in most of the cases, the properties of this high-temperature component is still poorly constraint observationally because of its low density and hence low emissivity (e.g., \citealt{Strickland09} and references therein). This means we have detected only a small fraction of the feedback material (the ``missing galactic feedback problem'', see \citealt{Wang10,Mineo12} and references therein), which prevents us from quantitatively understanding the galactic feedback.

\emph{Radiative cooling timescale}: There are two possible fates of the coronal gas, either blown out into the intergalactic space or fall back to the galactic disk. The radiative cooling timescale ($t_{cool}$) and the dynamical timescale of the global gas flows describe the relative importance of these two possibilities. The speed of the galactic outflow, if there is any, cannot be directly measured with the existing X-ray data. However, in simulations of hot gas outflows around starburst galaxies \citep{Strickland00a} or low/intermediate-mass galactic spheroids \citep{Tang09a,Tang09b}, the outflow velocity can often reach several $10^2$ or even $\sim10^3\rm~km~s^{-1}$. This velocity corresponds to a dynamical timescale of $\lesssim10^7\rm~yr$, assuming a corona size of $\sim 10\rm~kpc$. Although $t_{cool}$ is proportional to $f^{1/2}$, it is \emph{unlikely} to be globally smaller than this dynamical timescale (the typical value of $t_{cool}$ is $\sim\rm Gyr$, Table~\ref{table:1Tspec}). Therefore, in starburst and/or low/intermediate-mass galaxies (most of our sample galaxies), outflow likely plays a dominant role. Nevertheless, the hot gas entrained in the outflow often has a broad velocity range. Even in the most energetic cases, a significant fraction of the hot gas may not have sufficiently high velocities (typically a few $\times10\rm~km~s^{-1}$) to overcome the galaxy's gravitational potential \citep{Strickland00a}. This gas tends to interact with the surrounding cool ISM and lost its energy radiatively, largely responsible for the observed soft X-ray emission. It is thus still possible that a significant fraction of the coronal gas may cool down within $t_{cool}$ and fall back to the galactic disk.

%\clearpage
\begin{deluxetable}{lccccc}
\centering
\tiny %\tiny\scriptsize\footnotesize\small\normalsize\large\Large\LARGE\huge\Huge
%\ptlandscape
  \tabletypesize{\tiny}
  \tablecaption{Correction Factors for Measured Hot Gas Properties}
  \tablewidth{0pt}
  \tablehead{
 \colhead{Name} & \colhead{$f_{L}$} & \colhead{$f_{n_e}$} & \colhead{$f_M$} & \colhead{$f_t$} \\
   & (1) & (2) & (3) & (4)
}
\startdata
IC2560 & 1.736 & 2.985 & 1.568 & 0.335 \\
M82 & 1.707 & 2.698 & 1.507 & 0.371 \\
NGC24 & 1.004 & 1.777 & 1.004 & 0.563 \\
NGC520 & 1.018 & 2.244 & 1.018 & 0.446 \\
NGC660 & 1.006 & 2.244 & 1.006 & 0.446 \\
NGC891 & 1.707 & 2.748 & 1.368 & 0.364 \\
NGC1023 & 1.017 & 2.244 & 1.017 & 0.446 \\
NGC1380 & 1.030 & 2.244 & 1.030 & 0.446 \\
NGC1386 & 1.010 & 2.244 & 1.010 & 0.446 \\
NGC1482 & 2.019 & 2.965 & 1.483 & 0.337 \\
NGC1808 & 2.834 & 3.280 & 1.734 & 0.305 \\
NGC2787 & 1.015 & 2.244 & 1.015 & 0.446 \\
NGC2841 & 1.021 & 2.244 & 1.021 & 0.446 \\
NGC3079 & 1.304 & 2.415 & 1.128 & 0.414 \\
NGC3115 & 1.062 & 2.244 & 1.062 & 0.446 \\
NGC3198 & 1.071 & 2.244 & 1.071 & 0.446 \\
NGC3384 & 1.757 & 2.591 & 1.242 & 0.386 \\
NGC3412 & 1.003 & 2.244 & 1.003 & 0.446 \\
NGC3521 & 1.300 & 2.244 & 1.300 & 0.446 \\
NGC3556 & 2.281 & 2.672 & 1.456 & 0.374 \\
NGC3628 & 1.583 & 2.510 & 1.488 & 0.398 \\
NGC3877 & 3.566 & 3.447 & 2.004 & 0.290 \\
NGC3955 & 1.007 & 2.244 & 1.007 & 0.446 \\
NGC3957 & 1.005 & 2.244 & 1.005 & 0.446 \\
NGC4013 & 1.608 & 2.692 & 1.305 & 0.371 \\
NGC4111 & 1.785 & 2.816 & 1.378 & 0.355 \\
NGC4217 & 3.733 & 3.204 & 1.879 & 0.312 \\
NGC4244 & 1.003 & 2.244 & 1.003 & 0.446 \\
NGC4251 & 1.008 & 2.244 & 1.008 & 0.446 \\
NGC4342 & 1.021 & 1.777 & 1.021 & 0.563 \\
NGC4388 & 1.651 & 2.717 & 1.312 & 0.368 \\
NGC4438 & 1.571 & 2.432 & 1.343 & 0.411 \\
NGC4501 & 4.301 & 3.263 & 2.090 & 0.306 \\
NGC4526 & 2.109 & 2.958 & 1.495 & 0.338 \\
NGC4565 & 1.233 & 2.244 & 1.233 & 0.446 \\
NGC4569 & 2.044 & 2.631 & 1.347 & 0.380 \\
NGC4594 & 1.897 & 2.540 & 1.636 & 0.394 \\
NGC4631 & 1.965 & 2.689 & 1.595 & 0.372 \\
NGC4666 & 3.214 & 3.082 & 1.830 & 0.324 \\
NGC4710 & 1.010 & 2.244 & 1.010 & 0.446 \\
NGC5102 & 1.036 & 2.244 & 1.036 & 0.446 \\
NGC5170 & 1.318 & 2.244 & 1.318 & 0.446 \\
NGC5253 & 1.683 & 2.671 & 1.293 & 0.374 \\
NGC5422 & 1.006 & 2.244 & 1.006 & 0.446 \\
NGC5746 & 1.337 & 2.244 & 1.337 & 0.446 \\
NGC5775 & 2.792 & 3.210 & 1.908 & 0.312 \\
NGC5866 & 1.525 & 2.637 & 1.267 & 0.379 \\
NGC6503 & 1.007 & 2.244 & 1.007 & 0.446 \\
NGC6764 & 7.976 & 4.084 & 2.762 & 0.245 \\
NGC7090 & 1.010 & 2.244 & 1.010 & 0.446 \\
NGC7457 & 1.002 & 2.244 & 1.002 & 0.446 \\
NGC7582 & 1.625 & 2.549 & 1.284 & 0.392 \\
NGC7814 & 4.906 & 3.748 & 2.225 & 0.267
\enddata
\tablecomments{\scriptsize All these correction factors should be multiplied to the measured hot gas properties (Table~\ref{table:1Tspec}) to conduct a uniform comparison, i.e., $f_L$ to $L_{hot}$ and $EM$, $f_{n_e}$ to $n_e$, $f_M$ to $M_{hot}$ and $E_{hot}$, $f_t$ to $t_{cool}$. See \S\ref{PaperIsubsec:spec} for details.}\label{table:correctionfactor}
\end{deluxetable}

\subsection{Other Complications}\label{PaperIsubsec:Contaminations}

Except for the systematical corrections already adopted in \S\ref{PaperIsubsec:DerivedParameters}, there are still some other uncertainties which can hardly be quantified. We discuss in this subsection how they may affect our results.

\subsubsection{Thermal and Chemical States}\label{PaperIsubsubsec:ThermalChemicalStates}

The multi-phase nature of the hot gas in nearby galaxies is suggested by both observations (e.g., \citealt{Konami11}) and simulations (e.g., \citealt{Strickland00a,Tang09b}). However, the temperature distribution of this multi-phase gas is usually robustly characterized with the low quality X-ray spectra (e.g., \citealt{Strickland04a,Li08,Owen09}). The fitted temperatures (often with a simple 1-T thermal plasma model) are thought to be significantly biased from the commonly used definitions of average temperature (e.g., luminosity-weighted or emission-measure-weighted temperature; \citealt{Mazzotta04,Vikhlinin06}). Therefore, we need to test how such possible bias may affect our results.

The multi-temperature thermal plasma is most likely produced by the interaction between the SN-driven superwind and the ambient cool ISM \citep{Strickland02}. We thus test a 2-T plasma model for late-type galaxies; early-type ones are typically poor in cool gas and are expected to have a relatively simple thermal state. Fig.~\ref{fig:spec2T} presents the spectra of 16 galaxies for which the 2-T plasma model gives significantly different results from the 1-T model, while Table~\ref{table:specpara2T} lists the fitted parameters.

\begin{figure}[!h]
\begin{center}
\psfig{figure=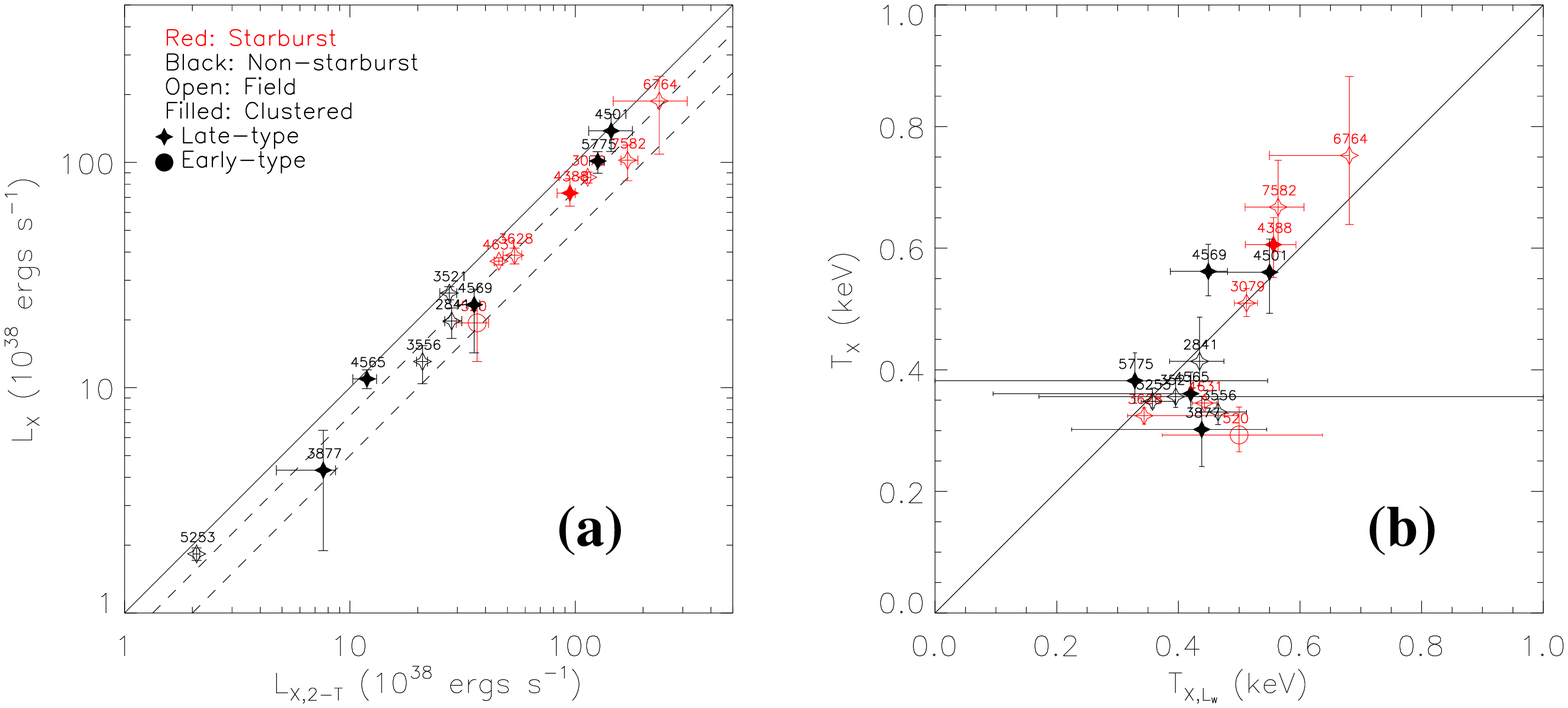,width=1.0\textwidth,angle=0, clip=}
\caption{(a) 0.5-2~keV luminosity of the thermal component of the 1-T model ($L_X$) compared to the total luminosity of the two thermal components of the 2-T model ($L_{X,2-T}$). The solid line marks where $L_X=L_{X,2-T}$, while the two dashed lines marks where $L_X=0.5~L_{X,2-T}$ and $L_X=0.75~L_{X,2-T}$. (b) The temperature measured from the 1-T model ($T_X$) plotted against the luminosity weighted temperature of the 2-T model ($T_{X,L_W}$). The solid line marks where $T_X=T_{X,L_W}$.}\label{fig:test2T}
\end{center}
\end{figure}

In Fig.~\ref{fig:test2T}a, we compare the \emph{luminosities} of the thermal emission obtained from the 1-T ($L_X$) and 2-T ($L_{X,2-T}$) models. While the two luminosities show a tight correlation, $L_{X,2-T}$ is systematically larger than $L_X$, because the power law component tends to be suppressed in a 2-T fit, especially at high energies (Fig.~\ref{fig:spec2T}). Nevertheless, this systematical difference is only about 25\%, typically comparable to or even less than the measurement uncertainty in the luminosity of a corona. Therefore, even if the 2-T plasma model represents a better description of the coronal thermal state, the luminosity obtained from the 1-T fit is still a good approximation of the coronal emission.

We also check the adequacy of the 1-T model in characterizing the \emph{thermal state} of a corona. We define the luminosity weighted temperature of the 2-T model as: $T_{X,L_W}=(T_{low}L_{X,low}+T_{high}L_{X,high})/(L_{X,low}+L_{X,high})$, where subscripts $low$ and $high$ refer to the low-T and high-T components. As shown in Fig.~\ref{fig:test2T}b, $T_{X,L_W}$ is generally consistent with $T_X$, showing no significant bias within the errors. The large errors of $T_{X,L_W}$ for some galaxies are due to the uncertainty in the decomposition of the two thermal components. Therefore, the temperature obtained from the 1-T model typically gives an adequate characterization of the mean thermal state of a corona with the limited quality of the data.

The galactic coronae, and even the IGM, are thought to be metal enriched by various types of stellar ejecta (e.g., massive stellar wind, planetary nebula, and SN ejecta). Therefore, the metal abundance pattern of the coronal gas can be affected by many factors, such as the competition of different metal sources and how different phases of gas mix with each other (e.g., \citealt{Gibson97,Kim04,Humphrey06}). Different abundance ratios (e.g., Fe/O) have been observed in different types of galaxies. Some of the ratios are significantly different from the solar values, indicating recent enrichment (e.g., \citealt{Martin02,Humphrey04,Li09,Ji09}). However, the absolute metal abundance of the coronal gas is typically poorly constrained because the continuum, dominated by the bremsstrahlung of free elections from hydrogens, can hardly be detected in soft X-ray at the low resolution of a typical CCD spectrum; grating spectra, as can be obtained for some relatively compact bright galaxies with deep \emph{XMM-Newton} RGS observations, cannot help much, because the continuum emission from point-like sources cannot be easily subtracted (e.g., \citealt{Liu12}). The Fe/O ratio (Fe and O are often employed as the most important products of Ia and CC SNe, representing the feedbacks from old and young stellar populations) can be readily measured; but its interpretation can still be complicated. For example, \citet{Ji09} show that the uncertainty in the thermal state (e.g., the need for a second temperature component to represent the temperature range of the hot gas) can significantly affect the measured iron abundance (the Fe-bias), but little on the oxygen abundance. In a typical 1-T model, this effect can increase the uncertainty of the measured Fe/O value by as large as $\sim30\%$.

\begin{figure}[!h]
\begin{center}
\psfig{figure=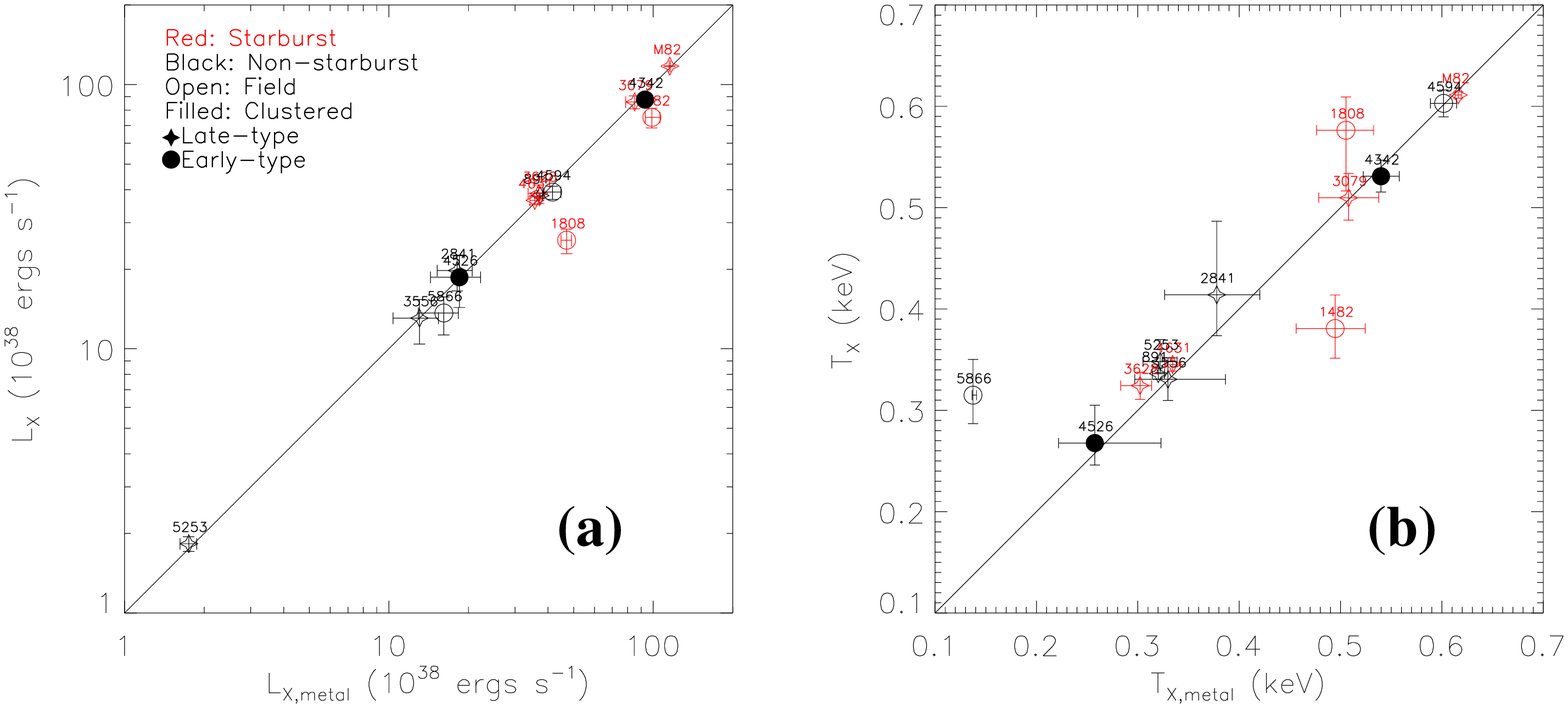,width=1.0\textwidth,angle=0, clip=}
\caption{(a) 0.5-2~keV luminosity of the thermal component of the 1-T model with ($L_X$) or without ($L_{X,metal}$) the abundance ratio fixed. The solid line marks where $L_X=L_{X,metal}$. (b) Similar to (a), but for the temperature.}\label{fig:testmetal}
\end{center}
\end{figure}

All the above uncertainties in the metal abundance measurement may strongly affect the estimation of other coronal properties (e.g., \citealt{Strickland04a,Grimes05,Tullmann06a}). We then check how the assumption of Fe/O ratio in the 1-T model (\S\ref{PaperIsubsubsec:SpectraAnalysis}) affects the measurements of the luminosity and temperature. Fig.~\ref{fig:testmetal} shows that the luminosity and temperature with or without the Fe/O ratio fixed are generally consistent with each other. The only exception is NGC~5866. For this galaxy, the temperature fitted with the abundance ratio fixed is substantially greater than that if it is allowed to vary. Indeed, the galaxy has been reported in a detailed study to have a strong Fe~L line bump while no detectable oxygen lines \citep{Li09}. More sensitive spectroscopic observations are needed to understand such an unusual abundance pattern in this cool-gas-rich S0 galaxy.

In addition to the uncertainty in the thermal and chemical states, a probably more important factor in the measurement of hot gas properties arises from the residual stellar contribution. The model of the stellar component (CV+AB, as well as the power law accounting for both the unresolved LMXB and the residual background) has both thermal and power law components with different relative contributions in different galaxies (\S\ref{PaperIsubsec:spec}). In addition, the specific contribution of faint stellar sources in soft X-ray (per stellar mass) may also vary in different types of galaxies (e.g., due to different SF history). This uncertainty can be as large as $\sim70\%$ \citep{Revnivtsev07,Revnivtsev08,Bogdan11,Boroson11}. Therefore, for galaxies in which a large fraction of the ``diffuse'' X-ray emission comes from unresolved stellar sources, the hot gas properties obtained from the spectral analysis can have large systematic errors (e.g., for NGC~3115, \S\ref{PaperIsec:Individual}, \citealt{David06,Li11}).

Our focus here is at characterizing the average properties of the coronae (luminosity, average temperature, and Fe/O ratio), instead of studying the specific thermal and chemical states (temperature distribution and absolute abundance of different elements). Therefore, we believe that a simplified 1-T model with fixed abundance ratio plus stellar contributions is an optimal choice for a unform, statistical analysis of the galactic coronae.

\subsubsection{Inclination Angle}\label{PaperIsubsubsec:Inclination}

\begin{figure}[!h]
\begin{center}
\epsfig{figure=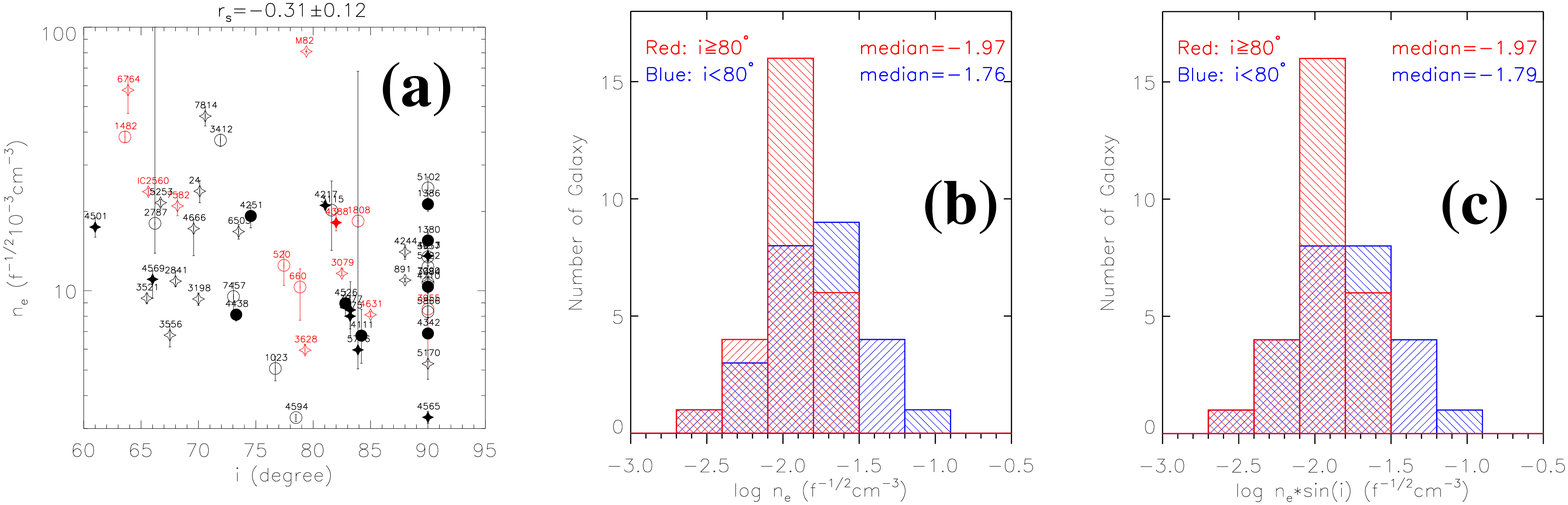,width=1.0\textwidth,angle=0, clip=}
\caption{(a) Hot gas density ($n_e$) vs. inclination angle ($i$), where the symbols have the same meanings as those in Fig.~\ref{fig:test2T}. (b) $n_e$ distributions of highly ($i\geq80^\circ$) and moderately ($i<80^\circ$) inclined galaxies, where the median value of $\log n_e$ for each subsamples is given at the upper right corner. (c) Similar to (b), but with a simple geometry correction for the volume used in the calculation of $n_e$.}\label{fig:test_incli}
\end{center}
\end{figure}

Our measurement may also depend on the inclination angle of a galaxy. The diverse of inclination of the present sample may affect the volume estimate, the filtering of the galactic disk, and so the correction of the measurement. We have checked the dependence of the measured hot gas properties on the inclination angle ($i$) to investigate any possible inclination bias. As an example (the most significant dependence on $i$), Fig.~\ref{fig:test_incli}a shows that $n_e$ seems to be slightly correlated with $i$; but the correlation is only marginal ($-0.31\pm0.12$) and probably dominated by only a few galaxies with high $n_e$ values. Fig.~\ref{fig:test_incli}b further indicates that the difference between highly and moderately inclined galaxies (divided by $i=80^\circ$) may be real, but a simple geometric correction cannot account for this difference (Fig.~\ref{fig:test_incli}c). This possible systematical bias, if not dominated by several extreme cases, could be attributed to the contamination from outer disk emission in moderately-inclined galaxies (\S\ref{PaperIsubsec:spatial}, \ref{PaperIsubsec:spec}). Nevertheless, such potential inclination biases are generally small, compared to the measurement uncertainty itself, as well as uncertainties caused by the intrinsic diversities in the coronal morphologies (e.g., the lopsidedness) or physical/chemical states. Therefore, we have made \emph{no} attempt to systematically correct for any inclination effects on our measurements. We will instead compare high and moderate inclined galaxies in the subsequent statistical analysis (Paper~II).

\subsubsection{AGN Effects}\label{PaperIsubsubsec:AGN}

We further consider the potential effects of AGN on the \emph{coronal} properties. AGN may affect our analysis in two different ways, either via the scattered light of the central X-ray bright point source, or via various forms of feedback (ejection or heating of the surrounding gas caused by the radiation, winds and/or jets, \citealt{Fabian12}). As presented in \S\ref{PaperIsec:SampleSelection}, none of our sample galaxy has nuclear source bright enough that its scattered light (wings of PSF and CCD readout steak) may dominate the truly diffuse coronal X-ray emission. In fact, we benefit from the large inclination angle in the sample selection that most of the X-ray emission from AGN may have already been blocked by the cool gas in the disk.

On the other hand, various forms of feedback from the AGN may affect the galaxy environment significantly (e.g., \citealt{Fabian12} and references therein), so potentially affect our X-ray analysis of the galactic coronae. Generally, there are two major modes of AGN feedback (e.g., \citealt{Ciotti10}). The \emph{radiative mode} (also known as the quasar or wind mode) mainly operates when the accreting black hole is close to the Eddington limit. Although it is sometimes suggested that the radiation from AGN could be responsible for some extended X-ray features even as far as several kpc from the nucleus, by the photo-ionization and fluorescence of the gas (e.g., \citealt{Young01}), this mode typically has the most significant effect on the dense cool gas in the inner part of a galaxy. Little evidence is currently revealed for the direct interaction between these relatively homogenous wind and the large scale hot gas. We thus neglect it from the direct effects on the corona.

The other mode is known as the \emph{kinetic mode} (or jet/radio mode). There are many observational evidence (e.g., X-ray jet and filaments) for the direct interaction between the jet and the hot halo (e.g., \citealt{Kraft00,Kraft02,Forman05,Forman07}). In addition, many AGN blowout superbubbles are also detected in the hot surrounding medium of \emph{massive} galaxies located in the center of cool core clusters, which plays an important role in shaping the circumgalactic environment (e.g., \citealt{Kraft07,Kraft09,David09}). There are also some lines of evidence that the diffuse X-ray morphology is correlated with the radio/X-ray luminosities of AGNs in normal elliptical galaxies, which may indicate that the coronal gas is disturbed by AGN and thus departs from hydrostatic equilibrium with the gravitational potential \citep{Diehl07,Diehl08}. Such a radio mode AGN feedback may even be energetic enough to unbind a significant fraction of the IGM in galaxy groups \citep{Giodini10}, so explain the breaking of self-similarity observed in the scaling relations between the gravitational mass and some hot gas properties of massive clusters of galaxies (e.g., \citealt{Ponman99,Ponman03,Vikhlinin02,Sanderson03a,Sanderson03b,Maughan12}).

In spite of these lines of observational evidence, which are mostly for massive elliptical galaxies, we do not find much evidence for prominent AGN feedback features in our sample of disk galaxies. There are several exceptions (\S\ref{PaperIsec:Individual}), such as the $\sim$kpc scale limb-brightened nuclear bubble of NGC~3079 \citep{Cecil02}, the small nuclear bubble of NGC~4438 \citep{Machacek04}, the coincidence between the radio bubble and diffuse X-ray emission of NGC~6764 \citep{Croston08}, etc. In most cases, such structures do not significantly affect our measurements of the global properties of the relevant individual coronae. Furthermore, AGN feedback is often \emph{energetically} less important (than SN feedback) in galaxies with small or intermediate mass bulges (e.g., \citealt{David06}), as is the case for most of the present sample galaxies (refer to \S\ref{PaperIsec:SampleSelection} for details). We thus believe that AGN feedback only have limited \emph{direct} effect on the coronal X-ray emission.

In addition to the \emph{direct} effect on the coronal X-ray emission, AGN feedback could also have strong effects on various galaxy properties, which may then affect the coronae \emph{indirectly}. For example, the intense flux of photons and particles produced by the AGN may sweep the cool gas in the galaxy, or at least keep the gas hot and reduce radiative cooling, therefore quench the SF (e.g., \citealt{Croton06,McNamara06,Schawinski09,Farrah11}). Strong AGN feedback can also reduce the accretion and stifle the fuel supply onto the supermassive BH, then terminates the nuclear activity (e.g., \citealt{Murray05,Debuhr11,Power11}). All these changing of galaxy properties (e.g., SF and nuclear activity) may have potential effects on the coronal properties, e.g., by changing the stellar feedback rate or the global accretion rate from the IGM (e.g., \citealt{Fabian12} and references therein). Nevertheless, here we only consider the direct interaction between AGN and the coronal gas (such as heating, mass injection, and compression, etc.; e.g., \citealt{Young01,Croston08,Giodini10}), while leaving the discussions of the scatters caused by various indirect effects to the correlation analysis between the coronal properties and various galaxy properties (Paper~II).

\begin{figure}[!h]
\begin{center}
\epsfig{figure=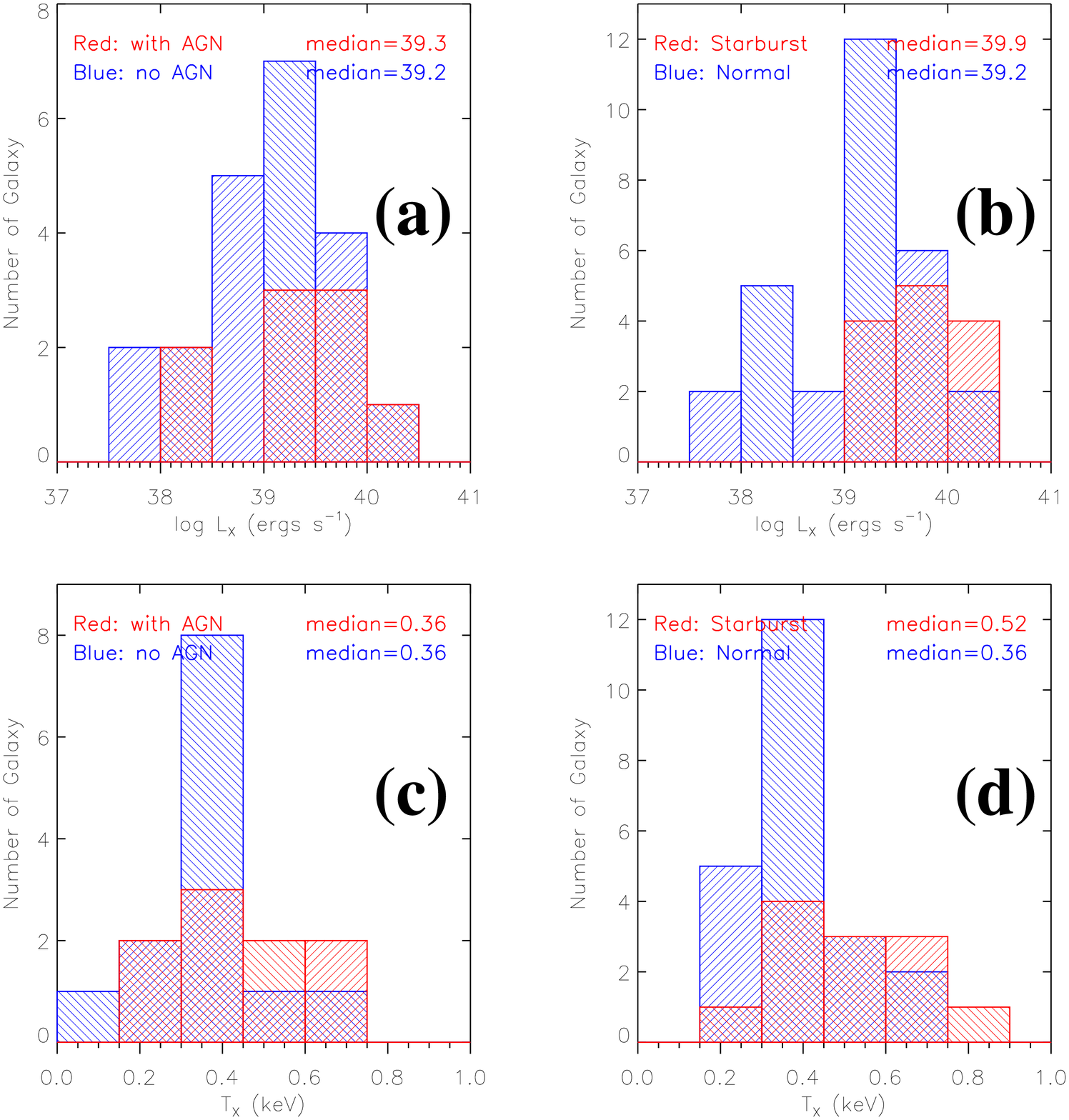,width=0.95\textwidth,angle=0, clip=}
\caption{Luminosity (a,b) and temperature (c,d) distributions of subsamples: (a,c) AGNs vs. non-AGN, as classified in \citet{Bettoni03}; (b,d) starburst vs. non-starburst. The median parameter values of each subsample is given at the upper right corner of the respective panel.}\label{fig:testAGN}
\end{center}
\end{figure}

It may be helpful to statistically compare the coronal X-ray luminosity and temperature measurements of the galaxies with or without AGN (Fig.~\ref{fig:testAGN}; the classification of the galaxies plotted in the figure is based on \citealt{Bettoni03}). We find that the median values of $L_X$ and $T_X$ for these two subsamples do not show significant differences ($\sim0.1\rm~dex$ higher in $L_X$ for galaxies with AGNs and no noticeable difference in $T_X$). In contrast, starburst and non-starburst galaxies are indeed distinguishable statistically (different by 0.7~dex in $L_X$ and 0.16~keV in $T_X$). We thus conclude that \emph{AGNs are generally less important than SF in regulating the coronal properties of our sample galaxies}.

\subsection{Possible Contribution from CXE}\label{PaperIsubsec:CXE}

There are growing lines of evidence that the CXE (also called charge transfer) makes a non-negligible contribution to the X-ray line emission in various sorts of astrophysical objects (see \citealt{Dennerl10} for a recent review). The CXE describes a process in which a highly ionized ion (e.g., \ion{O}{8}) takes one or more electrons from another atom (like H, $\rm H_2$, or He), which can be represented by: $A^{q+}+N\rightarrow A^{(q-1)+*}+N^+$. Such a process will produce an excited ion $A^{(q-1)+*}$, which emits X-ray photons as it decays to the ground state. Extensive studies have been carried out on the solar wind CXE and its contribution to the X-ray emission of some solar system objects, as well as to the Galactic soft X-ray background (e.g., \citealt{Snowden04,Koutroumpa11}). Indeed, the CXE has further been speculated to be important in many other types of astrophysical objects (e.g., \citealt{Lallement04}). According to the diffuse X-ray emission from nearby galaxies, it has been shown that a substantial fraction of it cannot arise from optically-thin thermal plasma, but may originate in the CXE (e.g., \citealt{Ranalli08,Liu10,Liu11,Liu12,Konami11}). This scenario is also consistent with the spatial correlation between some cool/hot gas features, especially in active SF galaxies (e.g., \citealt{Wang01,Strickland07,Liu12}).

Existence of the CXE will not only affect the spectral modeling of specific line emissions, but more important to this work, also potentially significantly enhance the global soft X-ray emission. The CXE produces only line emission, which cannot be well separated from the thermal emission in a low resolution CCD spectrum. For example, the relative strengthes of the forbidden ($0.5611\rm~keV$) and inter-combination ($0.5686\rm~keV$) lines to the resonance line ($0.5740\rm~keV$) of the \ion{O}{7}~K$\alpha$ triplet is often adopted as a spectral diagnostic for the presence of the CXE \citep{Liu10,Liu11}. While the CXE is closely related to the presence of cool gas, the amount of cool gas cannot be used as a direct diagnostic parameter of the CXE strength. This is because the amount of cool gas is often tightly correlated with the SFR (e.g., Fig.~\ref{fig:SFtracer}b), or other cool-hot gas interaction processes, such as mass loading and turbulent mixing, which all enhance radiative cooling (e.g., \citealt{Li11}). Nevertheless, since all these processes are more or less related to SF, we may expect that some enhanced cool/hot gas features, especially in low-density environments (where the thermal emission is relatively weak), may originate in the CXE.

A few apparent candidates for significant CXE contributions present in our sample. The analysis of the \emph{XMM-Newton} RGS spectra of M82 has shown that the contributions of the CXE are about 90\%, 50\%, and 30\% to the \ion{O}{7}, \ion{Ne}{4}, and \ion{Mg}{6} triplets, respectively \citep{Liu11}. Although the grating spectral analysis is focused chiefly on the central region (typically $\lesssim1^\prime$ from the nucleus), it is still expected that a considerable fraction of the coronal soft X-ray line emission may arise from the CXE. In particular, M82 is located within a complex of \ion{H}{1} gas, formed apparently from the tidal interaction among members of the M81 group \citep{Yun94}; a large number of interfaces, generated by expected strong interaction between the gas and the galactic superwind, may lead to strong CXE emission. Similarly, for the disk-wide starburst galaxy, NGC~4631, the measured line ratio is also consistent with a CXE origin \citep{Liu12}.

Another possible example of a significant CXE contribution is NGC~4438. This galaxy is unusually X-ray bright ($L_X\sim5\times10^{39}\rm~ergs~s^{-1}$, compared to $\sim7\times10^{39}\rm~ergs~s^{-1}$ of M82 and $\sim2\times10^{39}\rm~ergs~s^{-1}$ of NGC~4594; Table~\ref{table:1Tspec}) for its moderate stellar mass ($\sim3\times10^{10}\rm~M_\odot$, compared to $\sim1.6\times10^{11}\rm~M_\odot$ of NGC~4594) and SFR ($\sim0.5\rm~M_\odot~yr^{-1}$, compared to $\sim7.7\rm~M_\odot~yr^{-1}$ of M82). It is located in the Virgo cluster and shows tidal interaction with a companion galaxy NGC~4435 (\S\ref{PaperIsec:Individual}). In addition, the cool/hot gas features are also spatially correlated \citep{Machacek04}. In contrast, galaxies undergoing tidal interaction and cool gas transfer from companions, but located in non-cluster environments (e.g., NGC~5775), do not have such well-matched cool/hot gas features (e.g., \citealt{Li08}). We thus speculate that the enhanced X-ray emission may at least partly arise from the
CXE between the stripped cool gas and the hot ICM.

Due to the scarcity of high-quality, high-resolution spectroscopic observations and the lack of suitable spectral models \citep{Liu12}, the contribution of CXE to the total diffuse X-ray luminosity can hardly be constrained in individual galaxies or in a sample. However, at least in some cases, this component appears to be non-negligible.

\section{Summary}\label{PaperIsec:summary}

We have conducted a systematical data reduction and analysis of the \emph{Chandra} data of 53 nearby disk galaxies, of which 20 are presented for the first time to study the \emph{diffuse} soft X-ray emission. This sample covers broad ranges of SFR ($\sim0.01-10\rm~M_\odot~yr^{-1}$), stellar mass ($\sim5\times10^8-1.5\times10^{11}\rm~M_\odot$), and other galaxy properties. The sample size is much larger than those of previous studies of similar types of galaxies (typically $\lesssim10$ galaxies). We have conducted uniform data calibration of all the galaxies. In particular, discrete point-like sources are detected and removed, while significant contributions from unresolved stellar sources are carefully estimated for different types of galaxies, allowing for a spectral analysis of the truly diffuse hot gas emission. Based on these calibrations, we have provided measurements of various diffuse soft X-ray properties of the galactic coronae, including both the directly measured parameters from the spatial and spectral analysis (scale height, luminosity, and temperature), and some derived parameters (density, mass, thermal energy, and radiative cooling timescale). Notable properties are highlighted for individual galaxies. Our products (images, spectra, brightness profiles, and measured hot gas parameters) form a comprehensive database for the study of the galactic coronae.

We have also discussed various corrections and complications in the measurements. In particular, we have investigated the effects caused by uncertainties in the thermal and chemical states of the hot gas. We find that a simple 1-T model with fixed abundance ratio plus stellar contributions is typically sufficient to characterize the overall properties of the coronal emission. The different inclination angles of our sample galaxies may cause a small systematical bias due to the potential contamination from the disk emission/absorption in moderately inclined galaxies. But this bias cannot play a significant role in determining the coronal properties. Several galaxies in our sample host AGNs; some of them may even be responsible for distinct diffuse X-ray features. But in general, AGN feedback is less important than SF feedback for our sample of disk galaxies; the presence of the diffuse X-ray features related to AGN does not significantly affect our measurements of the global coronal properties. The CXE may significantly contribute to the soft X-ray emission in some of the galaxies, either active in SF or not. In particular, the unusually high X-ray luminosity of the tidally disturbed gas-rich Virgo cluster member NGC~4438 is most likely due to the CXE. Further insights into the origin of the coronae can be obtained from a statistical analysis of the measurements, together with other galaxy parameters, which is the topic of Paper~II.

\acknowledgements

We thank the referee for his/her constructive comments and suggestions, which are very helpful in the revision of the paper. This work is supported by NASA through the CXC/SAO grant AR0-11011B, and the ADAP grant NNX12AE78G.

\end{document}